\documentclass[fleqn,usenatbib]{mnras}

\usepackage{newtxtext,newtxmath}
\usepackage{bm}

\usepackage[T1]{fontenc}

\DeclareRobustCommand{\VAN}[3]{#2}
\let\VANthebibliography\thebibliography
\def\thebibliography{\DeclareRobustCommand{\VAN}[3]{##3}\VANthebibliography}

\usepackage{graphicx}	
\usepackage{amsmath}	
\usepackage{amssymb}	

\title[X-ray Pulsars Polarisation I]{Polarisation of Accreting X-ray Pulsars. I. A New Model}

\author[I. Caiazzo \& J. Heyl]{
Ilaria Caiazzo,$^{1,2}$\thanks{E-mail: ilariac@caltech.edu}
Jeremy Heyl,$^{2}$
\\
$^{1}$TAPIR, Walter Burke Institute for Theoretical Physics, Mail Code 350-17, Caltech, Pasadena, CA 91125, USA\\
$^{2}$Department of Physics and Astronomy, University of British Columbia, Vancouver, BC V6T 1Z1, Canada
}

\date{Accepted XXX. Received YYY; in original form ZZZ}

\pubyear{2020}

\begin{document}
\label{firstpage}
\pagerange{\pageref{firstpage}--\pageref{lastpage}}
\maketitle

\begin{abstract}
A new window is opening in high-energy astronomy: X-ray polarimetry. With many missions currently under development and scheduled to launch as early as 2021, observations of the X-ray polarisation of accreting X-ray pulsars will soon be available. As polarisation is particularly sensitive to the geometry of the emission region, the upcoming polarimeters will shed new light on the emission mechanism of these objects, provided that we have sound theoretical models that agree with current spectroscopic and timing observation and that can make predictions of the polarisation parameters of the emission. We here present a new model for the polarised emission of accreting X-ray pulsars in the accretion column scenario that for the first time takes into account the macroscopic structure and dynamics of the accretion region and the propagation of the radiation toward the observer, including relativistic beaming, gravitational lensing and quantum electrodynamics. In this paper we present all the details of the model, while in a companion paper, we apply our model to predict the polarisation parameters of the bright X-ray pulsar Hercules X-1.
\end{abstract}




\section{Introduction}



Accreting X-ray pulsars are highly magnetised neutron stars that live in a binary and accrete material from a companion star. The material, mostly ionised hydrogen, becomes unbound from the companion (either because the star exceeds its Roche lobe, or because of strong winds) and becomes gravitationally bound to the neutron star. As the material gets closer to the compact object, it forms an accretion disk, and when it reaches the surface of the neutron star, the kinetic energy of the accretion flow is converted into X-ray emission. The pulsating nature of the X-ray emission was interpreted quickly after their discovery \citep{1962PhRvL...9..439G,1971ApJ...167L..67G,1972ApJ...174L.143T} as resulting from the channelling along magnetic field lines of accretion gas onto the magnetic poles of the neutron star \citep{1970Ap......6..214A,1972A&A....21....1P}. However, it was immediately clear that the high pulse fraction detected was impossible to explain merely by the presence of isotropically emitting hot spots on the surface of the rotating neutron star, and that a strong beaming of the radiation was required \citep{1973A&A....25..233G}. A possible beaming mechanism is naturally provided by the presence of a strong magnetic field: the cross-sections of the elementary processes of interaction between radiation and matter have a strong dependence on the angle between the magnetic field and the propagation direction of the photons, and at small angles with respect to the direction of the magnetic moment one can see deeper in the atmosphere. If the kinetic energy of the in-falling material is deposited deep in the atmosphere, then the emission from the hot spots will have a characteristic ``pencil'' beam pattern \citep{1973A&A....25..233G,1975PASJ...27..181D,1975A&A....42..311B}. An alternative model invokes the presence of a radiative shock above the surface of the neutron star, in which the in-falling gas is slowed down considerably by radiation before reaching the surface and an accretion column is formed above the magnetic pole in which the ionised gas is slowly sinking. In this second scenario, the photons escape from the walls of the column and the emission has a ``fan'' beam pattern \citep{1973ApJ...179..585D,1976MNRAS.175..395B,1991ApJ...367..575B,2007ApJ...654..435B}. Due to the low resistivity, the depth to which the plasma penetrates into the dipole field is small compared to the magnetospheric radius, and the accretion channel could be similar to a thin wall of a funnel with a large radius or a solid, axisymmetic column with a small radius. Both the hot-spot and the column scenario could be present in different pulsars, and the discriminant is thought to be the X-ray luminosity: for low luminosity, the gas can free-fall all the way to the surface of the star, creating a hot-spot, while above a critical luminosity, radiation pressure is able to slow the gas above the surface, and hence the column is formed \citep{1976MNRAS.175..395B}. In a recent work, \citet{2015MNRAS.447.1847M} demonstrated that the critical luminosity is a non-monotonic function of the magnetic field strength \citep[see also][]{2015MNRAS.454.2539M}, while \citet{2017MNRAS.466.2143D} detected what could possibly be a transition between the two geometries in the accreting pulsar V 0332+53. In this work, we focus on the accretion column scenario.

The continuum X-ray emission of accreting X-ray pulsars is often described by phenomenological models, including an absorbed power law extending up to $\sim100$ keV with a roll-over at $\sim30-50$ keV or a broken power law \citep{2006AdSpR..38.2742O,2016A&A...591A..29F}. Several attempts have been made to develop spectral models that link the X-ray emission to the accretion physics \citep{1973ApJ...179..585D,1985ApJ...298..147M,1985ApJ...299..138M,1980ApJ...236..911Y,1981ApJ...251..278N,1987pasj...39..781k,1996ApJ...457L..85K}, but the modelling is complicated by the fact that the accretion regions are radiation-dominated, which means that the radiation transfer is coupled with the hydrodynamics of the flow; by the presence of a relativistic bulk motion in the in-falling gas, which in turns makes the modelling of Compton up-scattering more difficult; and by the strong magnetic field, which changes all the cross sections for scattering and absorption. All these complications should be addressed self-consistently and the study of the polarisation parameters should be tailored to the spectral formation model. The polarisation of X-ray photons provides two additional observables, polarisation degree and angle, which are extremely sensitive to the geometry of the emission regions and to the structure of magnetic fields, and can therefore be a powerful tool to understand X-ray pulsars.

Among these attempts of calculating the spectral formation from a physical accretion model, \citet{1982Ap&SS..86..249K}, \citet{1981ApJ...251..278N,1981ApJ...251..288N}, \citet{1980ApJ...238.1066M} \citet{1985ApJ...298..147M,1985ApJ...299..138M},  \citet{1986PASJ...38..751K} and \citet{1987pasj...39..781k} have addressed the problem of polarisation. These models solve the problem of radiative transfer separately for the two polarisation modes (parallel and perpendicular to the magnetic field) and therefore calculate at the same time the flux and the polarisation degree of the emitted radiation. Their calculations assume a static, homogeneous atmosphere (with constant density, temperature and magnetic field) and two possible geometries: a slab, with the magnetic field perpendicular to the surface, and a column, with the field parallel to the walls. In order to calculate the spectrum of the outgoing polarisation, they solve the approximate radiative transfer equations separately for the two polarisation modes following the so-called Feautrier method \citep{1978stat.book.....M}, including vacuum, thermal and incoherent scattering effects. In these models, photons are mainly produced by thermal bremsstrahlung, and the polarisation of the X-ray signal is driven by the difference in opacities between the two polarisation modes. However, the models do not include relativistic effects and assume a static atmosphere, even though the ionised plasma is expected to reach the surface of the neutron star at a considerable fraction (up to $\sim0.5$) of the speed of light. Moreover, the spectral shape obtained \citep[][for example]{1985ApJ...298..147M} fails to describe the more recent observations of luminous X-ray pulsars \citep[e.g.][]{2016ApJ...831..194W}, especially the flattening of the spectrum at low energies.

The model of spectral formation developed by \citet{2005ApJ...621L..45B,2007ApJ...654..435B}, hereafter the B\&W model, has been the most successful at reproducing the observed phase-averaged spectra from accreting X-ray pulsars starting from an analysis of the accretion physics. In their model, which considers the emission from an accretion column in the ``fan beam'' context, the directional dependence of electron scattering is treated in terms of mode-averaged cross sections, and therefore the problems of polarisation and of radiative transfer are considered separately, and no information is given on the polarisation of light. The model, however, calculates for the first time the X-ray spectrum from the solution of a coupled radiation and hydrodynamic transport equation, and predicts a spectrum that fits very well the observed profiles and provides insights on the physical properties of the accretion flow, such as the width of the column, the temperature of the gas and the strength of the magnetic field. \citet{2017ApJ...835..129W,2017ApJ...835..130W} refined the B\&W model with the implementation of a realistic dipole geometry, instead of using a cylinder with straight walls, and a self-consistent treatment of the energy transfer between electrons, ions, and radiation to calculate the temperature profile of the electrons in the column. 

A model that agrees with current observations and that predicts the expected polarisation characteristics of accreting X-ray pulsars is needed, as X-ray polarimetry will open a new window on compact objects very soon. Several observatories with an X-ray polarimeter on board are now at different stages of development: in the 1--10~keV range, the NASA SMEX mission \textit{IXPE} \citep{2016SPIE.9905E..17W}, scheduled to fly in 2021, and the Chinese--European \textit{eXTP} \citep{2016SPIE.9905E..1QZ}, launching in 2025; in the medium range, 5-30 keV, the Indian \textit{POLIX}, scheduled for launch in 2021~\citep{2016cosp...41E1533P,2011ASInC...3..165V}; in the hard-X-ray range, 15--150 keV, the balloon-borne \textit{X-Calibur}~\citep{2014JAI.....340008B} and \textit{PoGO+} \citep{2018MNRAS.tmpl..30C}; and, in the sub-keV range, the narrow band (250 eV) \textit{LAMP}~\citep{2015SPIE.9601E..0IS} and the broad band (0.2--0.8 keV) rocket-based \textit{REDSox} \citep{SPIE_REDSoX}. In addition, the broad-band 0.2-60~keV polarimeter XPP, still at the concept stage, will be an incredible tool to study X-ray pulsars \citep{2019arXiv190409313K,2019arXiv190710190J}. Many of these missions will employ gas pixel detectors, which have been recently used in \textit{PolarLight} \citep{2019ExA....47..225F}, a proof of concept polarimeter on board of a CubeSat that was able to measure the X-ray polarisation of the Crab Nebula \citep{2020NatAs.tmp..100F}.

This is the first in a series of papers in which we present a new model for the polarised emission of X-ray pulsars in the context of the B\&W model of spectral formation. Contrary to previous models, who solve the radiative transfer equations separately for the two polarisation modes (parallel and perpendicular to the magnetic field), assuming that they stay distinct as the photons propagate out of the column, we keep track of the full Stokes vector and we calculate the polarisation parameters independently of the radiative transfer solution. In this way, we can keep our calculation fully analytical, and we can make predictions on the polarisation degree and angle of the X-ray emission as a function of the geometry of the system and of a few parameters that can be obtained by spectral fitting in the context of the B\&W model. We will therefore be able to simultaneously fit our model to the spectral and polarisation signals, knowing that the spectral shape agrees with current observations.
Furthermore, our model includes the structure and dynamics of the column and the propagation effects for the first time.

In \S~\ref{sec:B&W}, we summarise the accretion geometry and spectral formation that we assume in the context of the B\&W model. We then explain the formalism at the base of our polarisation model (\S~\ref{sec:PolB}) and the details of our calculations, in particular how we calculate the polarisation state of radiation inside the column (\S~\ref{sec:PSin}), how we include the effects of special and general relativity (\S~\ref{sec:beaming} and~\ref{sec:lb}) and how we include the effect of vacuum birefringence (\S~\ref{sec:QED}). In \S~\ref{sec:comp} we compare our results with earlier models and in \S~\ref{sec:discussion} we summarise how the different effects affect the polarisation signal. In the accompanying paper (Caiazzo \& Heyl 2020b, hereafter Paper II), we apply our model to predict the polarisation signal of the X-ray pulsar Hercules X-1.

\section{The Model}
Our calculations are performed under the assumption that accretion occurs through the formation of one or two accretion columns at the magnetic poles of the neutron star. In particular, we assume the structure of the column to be as described in the B\&W model, that we introduce below. Our calculation of the polarisation parameters is done independently of radiative transfer; however, the results depend on the structure of the accretion column, specifically on the radius and height of the accretion column, on the strength of the magnetic field, on the velocity profile of the gas inside the column and on the optical depth of the column. All these parameters can be obtained by spectral fitting within the context of the B\&W model. Because this calculation is performed independently of the radiative transfer, we calculate and present the relative intensity $I$ as a function of direction and polarisation, and the final step in fitting the pulsed-averaged spectrum of a particular source using the B\&W model determines the normalisation of the intensity as a function of energy.

\subsection{The Becker and Wolff model}
\label{sec:B&W}

\begin{figure}
    \centering
    \includegraphics[width=0.9\columnwidth]{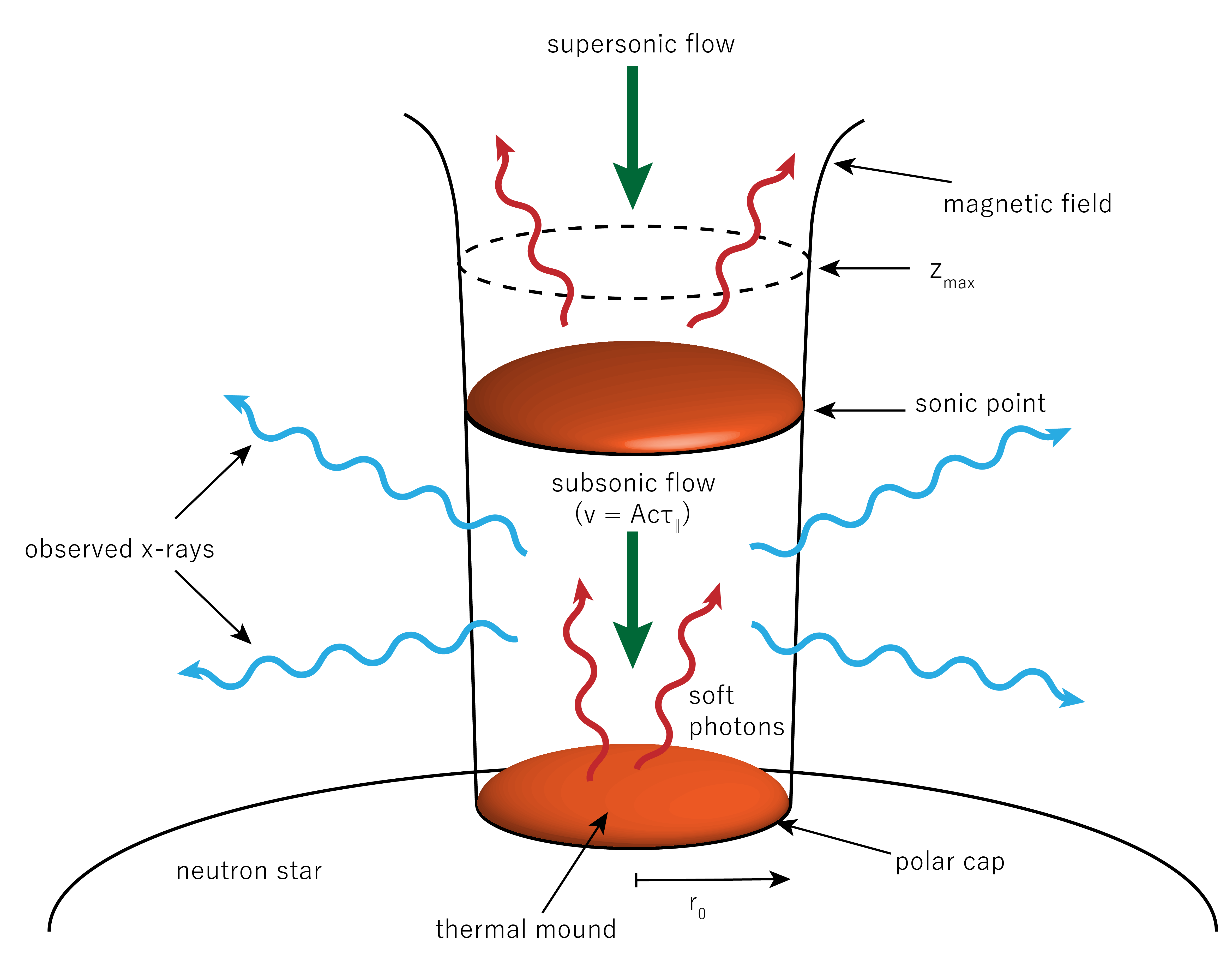}
    \caption{The structure of the accretion column in the B\&W model \citep[adapted from][]{2007ApJ...654..435B}.}
    \label{fig:B&Wcol}
\end{figure}

In their 2007 paper, \citet{2007ApJ...654..435B} proposed a new model for spectral formation in luminous X-ray pulsars that quite successfully reproduces the phase-averaged spectrum of bright X-ray pulsars as Hercules X-1 (Her X-1). In the model, the ionised gas accreted from the companion star is funneled inside a column at the polar caps of the neutron star. The strong magnetic field keeps the gas confined inside the column as in a  ``pipe'', which is however transparent for radiation. Fig.~\ref{fig:B&Wcol} shows the geometry of the accretion column: the ionised gas freely falls from the accretion disk along the field lines to the top of the column, where the speed of the flow is supersonic; inside the column, radiation pressure slows down the gas until it comes to rest at the bottom of the column. Seed photons in the column are produced by a combination of bremsstrahlung, cyclotron and blackbody radiation, and are scattered by electrons through Compton scattering. Blackbody photons are emitted by a thermal mound at the bottom of the column, so that the mound's surface represents the photosphere for creation and absorption of photons and the opacity in the rest of the column is given by electron scattering only. Bremsstrahlung and cyclotron photons are emitted throughout the column. The observed radiation comes from the walls of the column, in a ``fan beam''.

The B\&W model compute the X-ray spectrum by solving a radiation transport equation that includes both the effects of \textit{bulk} Comptonisation, for which photons are upscattered in energy through a first-order Fermi energisation, and of thermal Comptonisation. The upstream flow above the column is composed of fully ionised hydrogen moving at supersonic speed, reaching about half the speed of light at the top of the column. Inside the radiation-dominated column, electrons slow down as they transfer energy to the radiation field and stop at the stellar surface. Instead of using the exact solution for the velocity profile derived by \citet{1998ApJ...498..790B}, the authors use a particular form for the velocity profile that approximates the exact solution and also makes the transport equation separable in energy and space: 
\begin{equation}
    v(\tau) = -A \tau_\parallel
    \label{eq:velprofile}
\end{equation}
where $A$ is a constant and $\tau_\parallel$ is the optical depth in the direction parallel to the column vertical axis (and the magnetic field). $\tau_\parallel$ increases vertically, and is equal to zero at the stellar surface (see eq.~\ref{eq:taupar} below). $A$ is calibrated by equating the velocity at the sonic point (which lies just below the top of the column) to the exact velocity, which yields
\begin{equation}
	A = 0.20 \left(\frac{M_*}{M_\odot} \right) \left( \frac{R_*}{10\,\rm{km}}\right)^{-1} \xi , \quad \xi = \frac{\pi r_0 m_p c}{\dot{M}(\bar{\sigma}_\parallel\bar{\sigma}_\perp)^{1/2}}
	\label{eq:velp2}
\end{equation}
where $M_*$ and $R_*$ are the mass and radius of the neutron star, $r_0$ is the radius of the column, $m_p$ is the mass of the proton, $c$ is the speed of light, $\dot{M}$ is the accretion rate, and $\bar{\sigma}_\parallel$ and $\bar{\sigma}_\perp$ are the scattering cross sections for photons travelling parallel and perpendicular to the magnetic field, averaged over photon energy and polarisation state. The dimensionless parameter $\xi$ determines the importance of the escape of photons from the accretion column in the radiation transfer equation.

Since radiation pressure is dominant in the column, the dynamical structure of the flow is closely tied to the spatial and energetic distribution of the radiation, making the coupled radiation-hydrodynamic problem extremely complex. For this reason, in B\&W, the directional dependence of the electron scattering is treated in terms of the constant, energy- and mode-averaged cross sections $\bar{\sigma}_\parallel$ and $\bar{\sigma}_\perp$. In particular, $\bar{\sigma}_\perp $ is set to $ \approx \sigma_T $, the Thomson cross section, while $\bar{\sigma}_\parallel$ is expressed in terms of the accretion rate $\dot{M}$,  the radius of the accretion column $r_0$, and  the dimensionless parameter $\xi$. Both $r_0$ and $\xi$ are free parameters, recovered by fitting the emergent spectrum. The expression used is given in eq.~83 of \citet{2007ApJ...654..435B}:
\begin{equation}
    \bar{\sigma}_\parallel = \left(\frac{\pi r_0 m_p c}{\dot{M}\xi} \right)^2\frac{1}{\sigma_T} \, ,
\end{equation}
and the averaged cross section can be used to calculate the optical depth in the vertical direction:
\begin{equation}
   \tau_\parallel= \frac{\rho \bar{\sigma}_\parallel}{m_p}z
   \label{eq:taupar}
\end{equation}
where $z$ is the vertical coordinate with origin at the bottom of the accretion column.

In B\&W, $\bar{\sigma}_\parallel$ and $\bar{\sigma}_\perp$ are the cross section for photons propagating in the parallel and perpendicular direction with respect to the magnetic field, averaged over the polarisation modes. We do not use these cross sections in our calculations, as we calculate the angle- and energy-dependent cross sections for each mode, so this notation should not be confused with the cross sections for the parallel or ordinary mode ($\sigma_o$) and for the perpendicular or extraordinary mode ($\sigma_x$) that we introduce in \S~\ref{sec:PolB}. We calculate the polarisation signal independently of the radiation transfer calculated in the B\&W model; however, we assume the structure of the accretion column to be as described in the B\&W model, and in particular, the parameters that influence our results are the radius and height of the accretion column, the strength of the magnetic field, the velocity profile of the gas inside the column and the optical depth of the column. This information can be obtained in the context of the B\&W model from the following parameters:
\begin{itemize}
    \item the radius of the accretion column $r_0$; 
    \item the centroid energy at the observer of the cyclotron resonance scattering frequency (CRSF) $E_{\rm{cyc}}$; 
    \item the dimensionless parameter $\xi$ defined in eq.~\ref{eq:velp2}; 
    \item the accretion rate $\dot{M}$;
\end{itemize}
Throughout the paper, we will show an example of our calculation that employs the parameters obtained by fitting the phase-averaged NuSTAR spectrum of for the accreting X-ray pulsar Hercules X-1 (Her X-1) by \citet{2016ApJ...831..194W}, specifically: $r_0 = 107$~m, $E_{\rm{cyc}} = 37.7$~keV, $\xi = 1.36$ and $\dot{M}=2.59\times10^{17}$~g/s. In addition, we employ a value of $M_*=1.5$~M$_\odot$ for the mass and of $R_*=10$~km for the radius of the neutron star and a distance to the source of $D=6.6$~kpc \citep{1997MNRAS.288...43R}.

\subsubsection{What is the height of the column?}
\label{sec:zmax}
The height of the accretion column $z_{\rm{max}}$ in the B\&W model is found by equating the gas velocity at the top of
the column to the local free-fall velocity
\begin{equation}
    \left(\frac{2GM_*}{R_*+z_{\rm{max}}}\right)^{1/2} = cA\tau_{\rm{max}} \quad \rm{where} \quad \tau_{\rm{max}} = \left(\frac{\bar{\sigma}_\parallel}{\bar{\sigma}_\perp}\right)^{1/4}\left(\frac{2z_{\rm{max}}}{A\xi r_0}\right)^{1/2}
\end{equation}
And therefore
\begin{equation}
    z_{\rm{max}}=\frac{R_*}{2}[(1+C_1)^{1/2} - 1] \quad {\rm{where}} \quad C_1 = \frac{4GM_*r_0\xi}{A c^2R^2_*}\left(\frac{\bar{\sigma}_\perp}{\bar{\sigma}_\parallel}\right)^{1/2}\,.
    \label{eq:zmax}
\end{equation}
Imposing the velocity profile to be equal to the free-fall velocity at the top of the column translates into assuming that the radiative shock extends to the entire column, where the electron velocity smoothly changes from free-fall to zero, without any discontinuity. A different assumption on the velocity profile or on the nature of the shock can lead to a different height. For example, another possibility for a stationary solution would be assuming that a strong, adiabatic shock forms in a thin layer at the top of the column, where the velocity of the flow is abruptly reduced, and that the flow is then slowed down by radiation to a rest at the bottom of the column. In an adiabatic, radiation-dominated shock, the velocity decreases by at least a factor of seven \citep{1998ApJ...498..790B}, and therefore in this case we would impose the velocity at the top of the column to be $1/7$ (or less) of the local free-fall velocity instead of being equal:
\begin{equation}
    \frac{1}{7}\left(\frac{2GM_*}{R_*+z_{\rm{max}}}\right)^{1/2} \gtrsim cA\tau_{\rm{max}}
    \label{eq:zmaxshort}
\end{equation}
which yields a smaller column. For the parameters used as an example in this paper, eq.~\ref{eq:zmax} yields $z_{\rm{max}}= 6.6$~km, while eq.~\ref{eq:zmaxshort} yields $z_{\rm{max}}\lesssim 1.4$~km. These two assumptions give different results for the polarisation parameters and pulse shape, as we will show  \S~\ref{sec:discussion}.

\subsection{Polarisation from scattering in a strong magnetic field}
\label{sec:PolB}
The magnetic field at the surface of accreting X-ray pulsars is very high, of the order of $10^{12}-10^{13}$~G. Strong magnetic fields affect the motion of electrons by forcing them to move mainly along the field lines, and this tendency strongly alters the interactions between photons and electrons, including bremsstrahlung and scattering. In the B\&W model, the majority of the seed photons inside the accretion column are generated by bremsstrahlung, and in the presence of a strong magnetic field, bremsstrahlung photons below the cyclotron energy are emitted with a sine squared distribution with respect to the magnetic axis and mostly in the ordinary mode (or O-mode), i.e. with the photon's electric field oscillating in the plane that includes the wavevector of the photon and the direction of the magnetic field, because the emitting electrons are bound to oscillate along the magnetic field \citep{1969PhRvL..22..415C,1969Natur.224...49S,1983JPhB...16.3673L}. However, as we will see in \S~\ref{sec:PSin}, before exiting the column, photons undergo many electron scatterings, which modify the polarisation state and angular distribution of the emission. We here present a formalism that will allow us to compute how the scattering of seed photons by free-falling electrons in a strong magnetic field modifies the polarisation state of radiation inside the accretion column.

\begin{figure}
    \centering
    \includegraphics[width=0.8\columnwidth]{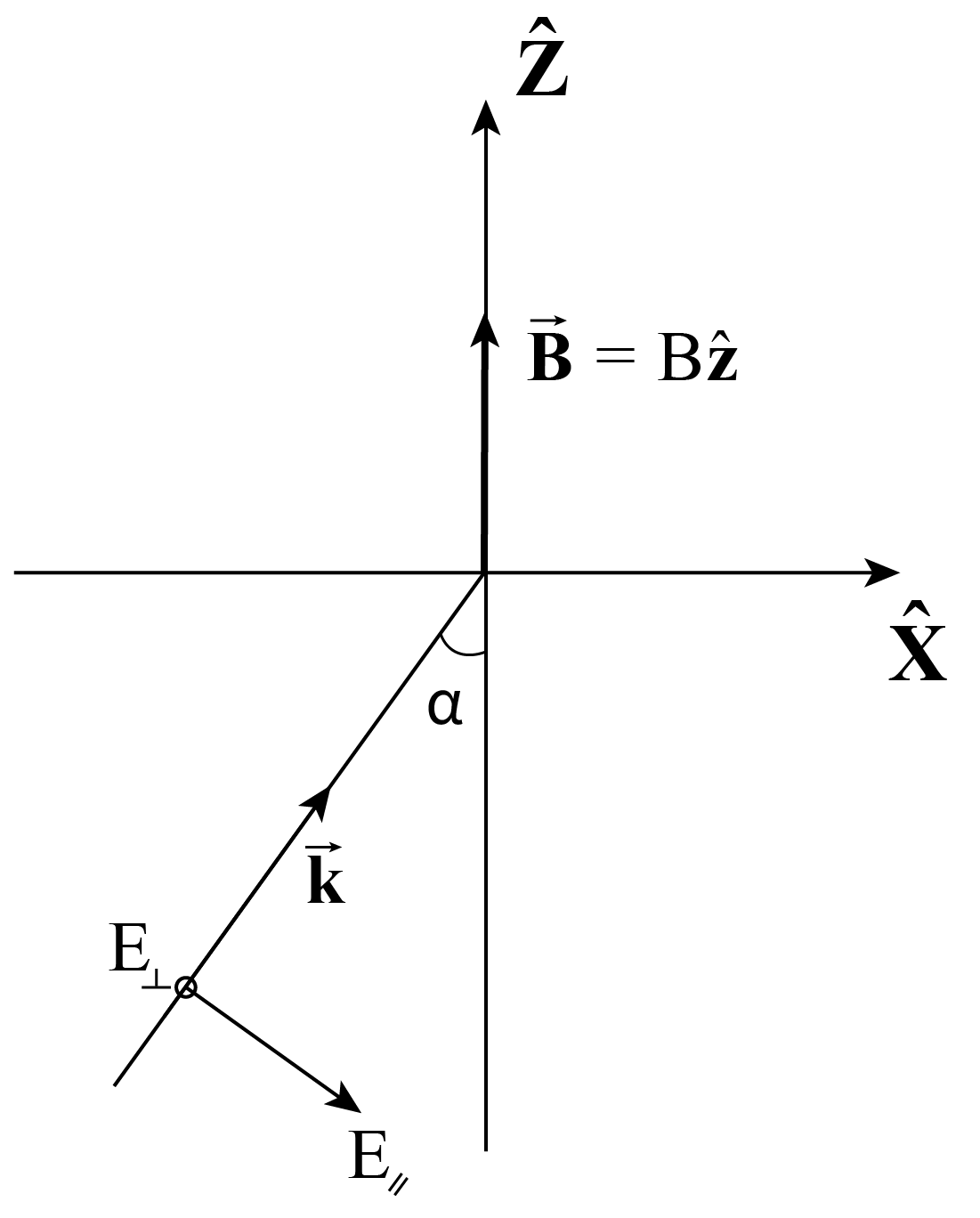}
    \caption{Thompson scattering in a magnetic field. The magnetic field direction is along the $\hat{Z}$-axis. The photon's electric field component in the the $X-Z$ plane, the plane that includes the field and the wavevector of the photon,$\bm{k}$, is called $E_\parallel$, while the component of the electric field perpendicular to the plane is called $E_\perp$.}
    \label{fig:Chou}
\end{figure}

\citet{1986Ap&SS.121..333C} introduced a formalism based on the Stokes parameters $(I,Q,U,V)$ and on the Mueller calculus to describe electron-scattering in the presence of a strong magnetic field. Fig.~\ref{fig:Chou} shows the geometry of a single electron scattering: the incoming photon's direction is at an angle $\alpha$ with respect to the magnetic field, which in the figure points along the $\hat{Z}$-axis, and the electron lies at the origin of the axes. The electric field of the photon is divided in two components: the component that lies in the same plane as the incoming photon's wave-vector and the magnetic field, the $X-Y$ plane, is the parallel component or $E_\parallel$, and the component perpendicular to the plane is $E_\perp$. In this frame, the Stokes parameters of the photons can be written as:
\begin{align}
\begin{split}
    I &= E_\parallel E^*_\parallel + E_\perp E^*_\perp \\
    Q &= E_\parallel E^*_\parallel - E_\perp E^*_\perp \\
    U &= E_\parallel E^*_\perp + E_\perp E^*_\parallel \\
    V &= i(E_\parallel E^*_\perp - E_\perp E^*_\parallel)
\end{split}
\end{align}

After scattering, the photon will now propagate at an angle $\theta$ with respect to the magnetic field direction and it will have acquired an azimuthal component as well, indicated by the angle $\phi$. The Stokes parameters after the scattering can be written in terms of the polar and azimuthal component of the electric field:
\begin{align}
\begin{split}
    I' &= E_\theta E^*_\theta + E_\phi E^*_\phi \\
    Q' &= E_\theta E^*_\theta - E_\phi E^*_\phi \\
    U' &= E_\theta E^*_\phi + E_\phi E^*_\theta \\
    V' &= i(E_\theta E^*_\phi - E_\phi E^*_\theta)
\end{split}
\end{align}

In the Mueller calculus, the Stokes parameters after one scattering, $(I',Q',U',V')$, can be expressed as a function of the incident Stokes parameters $(I,Q,U,V)$ in matrix form:
\begin{equation}
    \begin{pmatrix}
        I' \\
        Q' \\
        U'  \\
        V' 
    \end{pmatrix} = \frac{r^2_e}{2D^2}
    \begin{pmatrix}
        M_{11} & M_{12} & 0 & M_{14} \\
        M_{21} & M_{22} & 0 & M_{24} \\
        0 & 0 & M_{33} & M_{34}  \\
        M_{41} & M_{42} & 0 & M_{44} 
    \end{pmatrix}
     \begin{pmatrix}
        I \\
        Q \\
        U  \\
        V 
    \end{pmatrix}
\end{equation}
where $r_e = e^2/m_ec^2$ is the classical electron radius, $e$ is the electron charge, $m_e$ is the electron mass, $c$ is the speed of light and $D$ is the distance from the observer to the location of the scattering. The matrix elements $M_{ij}$ are functions of the direction of the incident radiation with respect to the magnetic field (described by the angle $\alpha$), of the direction of the outgoing radiation (described by the polar and azimuthal angles $\theta$ and $\phi$ with respect to the magnetic field direction), of the energy of the photon and of the strength of the magnetic field, and their full expressions can be found in \citet{1986Ap&SS.121..333C}.

As we are interested in an axisymmetric configuration (an accretion column) we can average over the azimuthal direction $\phi$, and therefore the matrix elements $M_{ij}$ become:
\begin{subequations}
\label{eq:elem}
\begin{align}
    M_{11} =\,& \frac{\zeta^2}{2}(1+x^2)(\cos^2\alpha+1)(\cos^2\theta+1) + \sin^2\alpha\sin^2\theta \\ 
    M_{12} =\,& \frac{\zeta^2}{2}(1+x^2)(\cos^2\alpha-1)(\cos^2\theta+1) + \sin^2\alpha\sin^2\theta  \\
    M_{14} =\,& -2\zeta^2 x \cos\alpha(1+\cos^2\theta) \label{eq:24} \\
    M_{21} =\,& \frac{\zeta^2}{2}(1+x^2)(\cos^2\alpha+1)(\cos^2\theta-1) + \sin^2\alpha\sin^2\theta \\
    M_{22} =\,& \frac{\zeta^2}{2}(1+x^2)(\cos^2\alpha-1)(\cos^2\theta-1) + \sin^2\alpha\sin^2\theta  \\
    M_{24} =\,& 2\zeta^2 x \cos\alpha\sin^2\theta \label{eq:24b}\\
    M_{33} =\,& 0 \\
    M_{41} =\,& -2\zeta^2 x (1+\cos^2\alpha)\cos\theta \\
    M_{42} =\,& 2\zeta^2 x \sin^2\alpha\cos\theta \\
    M_{44} =\,& 2\zeta^2 (1+x^2)\cos\alpha\cos\theta
\end{align}
\end{subequations}
where $x=\omega_c/\omega$ is the ratio between the cyclotron frequency $\omega_c=eB/m_ec$ and the photon frequency $\omega$, $B$ is the magnetic field and $\zeta=1/(x^2-1)$.
After averaging over $\phi$, we find that all the matrix elements that involve $U$ are equal to zero. We can therefore reduce the matrix to a $3\times3$ matrix where the third element corresponds to the circular polarisation parameter $V$:
\begin{equation}
    \begin{pmatrix}
        I' \\
        Q' \\
        V' 
    \end{pmatrix} = \frac{r^2_e}{2D^2}
    \begin{pmatrix}
        M_{11} & M_{12} & M_{13} \\
        M_{21} & M_{22} & M_{23} \\
        M_{31} & M_{32} & M_{33} 
    \end{pmatrix}
     \begin{pmatrix}
        I \\
        Q \\
        V 
    \end{pmatrix}
    \label{eq:scmatrix}
\end{equation}
where the third position corresponds to the fourth position in eq.s~\ref{eq:elem}, i.e. $M_{13} = M_{14}$ of eq.~\ref{eq:24}, $M_{23} = M_{24}$ of eq.~\ref{eq:24b} and so forth.

The angular dependence of the incoming and outgoing radiation, as well as of the matrix elements, can be expanded in a series of orthonormal functions in $\alpha$ and $\theta$. As we mentioned above, the angular distribution of the seed radiation inside the column is mainly $\sin^2\alpha$ for bremsstrahlung photons, and since the matrix elements are only functions of $\cos\alpha$, $\cos^2\alpha$ and $\sin^2\alpha$, and the same for $\theta$, the only important functions for the expansion are given by the orthonormal basis
\begin{align}
    f_1(\alpha) &= \frac{\sqrt{15}}{4}\sin^2\alpha;\\
    f_2(\alpha) &= \frac{\sqrt{6}}{2}\cos\alpha;\\
    f_3(\alpha) &= \frac{5\sqrt{3}}{4}\left(\cos^2\alpha-\frac{1}{5}\right)
\end{align}
and the same ones for $\theta$. We can write the Stokes vectors in this new basis:
\begin{align*}
    I &= l_1\times f_1(\alpha) + l_2\times f_2(\alpha) + l_3\times f_3(\alpha) \\
    Q &= l_4\times f_1(\alpha) + l_5\times f_2(\alpha) + l_6\times f_3(\alpha) \\
    V &= l_7\times f_1(\alpha) + l_8\times f_2(\alpha) + l_9\times f_3(\alpha) \\
    I' &= l'_1\times f_1(\theta) + l'_2\times f_2(\theta) + l'_3\times f_3(\theta)\\
    Q' &= l'_4\times f_1(\theta) + l'_5\times f_2(\theta) + l'_6\times f_3(\theta)\\
    V' &= l'_7\times f_1(\theta) + l'_8\times f_2(\theta) + l'_9\times f_3(\theta)
\end{align*}
where the angular dependence is carried by the basis functions, so that the $l_i$ are coefficients that depend on energy and magnetic field but not on angles. We can thus rewrite the scattering matrix in eq.~\ref{eq:scmatrix} as a $9\times9$ matrix in this new basis
\begin{equation}
    \begin{pmatrix}
        l'_1 \\
        l'_2 \\
        \vdots \\
        l'_{9} 
    \end{pmatrix} = \frac{r_e^2}{2D^2}
    \begin{pmatrix}
        a_{1,1} & a_{1,2} & \cdots & a_{1,9} \\
        a_{2,1} & a_{2,2} & \cdots & a_{2,9} \\
        \vdots  & \vdots  & \ddots & \vdots  \\
        a_{9,1} & a_{9,2} & \cdots & a_{9,9} 
        \end{pmatrix}
     \begin{pmatrix}
        l_1 \\
        l_2 \\
        \vdots \\
        l_{9} 
    \end{pmatrix}
    \label{eq:finalmatrix}
\end{equation}
where the matrix elements are just functions of the energy and magnetic field through $x$ and $\zeta$, and the angle dependence is conveyed by the $f$ functions. In this way, we can efficiently compute the effects of the scattering in the strong magnetic field on the seed radiation. The relation expressed in eq.~\ref{eq:finalmatrix} is valid in the instantaneous rest frame of the electrons; if the motion of the electrons is relativistic, we will have to consider beaming effects. Also, we have not yet considered any energy transfer between the electron and the photon, which we will have to include in the case of Compton scattering.

\begin{figure}
    \centering
    \includegraphics[width=\columnwidth]{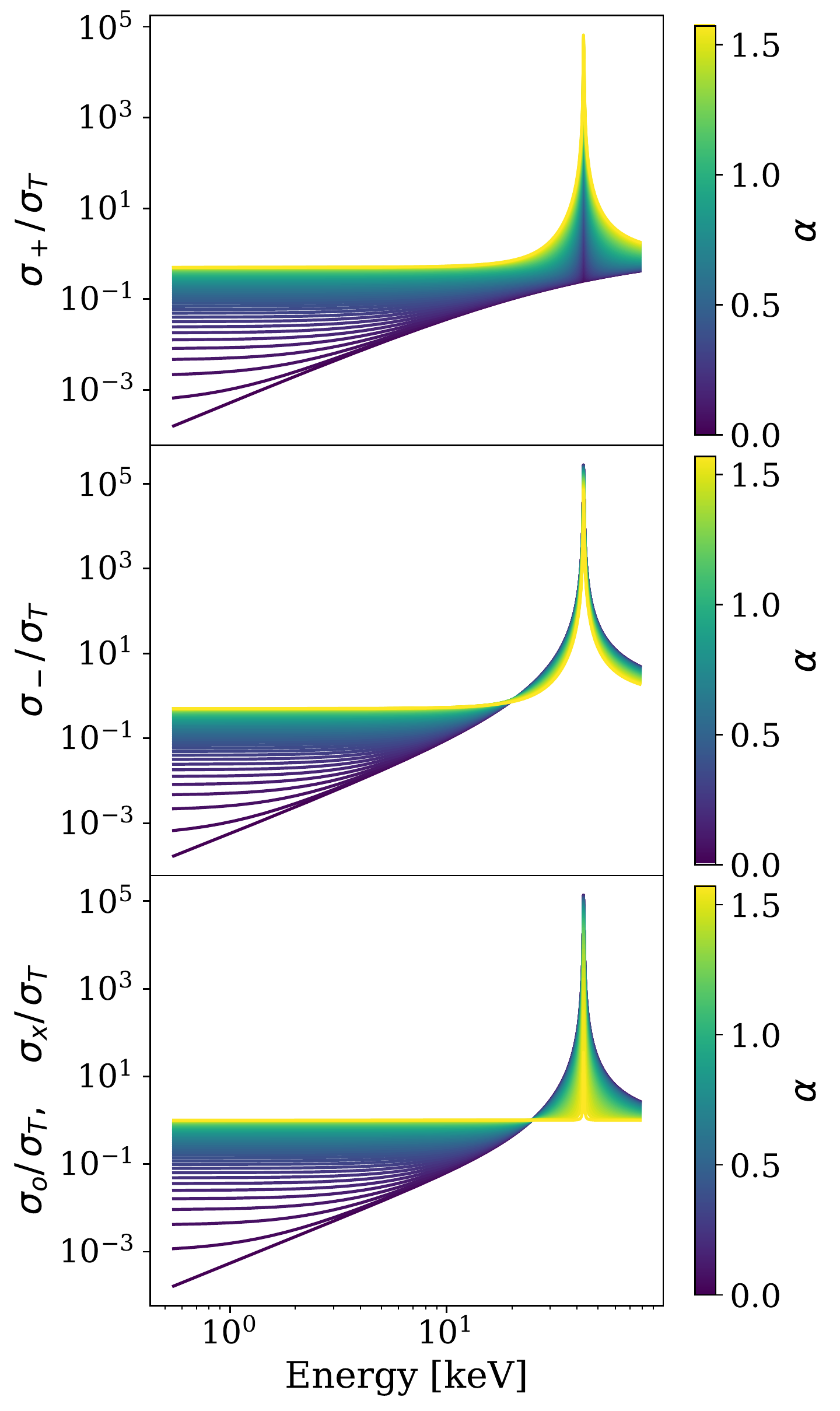}
    \caption[Cross sections for the different polarisation modes.]{Dependence of the cross sections on the incidence angle $\alpha$ and on the photon energy. $\sigma_+/\sigma_T$ (top) and $\sigma_-/\sigma_T$ (middle) are colour-coded with respect to $\alpha$. In the bottom panel, $\sigma_x/\sigma_T$, which does not depend on the incident angle, is shown as the solid black line, while $\sigma_o/\sigma_T$ is colour-coded with respect to $\alpha$ and when $\alpha=0$, $\sigma_o = \sigma_x$.}
    \label{fig:crosssections}
\end{figure}

From eq.~\ref{eq:finalmatrix}, we can calculate the angle- and energy-dependent cross sections for scattering for each polarisation mode by applying the matrix to a polarisation vector completely polarised in the mode under consideration and by averaging the emerging intensity over the outgoing angle $\theta$
\begin{equation}
    \sigma = 2\pi D^2 \int_1^{-1} I'(\alpha,\theta) d(\cos\theta) \,.
\end{equation}
In particular, it is interesting to focus on the ordinary mode (O-mode) $(I,Q,V)=(1,1,0)$, for which the incoming photon's electric field lies in the plane parallel to the magnetic field and the wavevector of the photon, the extraordinary mode (X-mode) $(1,-1,0)$, with the electric field perpendicular to the magnetic field, and two perpendicular circular polarisation modes, that we identify with the symbols $+$ and $-$: $(1,0,1)$ and $(1,0,-1)$. 
The corresponding cross sections are given by \citep[see also][]{1979PhRvD..19.2868H}:
\begin{subequations}
\begin{align}
    \sigma_o &=\sigma_T \left[ \sin^2\alpha + \cos^2\alpha\frac{x^2+1}{(x^2-1)^2} \right]\\
    \sigma_x &= \sigma_T \frac{x^2+1}{(x^2-1)^2}\\
    \sigma_+ &= \frac{1}{2}\sigma_T \left[\sin^2\alpha + \frac{(x^2+1)(1+\cos^2\alpha) - 4x\cos\alpha}{(x^2-1)^2} \right] \\
    \sigma_- &= \frac{1}{2}\sigma_T \left[\sin^2\alpha + \frac{(x^2+1)(1+\cos^2\alpha) + 4x\cos\alpha}{(x^2-1)^2} \right]
\end{align}
\label{eq:crosssections}
\end{subequations}

\noindent where $\sigma_T$ is the Thomson cross section. The dependence on energy and incidence angle of the different cross sections is shown in Figure~\ref{fig:crosssections}. They all seem to diverge at the cyclotron energy; however, for $x$ very close to 1, the energy transfer from photons heats up the electrons and damping effects become important \citep{1992hrfm.book.....M}. The lower panel depicts both $\sigma_x$ (black solid line) and $\sigma_o$ (colour-coded with $\alpha$). At energies lower than the cyclotron energy, because the electrons are forced to oscillate along the magnetic field axis, the cross-section for the extraordinary mode, for which the electric field is perpendicular to the magnetic field, is drastically reduced with respect to the field-free Thomson cross-section, and the cross-section for the ordinary mode ($\sigma_o$) diminishes for low $\alpha$ to become equal to that of the extraordinary mode ($\sigma_x$) when $\alpha = 0$, i.e. when photons propagate along the magnetic field, as expected.
\subsection{Polarisation state inside the column}
\label{sec:PSin}
As mentioned above, the majority of the seed photons inside the accretion column are generated by bremsstrahlung, and therefore are emitted mostly in the ordinary mode and with a $\sin^2\alpha$ distribution with respect to the magnetic axis \citep{1969PhRvL..22..415C,1969Natur.224...49S,1983JPhB...16.3673L}. Before exiting the accretion column, seed photons are scattered by the in-falling electrons, and in each scattering, the polarisation vector changes according to Eq.~\ref{eq:finalmatrix}. In order to calculate the average number of scatterings per photon, we need to calculate the density profile in the column, which depends on the accretion rate $\dot{M}$:
\begin{equation}
    \dot{M}=\pi r_0^2 \rho |v| = \pi r_0^2 \rho A \tau_\parallel c
\end{equation}
where $\rho$ is the density of the gas and where we used Eq.~\ref{eq:velprofile} to express the electrons velocity $v$. We can therefore calculate the optical depth for photons travelling perpendicularly to the column axis, knowing that the opacity is dominated by electron scattering:
\begin{equation}
    \tau_\perp = \frac{r_0 \rho \bar{\sigma}_\perp}{m_p} \sim \frac{r_0 \rho \sigma_T}{m_p} = \frac{\dot{M}\sigma_T}{m_p \pi r_0 A \tau_\parallel c}
\end{equation}
where we have employed the fact that the scattering cross section of photons propagating perpendicularly to the magnetic field is close to the Thomson cross section $\sigma_T$ at all energies except very close to the cyclotron energy, where it is even higher (see Figure~\ref{fig:crosssections}). $\tau_\parallel$ increases in the vertical direction and is of the order 1 at the top of the column. Employing the fitted parameters from \cite{2016ApJ...831..194W}, we find that $\tau_\perp\sim 500$ at the top of the column, and therefore greater than 500 throughout the column. This yields an average number of scatterings per photon
\begin{equation}
N_{\rm{sc}}\sim \tau_\perp^2 \gtrsim 250,000\,.
\label{eq:XPnscat}
\end{equation}
For Compton scattering, the energy transfer to the photon for a single scattering is given by \citep{1986rpa..book.....R}
\begin{equation}
    \Delta\epsilon/\epsilon \sim (\gamma^2-1) \lesssim 0.15
    \label{eq:deltaeps}
\end{equation}
where $\gamma = 1/\sqrt{1-\beta^2}$ is the Lorentz factor of the scattering centre, the electron, and we have used $\beta \sim 0.5$, which is the free-fall velocity of the electrons at the top of the column, and therefore the highest. 

From the estimates in Eqs.~\ref{eq:XPnscat} and~\ref{eq:deltaeps}, it is clear that an average photon has to undergo many scatterings before it can escape the column and that the energy transferred per scattering is very small. In the final tens of scatterings, the energy of the photon is therefore very close to its final energy, i.e. its energy when it finally manages to escape from the column. Thus, during the final tens of scatterings of each photon, the elements of the scattering matrix in eq.~\ref{eq:finalmatrix} will remain approximately unchanged. 

Multiplying a vector by the same matrix many times brings the vector close to the matrix's eigenstate with the largest eigenvalue, unless the vector itself is in an orthogonal eigenstate. Depending on the magnitude of the ratio between the largest eigenvalue and the rest, this process takes relatively few interactions. It is easy to see that, except for energies very close to the cyclotron energy, the largest eigenvalue of the scattering matrix is orders of magnitude larger than the other eigenvalues. For this reason, we can safely assume that, independently of the initial polarisation state of the photon, its Stokes vector will be in the matrix's predominant eigenstate just after a few scatterings.  
\begin{figure*}
    \centering
    \includegraphics[width=0.393\textwidth]{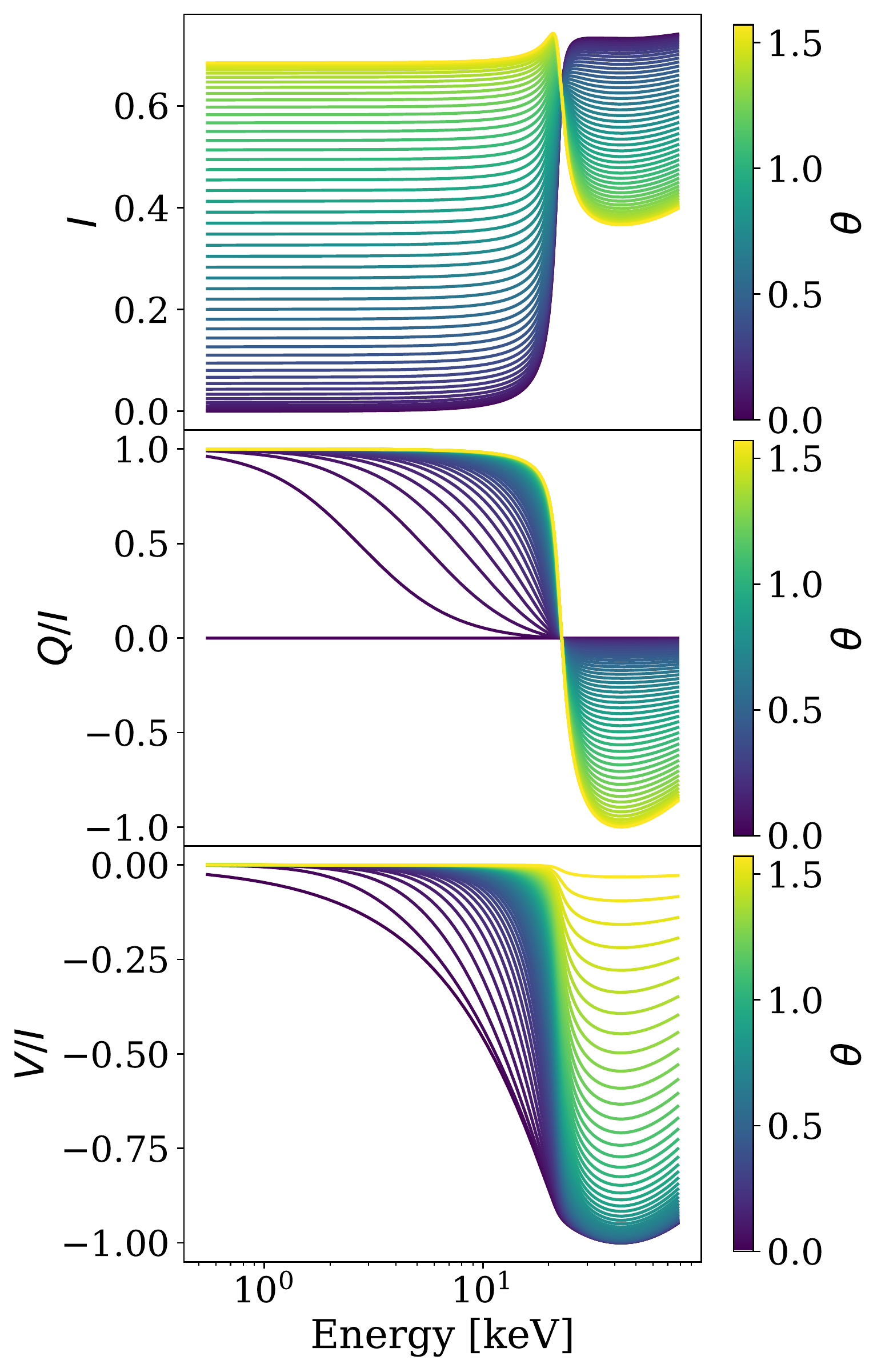}
    \includegraphics[width=0.38\textwidth]{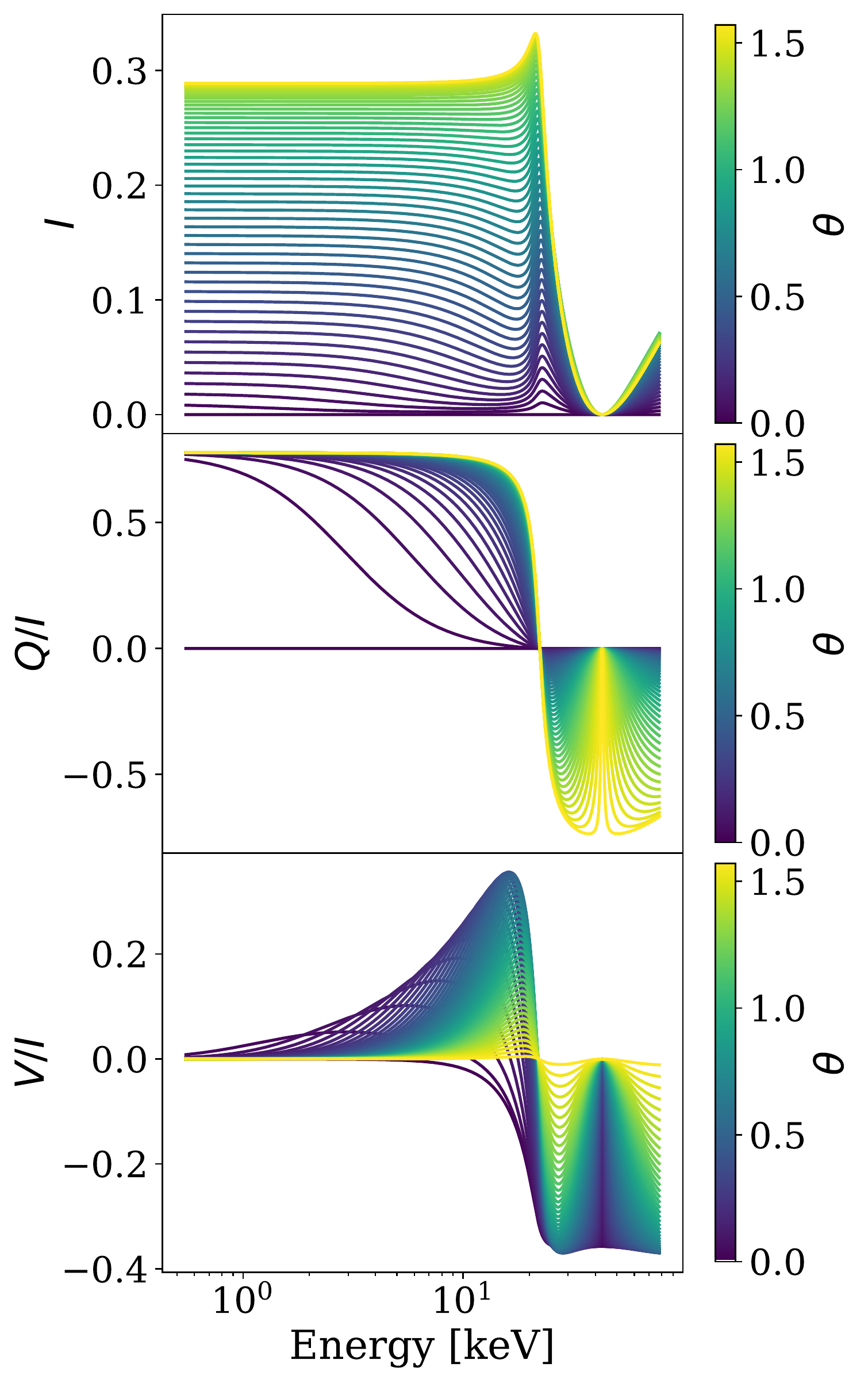}
    \caption{Average Stokes parameters. The left panels depict the polarisation parameters of radiation inside the accretion column, while panels on the right represent the polarisation after the photons have gone through the region of last scattering and left. The calculation does not yet include beaming effects. From top to bottom: intensity $I$ (arbitrary units), linear polarisation fraction $Q/I$ and circular polarisation fraction $V/I$ against the energy of the photons. The colour code represents the angle with respect to the magnetic field $\theta$.}
    \label{fig:in&out}
\end{figure*}

Therefore, the predominant eigenstate of the scattering matrix in Eq.~\ref{eq:finalmatrix} represents the polarisation state of radiation inside the column, in the rest frame of the electrons, regardless of the initial polarisation parameters of the seed photons. In the left panels of Figure~\ref{fig:in&out} the Stokes parameters for the Comptonised radiation inside the column are shown a as a function of energy and angle $\theta$, which is the angle with respect to the magnetic axis, $\hat{z}$. The results are shown for $\theta <\pi/2$ and are specular for $\theta>\pi/2$. As expected, except for very small angles, photons are always nearly linearly polarised in the ordinary mode (positive $Q$) and the intensity peaks at $\theta=\pi/2$ for energies lower than the cyclotron energy ($\sim 40$~keV), because the scattering centres are bound to oscillate along the magnetic field, while photons around the cyclotron energy present a mixture of extraordinary and circular polarisation. At the cyclotron energy, photons propagating along the magnetic field (at small $\theta$) have no linear polarisation (the two linear polarisation modes are equally perpendicular to the magnetic field), and have a strong circular polarisation due to resonant cyclotron scattering (electrons can be excited to the second Landau level). Photons near the cyclotron energy propagating at $\theta\sim\pi/2$, on the other hand, can resonantly scatter only if in the X-mode, and therefore $Q/I$ is equal to $-1$. 

This picture represents the polarisation state of photons propagating inside the column, but in order to find the average polarisation parameters of photons leaving the column, we have to take into account the difference in the cross sections for the different polarisation modes, and therefore the difference in the volume of the optically thin region close to the walls of the column. Since the cross section is much smaller, we expect the volume of the region of last scattering for extraordinary photons to be much larger than for ordinary photons, reducing the extent of linear polarisation at all energies.

If we now indicate with $(I,Q,V)$ the average polarisation state inside the column and with $(I',Q',V')$ the polarisation of the outgoing radiation, we can write
\begin{equation}
    Q= O - X
\end{equation}
where $O$ and $X$ are the intensities of the ordinary and the extraordinary modes inside the column. The outgoing intensities will be
\begin{align}
    O' &= \frac{1}{2}(Q+I)V_\parallel\\
    X' &= \frac{1}{2}(I-Q)V_\perp
\end{align}
where $V_i \propto \sin\theta/\sigma_i$ is the volume of the region of last scattering for each mode.
We can therefore write the Stokes parameters of the radiation coming out of the column as
\begin{align}
I' &= \frac{1}{2}(Q+I)V_\parallel + \frac{1}{2}(I-Q)V_\perp\\
Q' &= \frac{1}{2}(Q+I)V_\parallel - \frac{1}{2}(I-Q)V_\perp \\
V' &= \frac{1}{2}(V+I)V_+ - \frac{1}{2}(I-V)V_-
\label{eq:escape}
\end{align}
The right panels of Figure~\ref{fig:in&out} show the average Stokes parameters after the radiation has gone through the region of last scattering, and therefore left the column. Comparing to the left panels, which depict the polarisation state inside the column, we can see that the linear polarisation is reduced at low energy even for high angles because the low value of $\sigma_x$ favours the emission of extraordinary photons. Intensity is drastically lowered at high energies because all the scattering cross sections become divergent close to the cyclotron energy. For the same reason, radiation at the cyclotron energy is completely unpolarised.

\subsection{Relativistic beaming}
\label{sec:beaming}
The previous calculations were performed in the instantaneous rest frame of the electrons. Electrons are flowing down the column with a velocity that goes from about 0.5~$c$ at the top to zero at the bottom in the case considered here. For this reason, the emission from the column, especially from the top, where most of the radiation emerges, will be beamed. For higher luminosity pulsars, the radiation pressure could slow down the in-falling gas at greater distance from the surface, where the free-fall velocity of the electrons would be lower. Beaming changes the emission angle and energy of the photons and the total intensity of the radiation:
\begin{align}
\theta' &= \cos^{-1}\left(\frac{\cos\theta-\beta}{1-\beta\cos\theta}\right)\\
E_p' &= \frac{E_p}{\gamma(1+\beta\cos\theta)}\\
I' &= I \left(\frac{E_p'}{E_p}\right)^3
\end{align}
where we now indicate with a prime the quantities after beaming, $E_p$ is the energy of the photons, $I$ is the specific intensity (i.e. per unit energy), and $\beta=A\tau_\parallel$ is the speed of the electrons divided by the speed of light.

\begin{figure}
    \centering
    \includegraphics[width=\columnwidth]{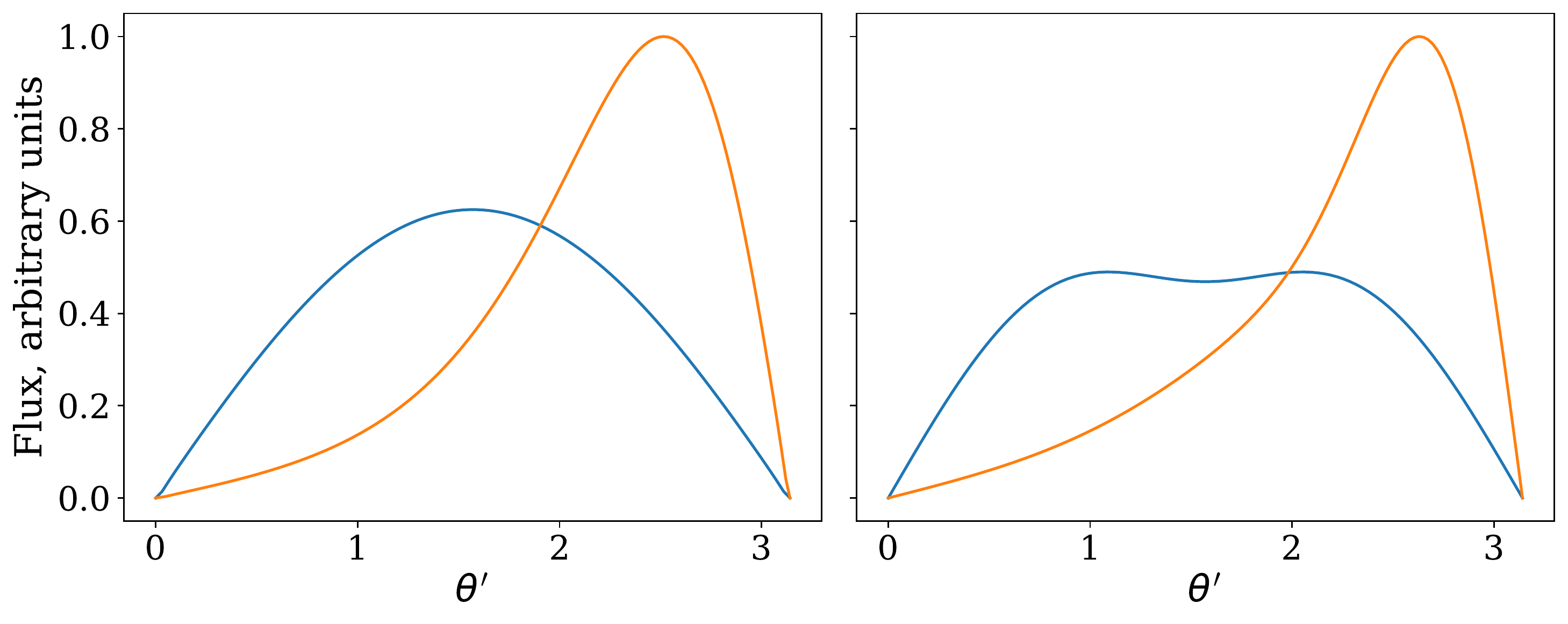}
    \caption{The effect of beaming on flux for $\beta=0.4$ and different photon energies. Solid blue line: flux, without beaming; solid orange line: flux, with beaming. Left panel: photon energy 1 keV; right panel: photon energy 27 keV. }
    \label{fig:beaming}
\end{figure}

\begin{figure*}
    \centering
    \includegraphics[width=0.38\textwidth]{Outtot.pdf}
    \includegraphics[width=0.39\textwidth]{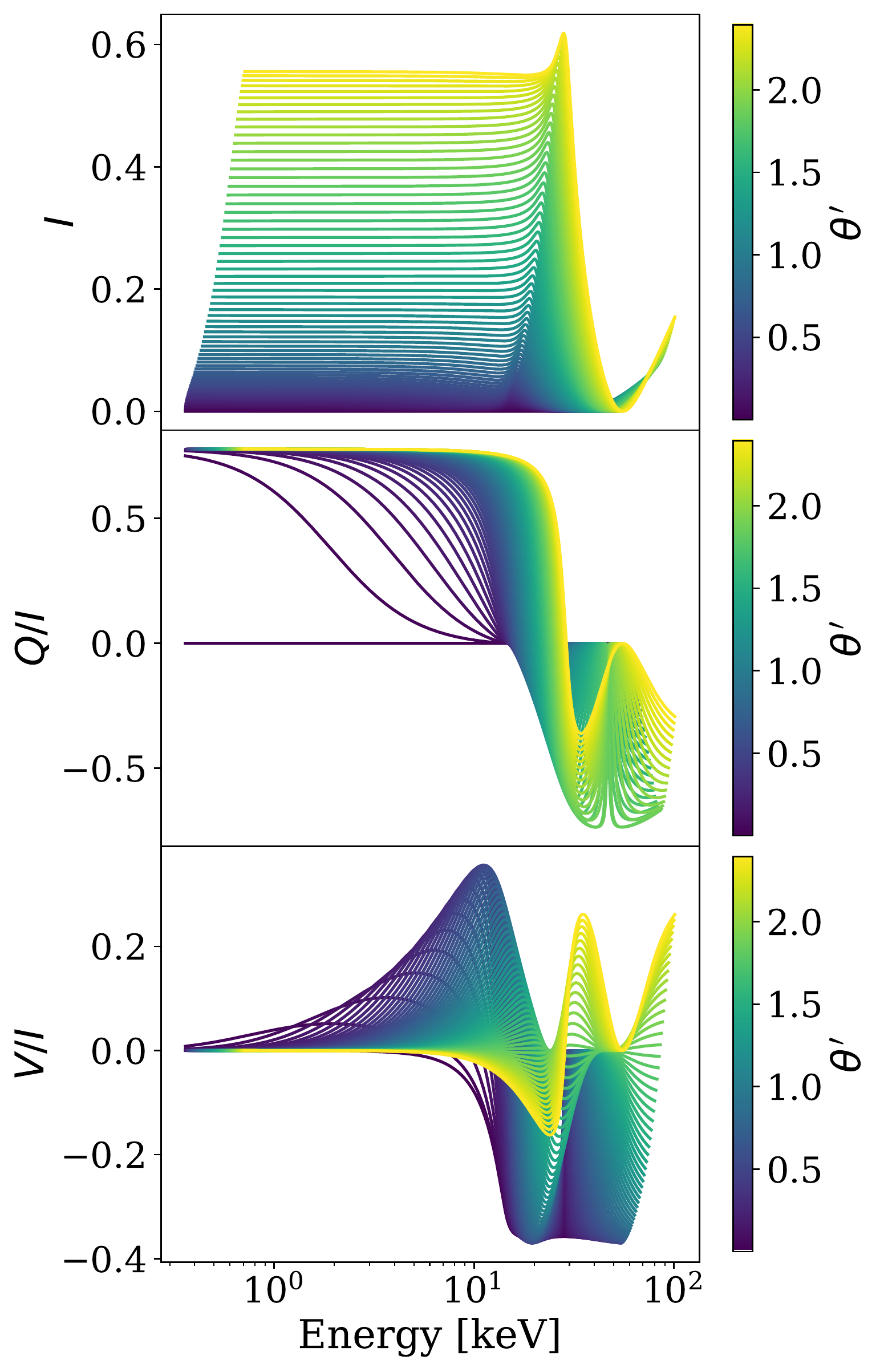}
    \caption[Average Stokes parameters before and after beaming]{Left: average Stokes parameters for the radiation coming out of the column in the frame of reference of the electrons, same as right panels in Fig.~\ref{fig:in&out}. Right: Stokes parameters after beaming for $\beta = 0.4$. Top: intensity $I$ (arbitrary units); middle: linear polarisation fraction $Q/I$; bottom: circular polarisation fraction $V/I$. The colour code represents the angle with respect to the magnetic field before beaming ($\theta$, right) and after beaming ($\theta'$, left). Please notice that $\theta$ goes from 0 to $\pi/2$, while $\theta'$ goes from 0 to the angle where the intensity peaks, at about 2.7 radians (see also Figure~\ref{fig:beaming}).}
    \label{fig:beamStoke}
\end{figure*}

Figure~\ref{fig:beaming} shows the effect of beaming on the angular distribution of flux for $\beta=0.4$ and for two photon energies, 1 keV and 27 keV: radiation is strongly beamed toward the surface of the star (high $\theta'$). The right panels of Figure~\ref{fig:beamStoke} show the average Stokes parameters after beaming, also for $\beta = 0.4$. Please notice that the angle $\theta'$ in the colour bar does not go from 0 to $\pi/2$, as in the previous plots, but from 0 to the angle where the intensity peaks, at about 2.7 radians (see also Figure~\ref{fig:beaming}). The main effect of beaming is to shift the angle where intensity peaks toward higher angles (and therefore the radiation is beamed toward the surface of the star) and to move the cyclotron absorption feature to higher energies at higher angles.

\subsection{Total emission at surface}
Because the optical depth is very high in the lower part of the column, most of radiation comes out from the top, where the beaming is strongest. The relation between flux, $\tau_\parallel$ and $z$ (vertical direction coordinate along the column) is
\begin{equation}
    \frac{L(z)}{L_{\rm{tot}}} = \left(\frac{z}{z_{\rm{max}}}\right)^{3/2} = \left(\frac{\tau_\parallel}{\tau_{\parallel\rm{max}}}\right)^{3} 
\end{equation}
In order to integrate the emission from the column, we divide the column in fractions of equal flux, where we calculate the beaming, and sum all the contributions. 

We first assume the presence of two columns at the two magnetic poles of the neutron star. We expect to see two columns if the magnetic field has a bipolar structure, except if the field is stronger at one pole, in which case the stronger pole would swipe the gas from the accretion disk at a radius outside the reach of the weaker pole. In this section, we assume an orthogonal rotator, with the magnetic field always lying in the same plane with the line of sight. From the orthogonal rotator, the results for any other rotational geometry can be easily calculated (see Appendix~\ref{sec:geo}). 

We are still focusing on the polarisation parameters at emission, and therefore we do not yet include any gravitational lensing effects. For this reason, the angle $\theta$ between the line of sight and the magnetic field is also the phase (we will now abandon the prime). At any phase, one column is at $\theta$ and one column is at $\pi-\theta$. Also, one of the columns is entirely visible, while the other one is partially covered by the star. If $z_{\rm{max}}$ is the height of the column, for each $\theta$ the part of the back column that we see is given by
\begin{equation*}
     z_{\rm{max}} - R_*\left(\frac{1}{\sin\theta} -1 \right) \leq  z \leq z_{\rm{ max}}\,.
\end{equation*}

\begin{figure}
    \centering
    \includegraphics[width=\columnwidth]{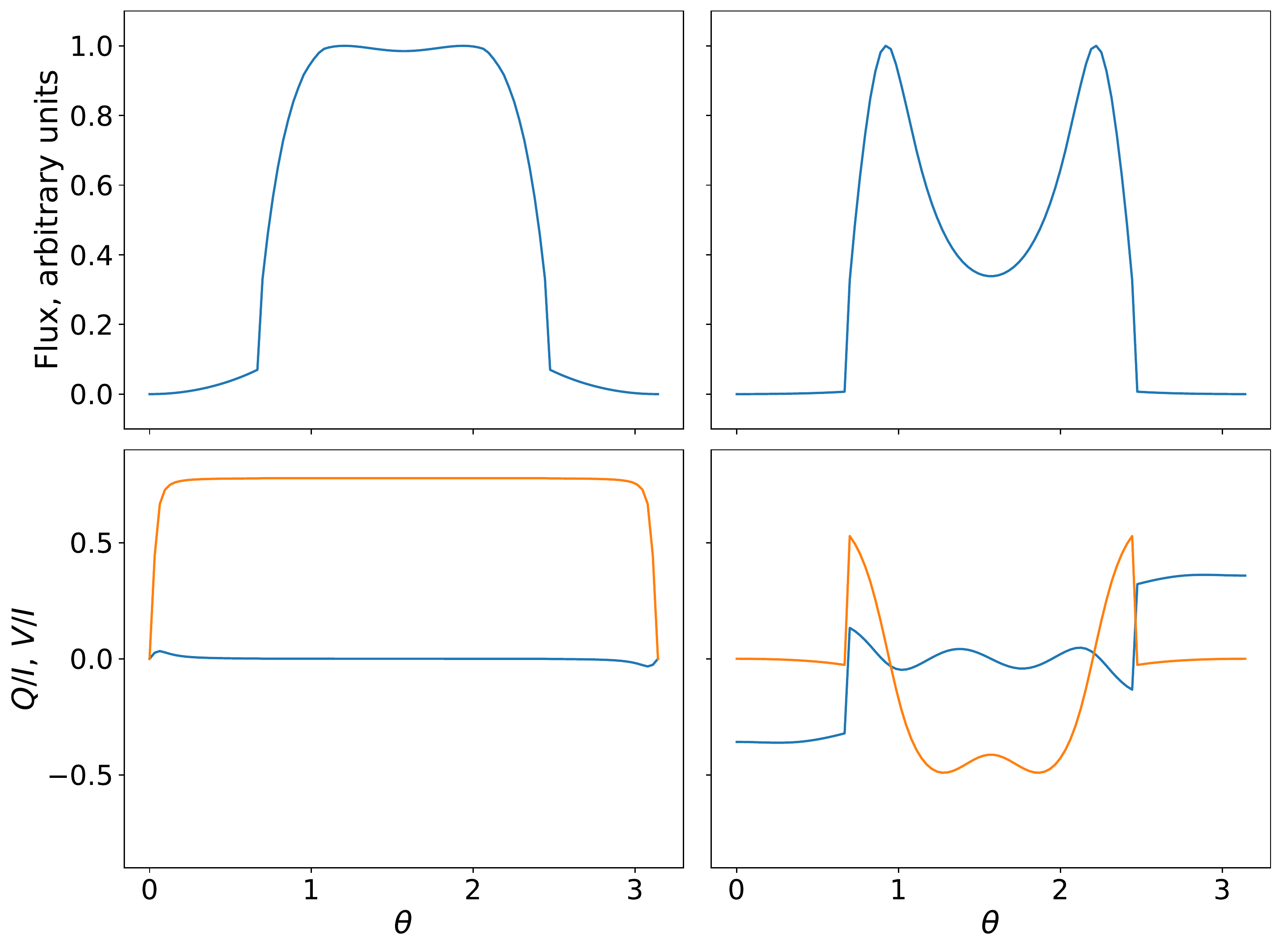}
    \caption{Phase pattern for intensity and polarisation fractions for two columns without light bending. Upper panels: intensity $I$ (arbitrary units). Lower panels: linear polarisation fraction $Q/I$ solid orange line, circular polarisation fraction $V/I$ solid blue line. Left panels: photon energy 1~keV; right panels: photon energy 27~keV.}
    \label{fig:phase2col}
\end{figure}

\begin{figure*}
    \centering
    \includegraphics[width=0.38\textwidth]{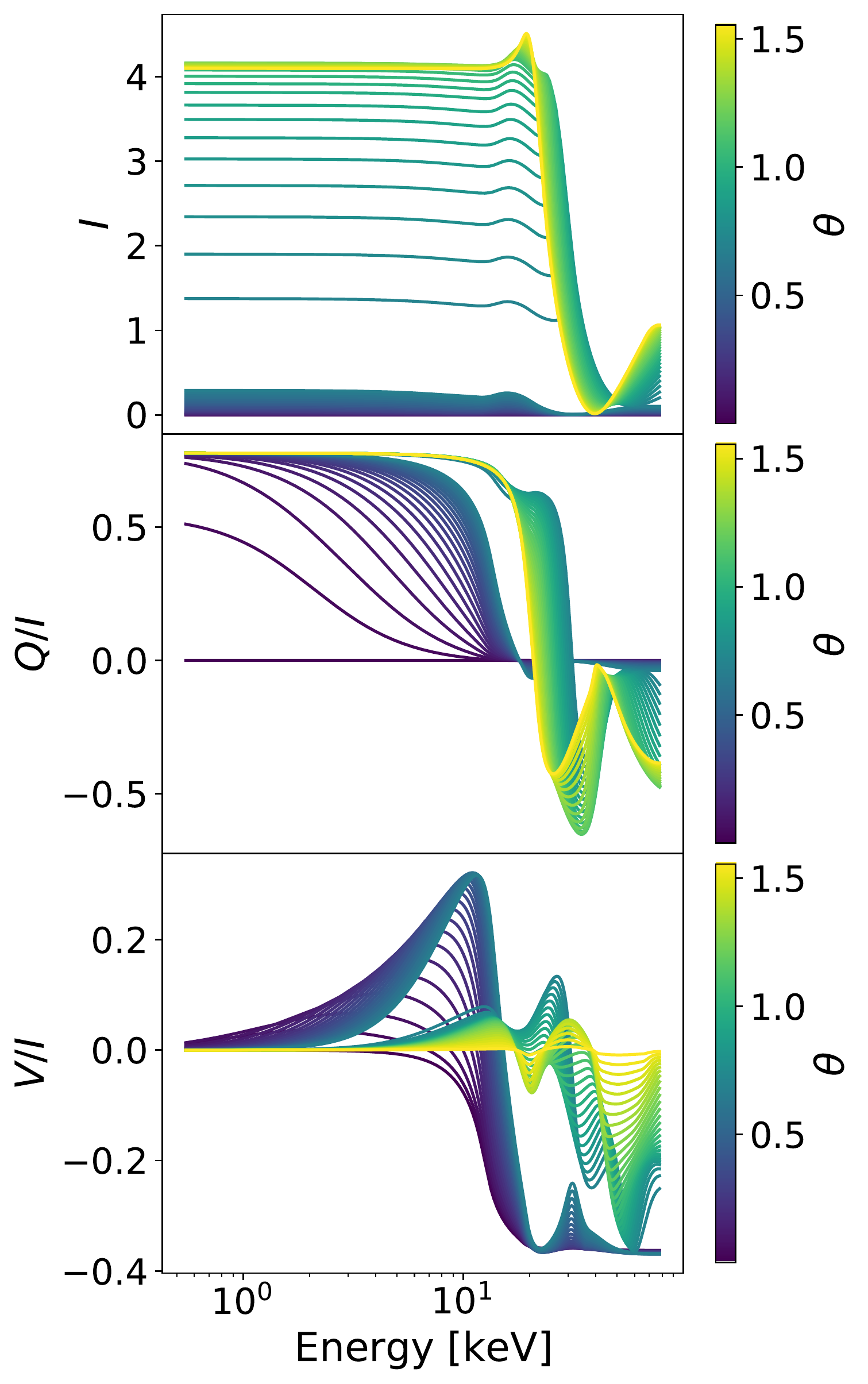}
    \includegraphics[width=0.39\textwidth]{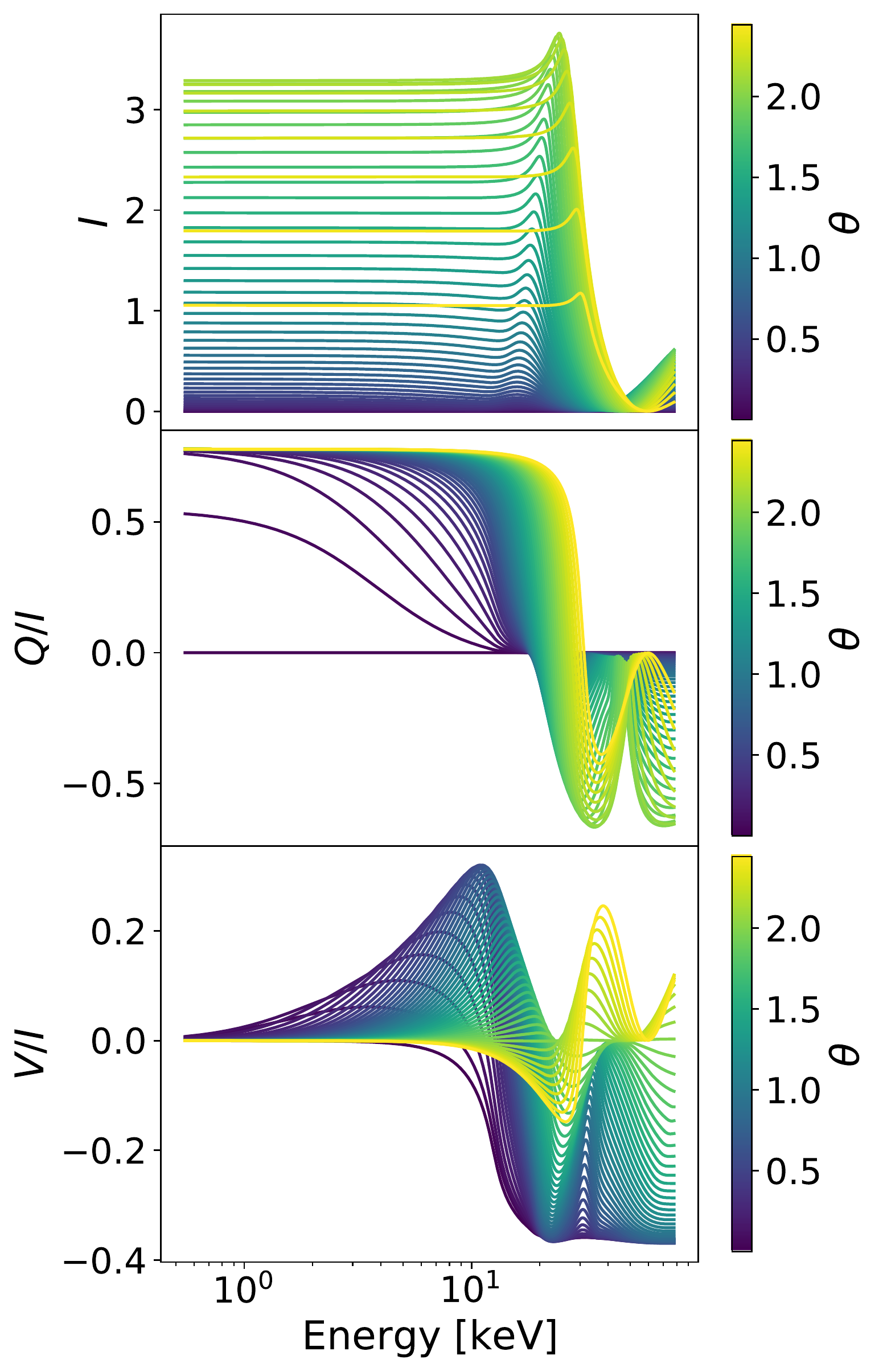}
    \caption{Average Stokes parameters for the emission from the two columns (left) and only one column (right). The colour code represents the phase angle $\theta$.}
    \label{fig:Stoke2col}
\end{figure*}

The radiation pattern with rotation phase that we obtain is shown in the upper plots of Figure~\ref{fig:phase2col} for two photon energies, 1 keV and 27 keV. The sudden rise in intensity at about $\pi/5$ is due to the fact that the top of the back column starts to be visible and that the emission is highly beamed toward the surface of the star, so the back column is the one dominating the emission. The same effect can be seen on the average Stokes parameters in the left panels of Figure~\ref{fig:Stoke2col}. The right panels of Fig.~\ref{fig:Stoke2col} show the Stokes parameters in case of only one column present: the results are quite similar to the ones depicted in the right panels of Fig.~\ref{fig:beamStoke}, because the bulk of the emission is coming from the top of the single column, where the in-fall velocity of the electrons is close to half the speed of light. 

\subsection{Gravitational lensing}
\label{sec:lb}
\begin{figure}
    \centering
    \includegraphics[width=\columnwidth]{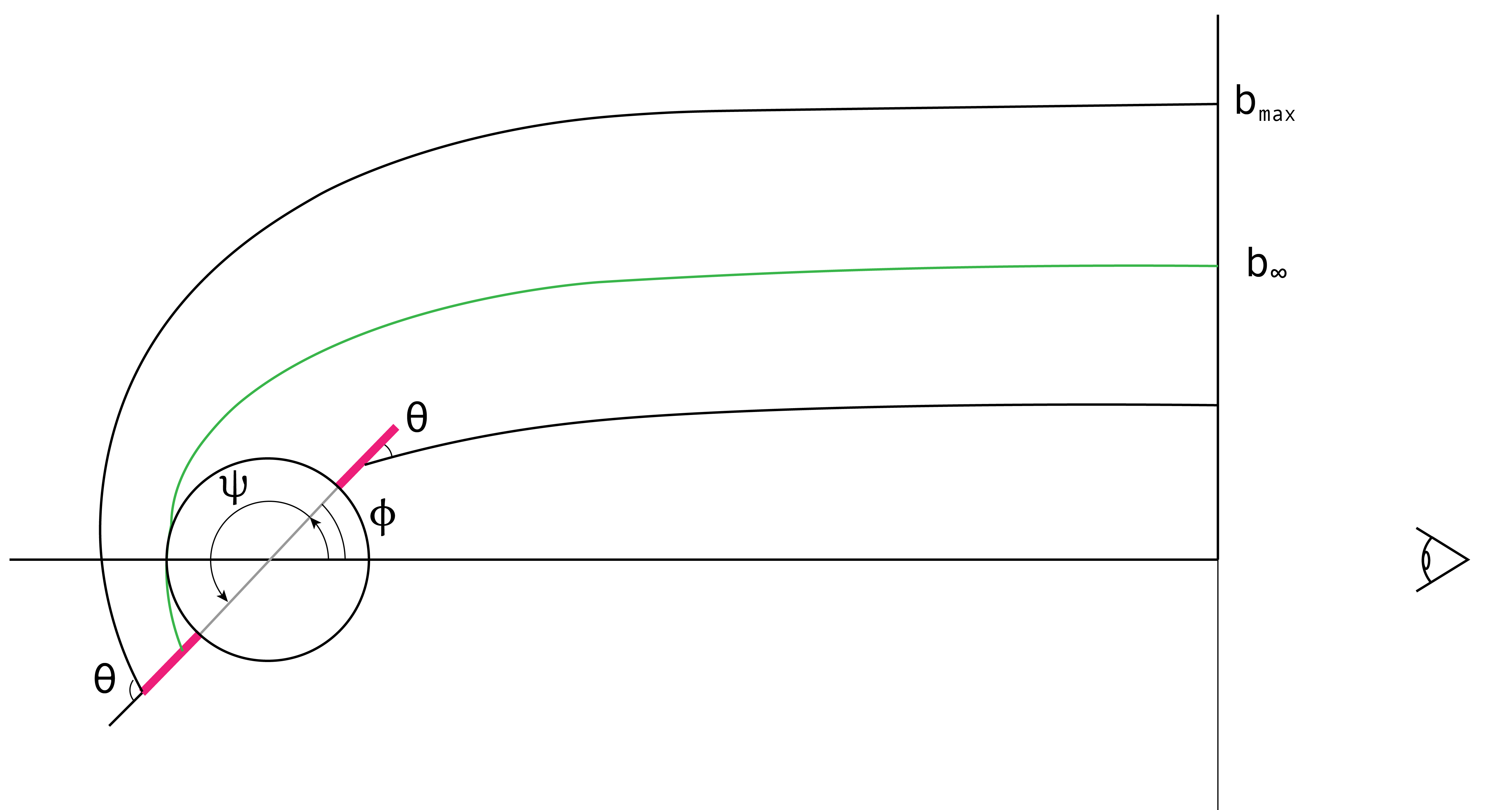}
    \caption{Light bending in the neutron star gravitational field. $\theta$ is the angle between the vertical direction of the column ($\hat{z}$) and the photon momentum at emission, $\psi$ is the angle between $\hat{z}$ and the line of sight , and $\phi$ is the rotation phase of the star.}
    \label{fig:lensingT}
\end{figure}

Because neutron stars are very compact, their strong gravitational field affects the propagation of light around them, and general relativity needs to be included when calculating the photons' path. Because of the strong fields, the path of light is bent, and therefore the image of the star results distorted at the observer: we can see the front of the star but also part of the back, depending on how compact the star is. Because of light bending, the angle between the magnetic field and the photon momentum at emission, which we call $\theta$, is now different from the phase, that from now on we will call $\phi$, and from the angle between the vertical direction of the column ($\hat{z}$) and the line of sight, that we will call $\psi$ (see Fig.~\ref{fig:lensingT}). 
The relation between $\theta$ and $\psi$ in general relativity is given by \citep{2002ApJ...566L..85B}  
\begin{align}
&\sin\theta = \frac{b}{R}\sqrt{1-\frac{R_g}{R}} \label{eq:impact}\\
&\psi = \int_R^\infty \frac{-u^\psi}{u^r}dr = \int_R^\infty \frac{dr}{r^2}\left[\frac{1}{b^2} - \frac{1}{r^2} \left(1-\frac{R_g}{r}\right) \right]^{-1/2}
\end{align}
where $R_g= 2GM_*/c^2$ is the gravitational radius of the neutron star, $b$ is the impact parameter and $R$ is the distance of the emission region from the centre of the star. In \citet{2002ApJ...566L..85B}, $R$ is the radius of the star, while, in the case of emission from a column, $R$ is the radius of the star plus $z$, the height along the column under consideration. For the column in the front of the star, we have to integrate between $R = R_* + z$ and infinity to get $\psi$. For the column in the back, we need to be more careful. For each light ray, labelled by the impact parameter $b$, we have to calculate the minimum distance from the centre of the star of the light path, defined by eq.~\ref{eq:impact} when $\sin\theta =1$, and integrate from $R = R_* + z$ to the minimum distance and then from the minimum distance to infinity. 

Depending on the compactness of the star, it is possible to see both sides of one column, from the front and from the back, because of light bending. Thus, for each phase $\phi$, we have to sum the contributions from the front column at $\psi = \phi$ and at $\psi = 2\pi - \phi$ and from the back column at $\psi = \pi - \phi$ and at $\psi = \pi + \phi$, making sure of excluding the parts of the columns that are blocked by the neutron star itself.
\begin{figure}
    \centering
    \includegraphics[width=\columnwidth]{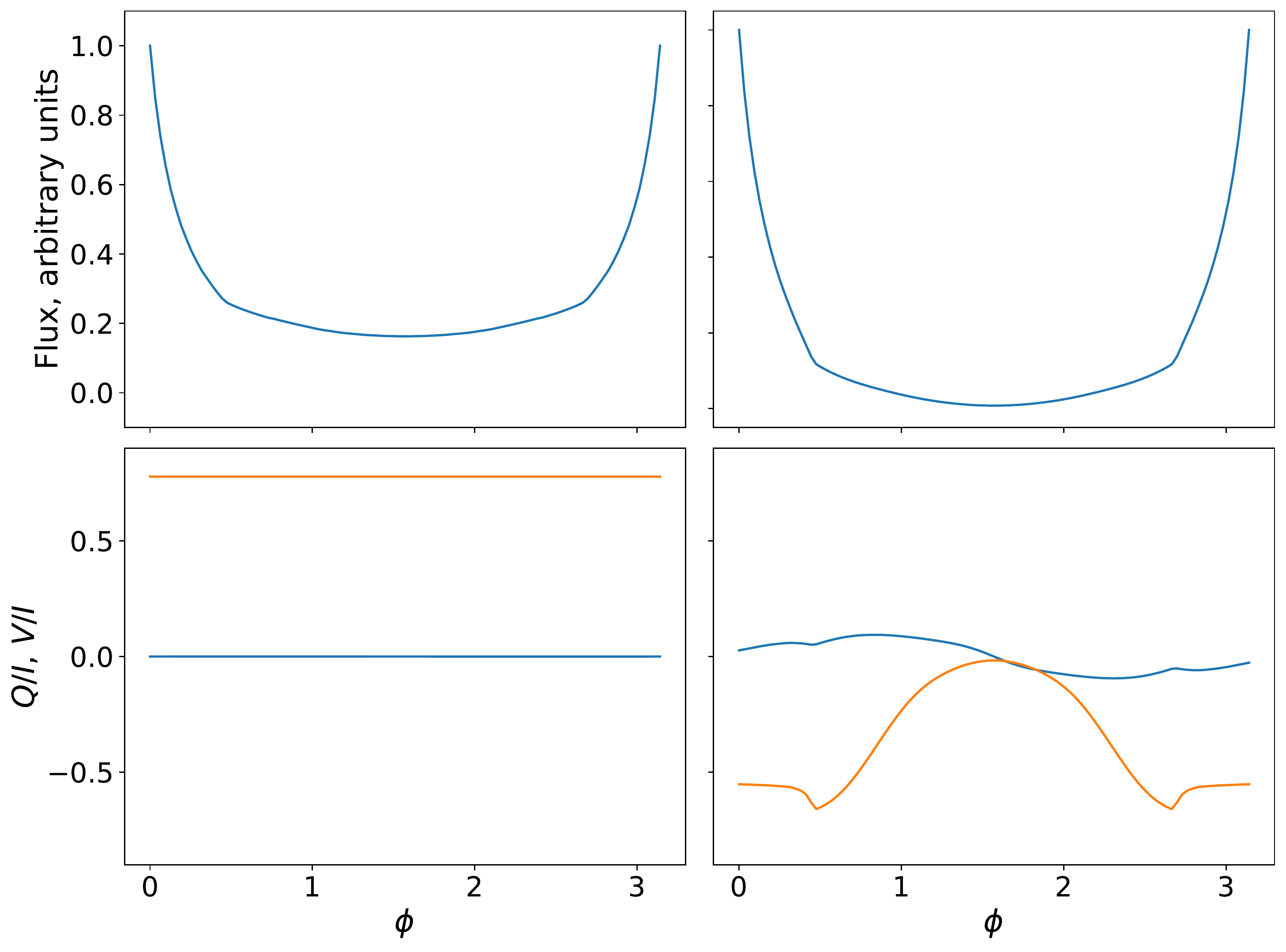}
    \caption{Phase pattern for intensity and polarisation fractions for 2 accretion columns with light bending. Upper panels: intensity $I$. Lower panels: linear polarisation fraction $Q/I$ solid orange line, circular polarisation fraction $V/I$ solid blue line. Left panels: photon energy 1~keV; right panels: photon energy 27~keV.}
    \label{fig:lb2col}
\end{figure}

\begin{figure*}
    \centering
    \includegraphics[width=0.383\textwidth]{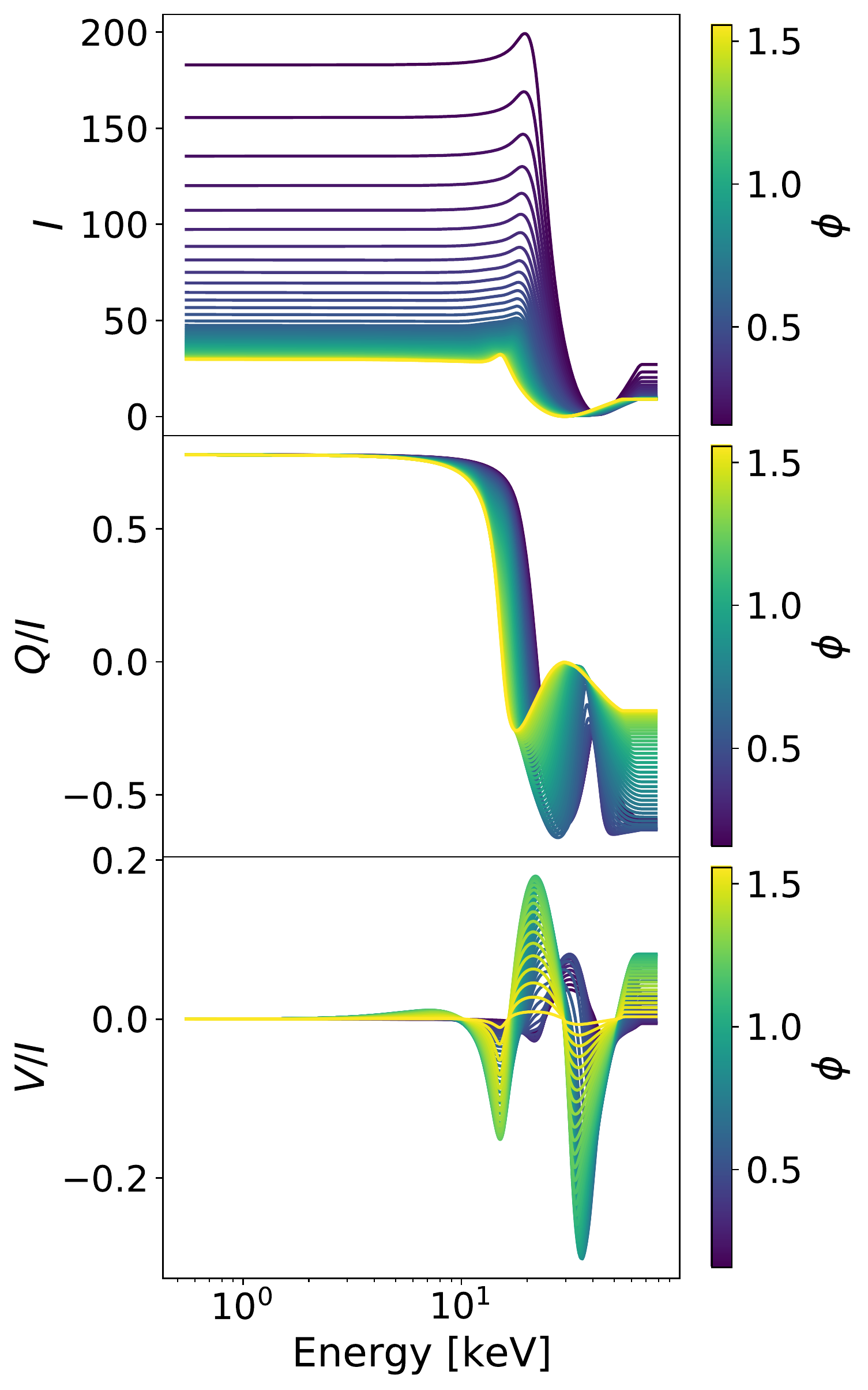}
    \includegraphics[width=0.38\textwidth]{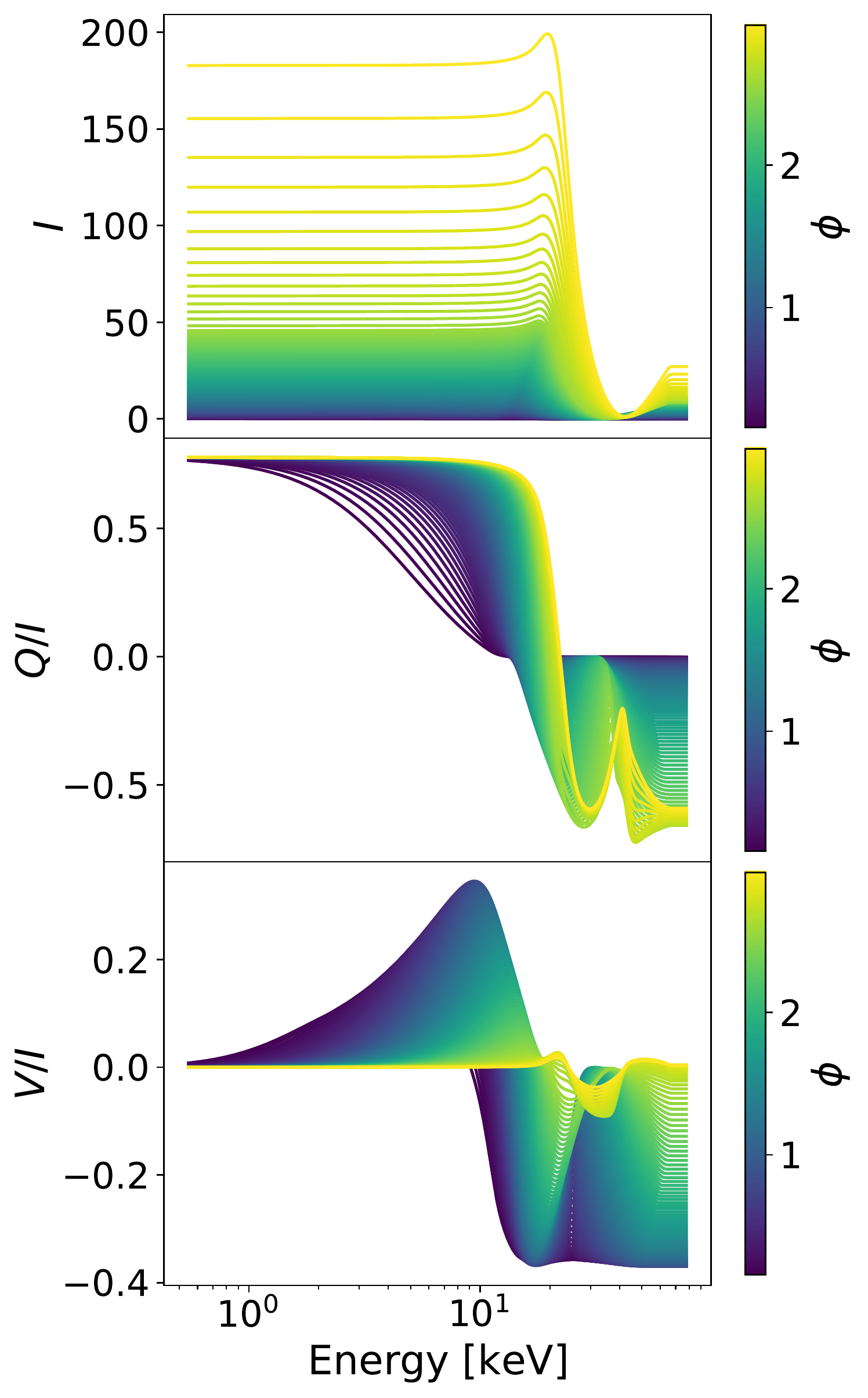}
    \caption{Average Stokes parameters for the emission at the observer for two columns (left) and only one column (right) with light bending. These figures do not yet include the effect of QED. The colour code represents the pulsar phase $\phi$, and it goes from 0 to $\pi$ for the one-column case and from 0 to $\pi/2$ for the two columns case.}
    \label{fig:StokeLB}
\end{figure*}

Furthermore, light bending and projection magnify the back column by a factor
\begin{equation}
{\cal M} = \frac{db}{dz} \frac{b}{\sqrt{z^2 \sin^2\psi + r_0^2}}
,
\end{equation}
where the added $r_0$ accounts for the finite radius of the column.  The peak magnification, which is of the order the ratio of the radius of the star and that of the column (about 100), is achieved when the column is pointed directly away from the observer.
The magnification, coupled with the fact that relativistic beaming, enhances the emission at high $\theta$, means that most of the emission comes from the back column. The huge magnification that we can see in Figures~\ref{fig:lb2col} and~\ref{fig:StokeLB} at $\phi\sim0$ and at $\phi\sim\pi$ hinges on the very particular geometry that we are considering: it requires the column to be pointing directly away from us, creating an Einstein ring around the star. The result of this effect is that very large pulsed fractions can be achieved by this model, just by varying the geometry of the star. In this particular geometry, the back column is magnified at $\phi\sim0$ (and at $\phi\sim\pi$), and the linear polarisation fraction at low angles is still high for low energies because it is dominated by the beamed emission from the back column, for which $\theta$ is high. The right panels of Fig.~\ref{fig:StokeLB} show the Stokes parameters for the one-column case, for which the magnification is achieved at $\phi=\pi$, when the only column is pointing directly away from us. 

An additional effect of general relativity is gravitational redshift: the energy of the photons is reduced as they travel out of the potential well of the star. The effect is different at different heights inside the column:
\begin{equation}
    E_o=E_p\sqrt{\frac{1-2GM_*}{c^2(R_*+z)}}
\end{equation}
where $E_p$ is the energy of the photon at emission, $E_o$ is the energy at the observer and $z$ is the height along the column. The x-axis in Fig.~\ref{fig:StokeLB} represents the energy at the observer.

\subsection{Including the effect of vacuum birefringence}
\label{sec:QED}
Before we can make a prediction for the polarisation parameters, we need to include an additional effect that comes from quantum electrodynamics (QED): vacuum birefringence. The presence of a strong magnetic field can make a medium birefringent, which means that the index of refraction in the medium depends on the angle between the polarisation of the photon and the magnetic field. In the case of the magnetised vacuum, the birefringence is caused by the interaction of photons with virtual electron-positron pairs: it is easier to excite virtual electrons along the direction parallel to the magnetic field than perpendicular to it, and thus photons in the ordinary mode travel slower than photons in the extraordinary mode \citep{1936...heisen...euler,1936...weisskopf,physrev.82.664}. 

The bulk of the accretion flow is confined to the accretion column, so we have two regions to consider: within the column, where the plasma density is high, and outside the column, where the plasma density is low. We can estimate the density within the column using the equation of continuity
\begin{equation}
    n m_p \pi r_0^2 v = {\dot M}.
    \label{eq:den}
\end{equation}
Because we consider that the radius of the column is constant and no mass is lost from the column, the density must increase as the velocity of the flow decreases down the column, so we will use the density at the top of column to provide a lower bound on the density. Using the parameters outlined in \S~\ref{sec:B&W} yields an electron number density at the top of the column of $3\times 10^{22}~\mathrm{cm}^{-3}$.  The results of \citet{1978SvAL....4..117G} indicate that at this density, the vacuum dominates the birefringence for photon energies above about 2~keV and below the cyclotron energy. In the vicinity of this critical energy, or vacuum resonance \citep[see][and references therein]{2019ASSL..460..301C}, the propagation modes through the magnetised plasma are approximately circular as opposed to approximately linear elsewhere. This could result in a modest depolarisation of the radiation near this energy, if a substantial depolarisation can occur between scattering events.  The typical scattering mean free path is about 50~cm at the top of the column, and decreases at lower heights, and the depolarisation length-scale is about 4~cm for photons travelling along the magnetic axis. Both of these values depend on angle, and therefore most photons will suffer significant depolarisation between scatterings and as they leave the column.

The effect is important only in a very narrow region in energy where the propagation modes are close to circular, centred around the vacuum resonance and the cyclotron resonances. Since the photon energy changes from scattering to scattering, this effect matters only if the energy of the photon after its final scattering within the column is very close to the resonance. We account for the depolarisation of the radiation after the final scattering as
\begin{equation}
Q'_{\textrm{QED}} = Q' \cos 2\theta_m . 
\end{equation}
where $Q'$ is calculated in Eq.~\ref{eq:escape} and $\theta_m$ measures how circular the propagation modes are and can be calculated using Eq.~12.29 of \citet{2019ASSL..460..301C}. Because the resonance lies at different energies along the column, the net effect of the vacuum resonance is modest. This prescription also accounts for the circularly polarised propagation modes near the cyclotron resonance. We do not perform a similar calculation for the circular polarisation, because as we show in the following, QED birefringence outside of the column suppresses the observed circular polarisation.

\begin{figure*}
    \centering
    \includegraphics[width=0.395\textwidth]{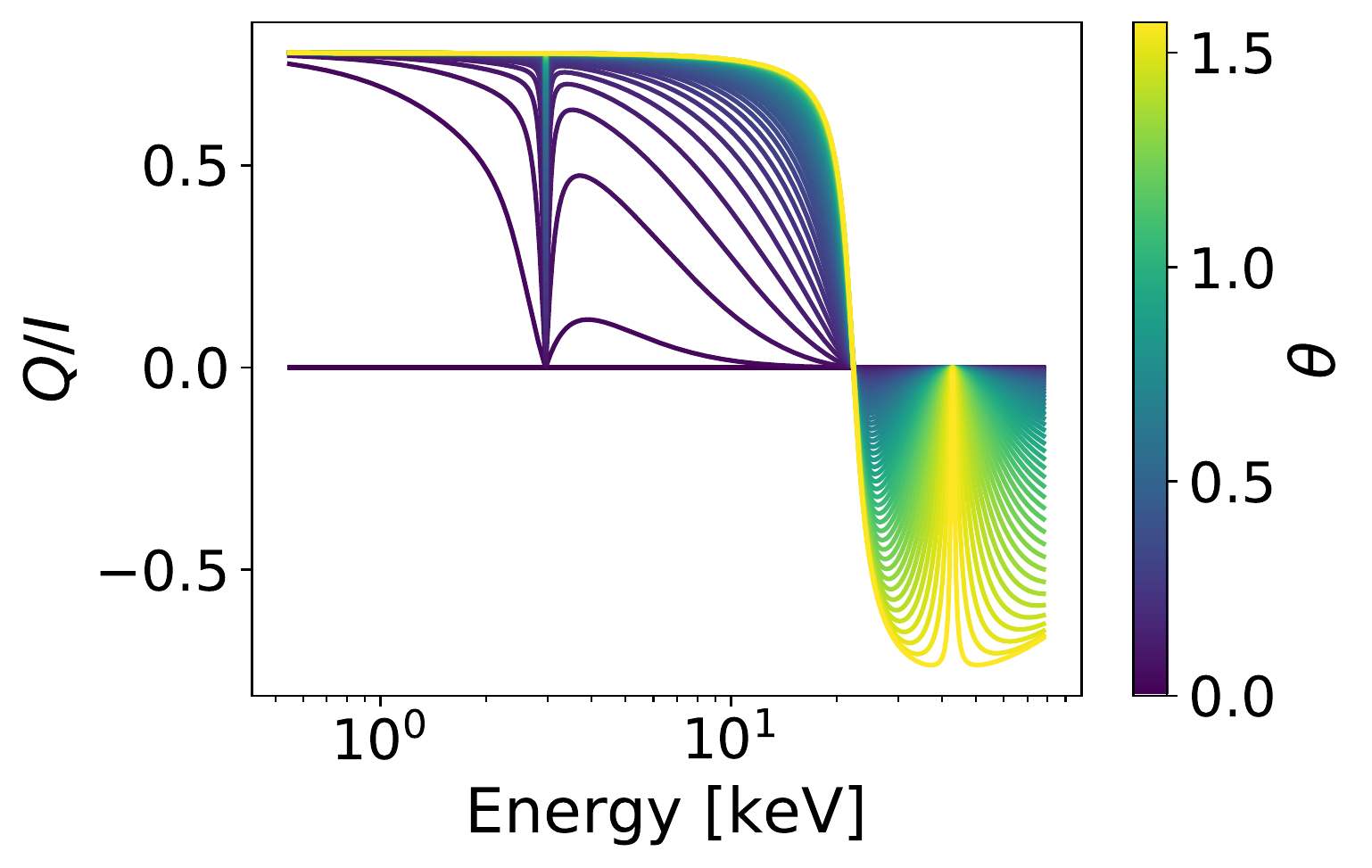}
    \includegraphics[width=0.39\textwidth]{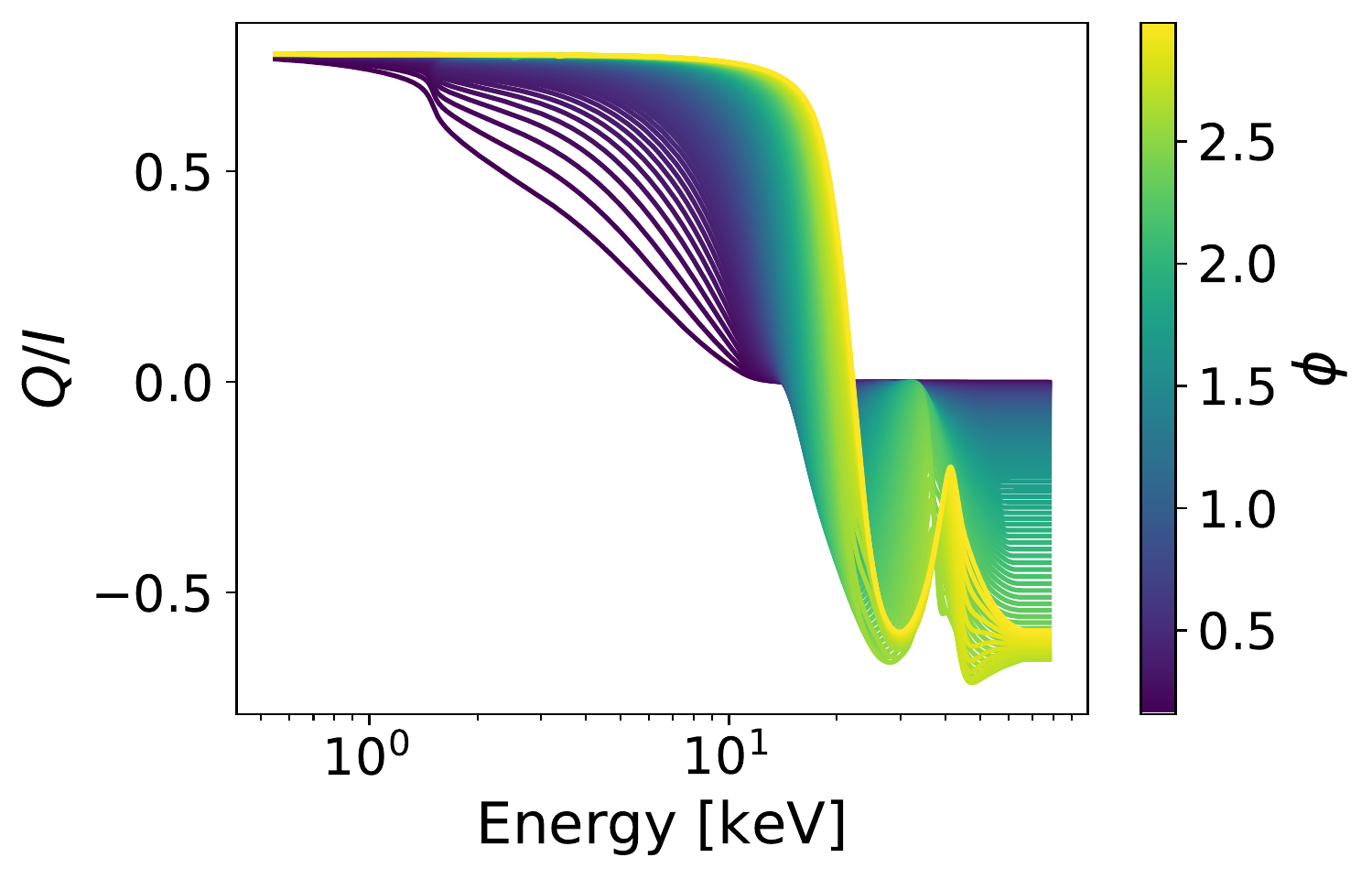}
    \caption{Average linear polarisation fraction for the one-column case. The left panel represents the linear polarisation fraction for radiation coming out of the column without the effects of relativistic beaming or general relativity (similar to the right central panel of Fig.~\ref{fig:in&out}), assuming only one value for the density in calculating the vacuum resonance. The right panel shows the linear polarisation fraction for the emission at the observer, including the effects of relativistic beaming and general relativity (similar to the right central panel of Fig.~\ref{fig:StokeLB}), and the vacuum resonance has been calculated using the density profile in eq.~\ref{eq:den}}
    \label{fig:withVP}
\end{figure*}

The left panel of Fig.~\ref{fig:withVP} shows how the vacuum resonance inside the column affects the polarisation spectrum of the radiation as it comes out of the column if a constant density is assumed. We can see that the plot is identical to the middle-right plot of Fig.~\ref{fig:in&out}, with the exception of a narrow range of energies close to the vacuum resonance at about 2~keV, where the polarisation fraction is lowered. The shape of the resonance feature in polarisation depends on angle as well, as the feature is narrower and shallower at higher angles. The density in the accretion column is not constant though, but increases as the gas sinks down the column. Since the feature is so narrow, adding the contributions from the different layers of the column washes out the feature and the only remaining effect is a small depolarisation at energies above a minimum energy, which corresponds to the energy of the vacuum resonance for the lowest density at the top of the column. This can be seen in the right panel of Fig.~\ref{fig:withVP}, where we show the linear polarisation fraction for the emission at the observer, including the effects of relativistic beaming and general relativity, for the one-column model, and where we have calculated the vacuum depolarisation assuming the density profile of eq.~\ref{eq:den}.  At high $\phi$, the polarisation in the right panel of Fig.~\ref{fig:withVP} is almost identical to the right central panel of Fig.~\ref{fig:StokeLB}, where QED is neglected, while at very low $\phi$ (viewing angles nearly parallel to the magnetic axis) we can see that the vacuum polarisation inside the column causes a small depolarisation at photon energies above about 2~keV.

Outside of the column the density of the plasma is assumed to vanish, so the propagation modes are linear corresponding to radiation polarised along and across the magnetic field. Because of birefringence, as photons propagate in the magnetosphere of neutron stars, their polarisation can still change even though the photons are propagating in vacuum. In general, for neutron stars, the effect of vacuum birefringence is to comb the polarisation direction to be aligned with the direction of the magnetic axis of the star and usually increases the observed total polarisation, with the effect being more dramatic the stronger the magnetic field \citep[][and references therein]{2018galax...6...76h,2019ASSL..460..301C}.  For radiation with a circularly polarised component, or a component linearly polarised at forty five degrees with respect to the magnetic field, the net effect is to destroy the polarisation. 

In the case of a thin accretion column, where the polarisation is already aligned with the magnetic field axis, the effect of vacuum birefringence in the propagation outside the column reduces to the so-called Quasi-Tangential (QT) effect: as photons travel inside the neutron star's magnetosphere, the polarisation of X-ray photons can change significantly when they cross the quasi-tangential point, where the photon momentum is nearly aligned with the magnetic field, and the net effect, when averaged over a finite emission area, is to decrease the fraction of linear polarisation.

\begin{figure*}
    \centering
    \includegraphics[width=0.395\textwidth]{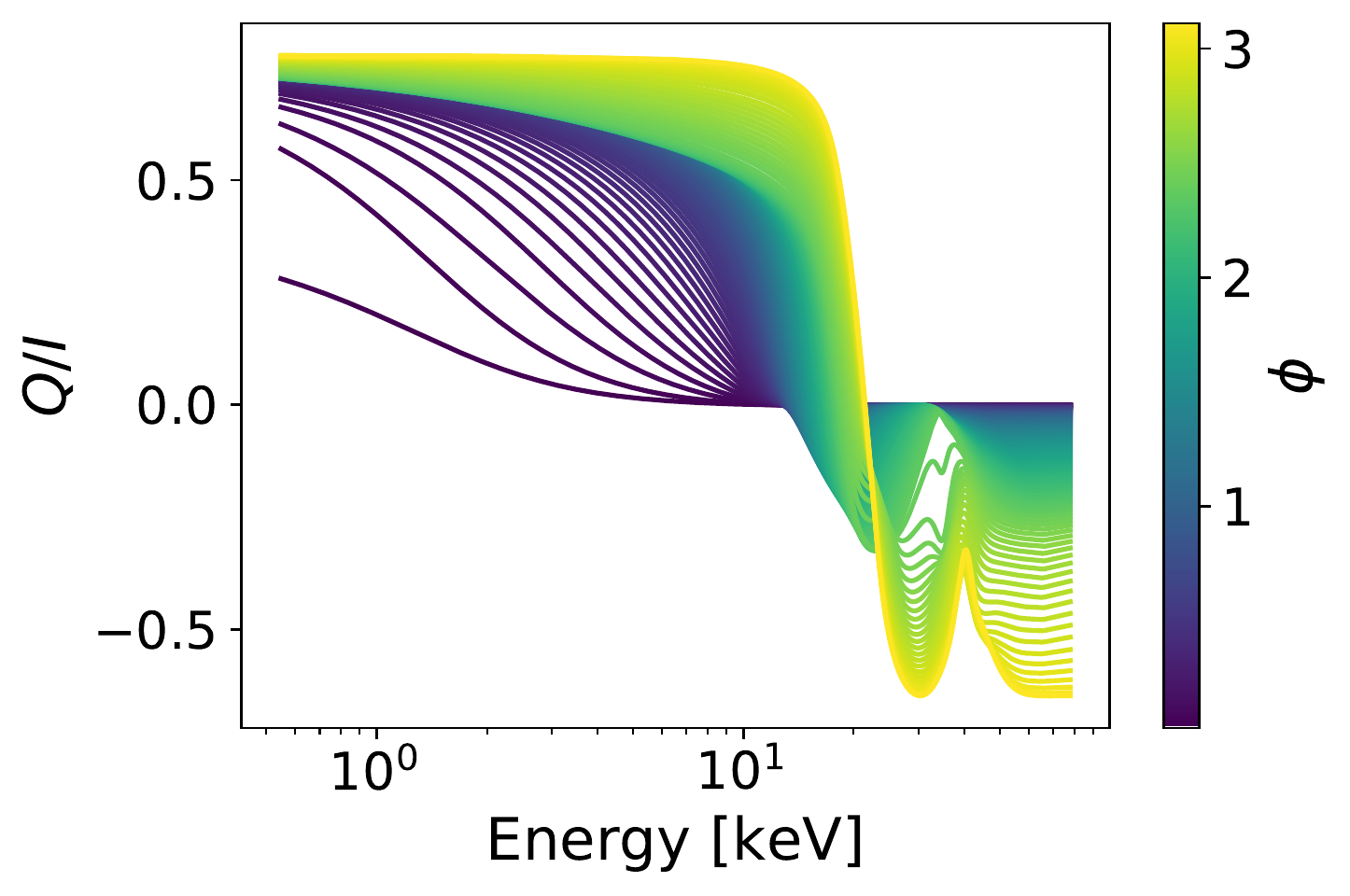}
    \includegraphics[width=0.395\textwidth]{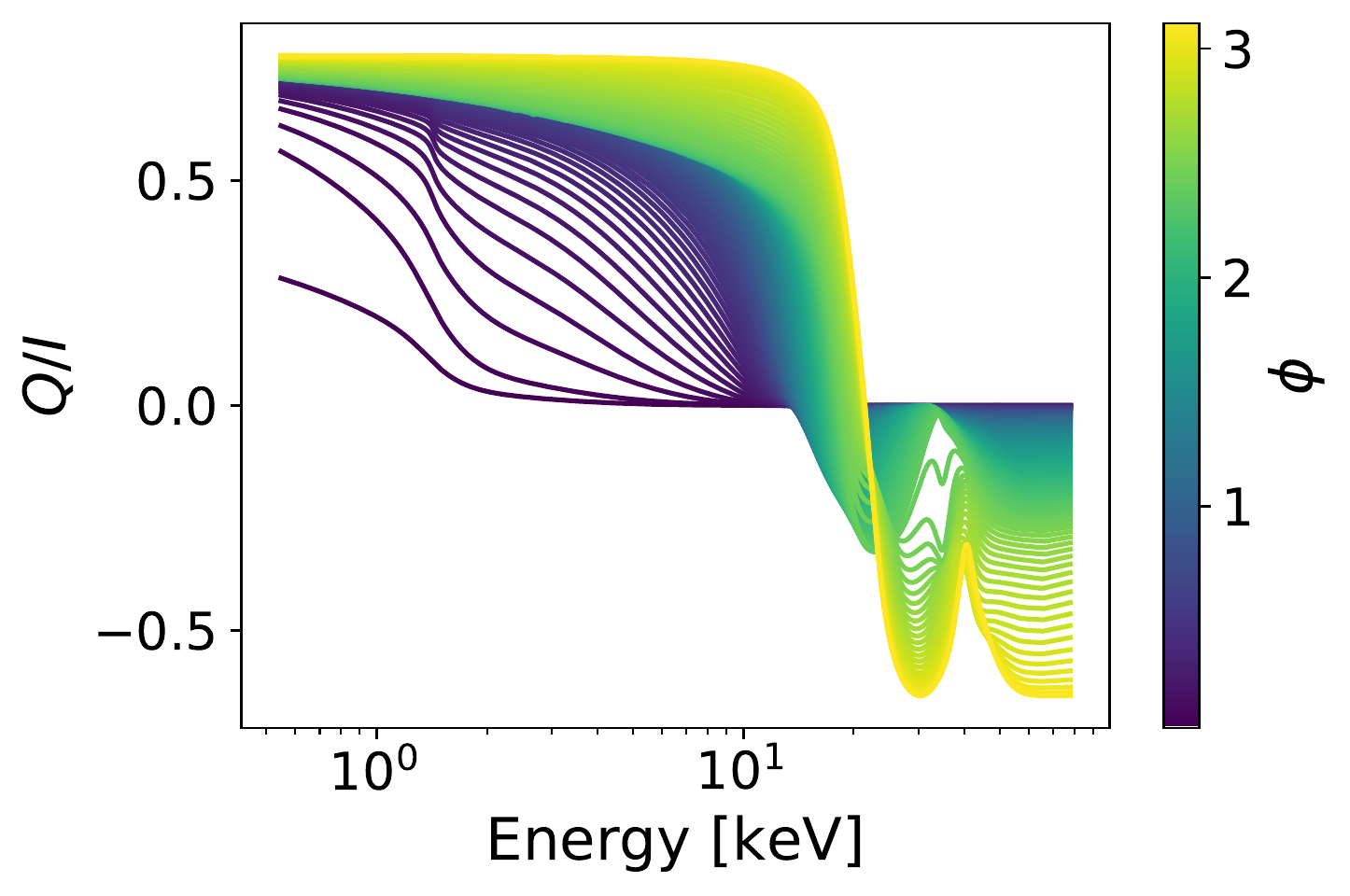}
    \includegraphics[width=0.395\textwidth]{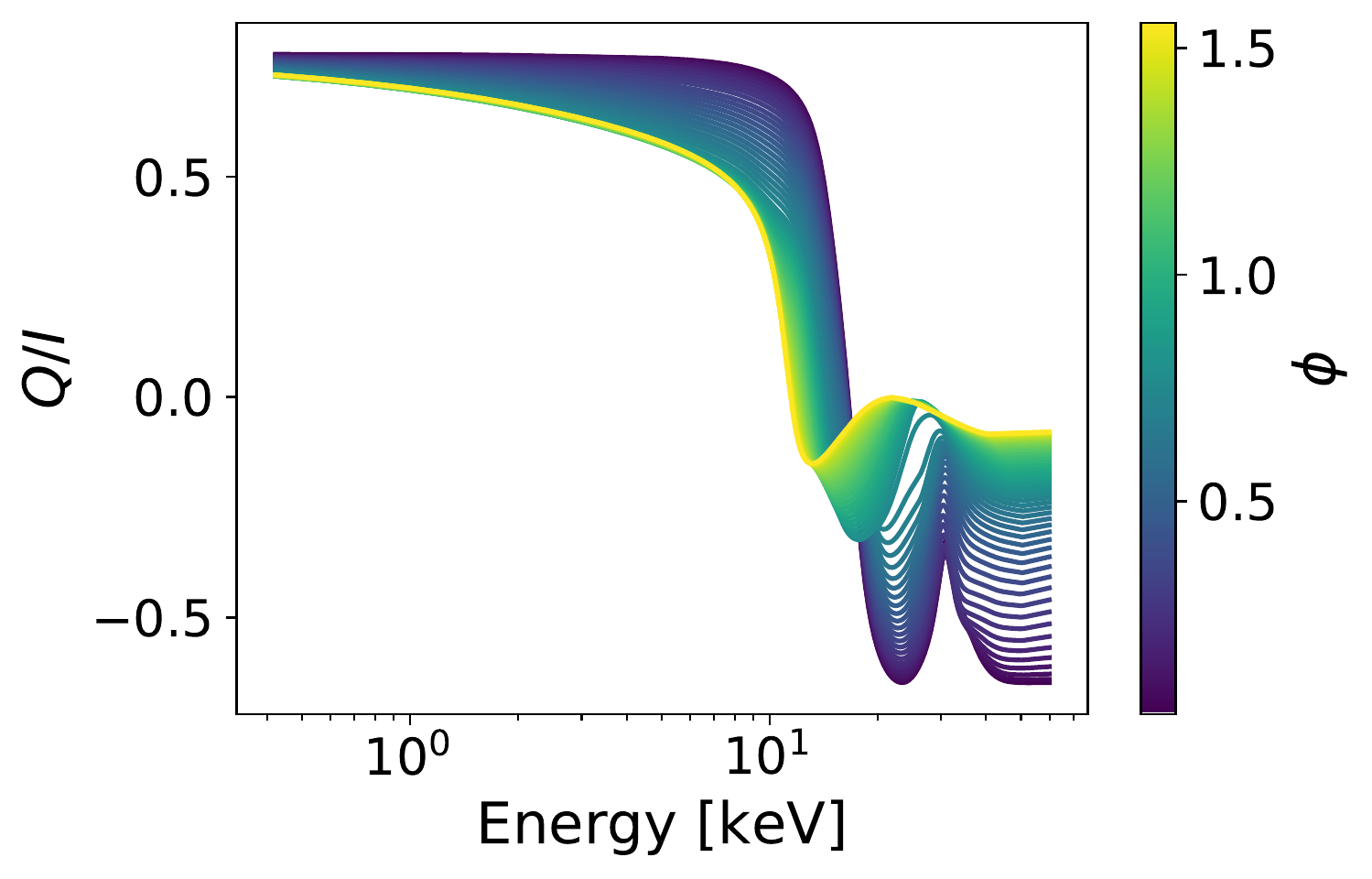}
    \includegraphics[width=0.395\textwidth]{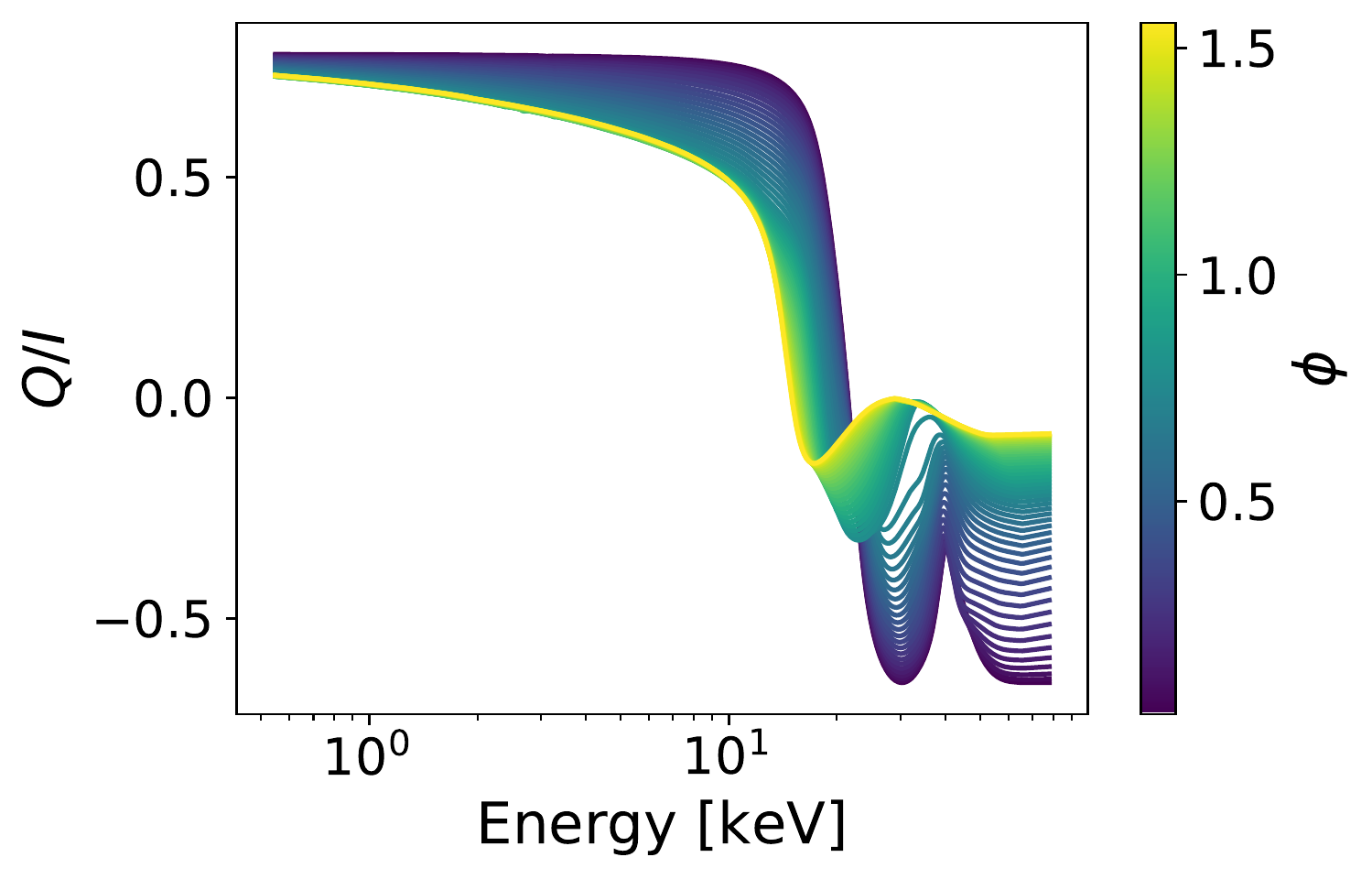}
    \caption{Average linear polarisation fraction at the observer with only the QT effect included (left) and with full QED (right). The upper panels depict the one column case and the lower panels the two column case. The colour code represents the pulsar phase $\phi$, and it goes from 0 to $\pi$ for the one column case and from 0 to $\pi/2$ for the two columns case.}
    \label{fig:finalQT}
\end{figure*}

\citet{2009mnras.398..515w} calculated the QT effect for radiation coming from a hot-spot on the surface on the neutron star. For this reason, they only considered emission from angles $\psi<\pi/2$. In Appendix~\ref{sec:QT}, we expand their calculation to higher angles to accommodate our case. We find that, while the intensity is not affected by the QT crossing, the linear and circular polarisation are. In particular, the circular polarisation of each photon receives a random rotation, and therefore the average circular polarisation of the emission is completely destroyed. The linear polarisation, on the other hand, is reduced by an amount that depends on where along the column the light is emitted, on the rotational phase of the star and on the energy of the photon. The effect on linear polarisation of the QT crossing is shown in the left panels of Fig.~\ref{fig:finalQT} for the one-column and the two-column cases, while the right panels include the effects of vacuum birefringence both inside and outside the column. If we compare the plots with the middle panels of Fig.~\ref{fig:StokeLB}, where QED is not included, we can see that QT crossing is the predominant QED effect, while the vacuum polarisation inside the column has only a very moderate effect for very low phases in the one-column model, when the column is pointing toward the observer. In the two-column model, the column facing away from the observer is always the one contributing the most, and therefore the effect of vacuum polarisation inside the column is negligible at all phases.

\section{Comparison with previous models}
\label{sec:comp}
The spirit of the model presented here is quite different from previous works \citep{1980ApJ...238.1066M,1981ApJ...251..278N,1981ApJ...251..288N,1982Ap&SS..86..249K,1985ApJ...298..147M,1985ApJ...299..138M,1986PASJ...38..751K,1987pasj...39..781k}. Over the past forty years since these models were developed, astronomers have acquired a vast amount of data about X-ray pulsars, but still no polarisation. We have built the model presented here to include most of physical processes that influence the observed X-ray emission: spectral formation, special and general relativity and quantum-electrodynamics. The focus has been to produce a model that can characterise the observations in detail while simplifying some theoretical details such as radiative transfer. Our model is built upon the B\&W picture for the formation of the spectra of X-ray pulsars and lacks the detailed radiative transfer of these earlier models, whereas the earlier models lack the details of the macroscopic structure and dynamics of the emission region and the propagation of the radiation to the observer, including general relativity and quantum-electrodynamics. These new ingredients strongly affect the emission that we observe, as can bee seen from Fig.~\ref{fig:beamStoke} onward.  

By design, our model accounts for the phase-averaged emission from X-ray pulsars through the B\&W model, whereas the earlier models do not fit the spectroscopic observations of X-ray pulsars from 0.5 to 50~keV. Thus, in our comparison, we will focus on how the emission varies as a function of angle, energy and polarisation in the different models. The earlier models calculate the emission from a small, uniform region with a slab or cylindrical geometry, so we will compare the results for a small region of our model, a single layer of the accretion column, which has a similar cylindrical geometry. Our results for the emission from a single layer are depicted in the right panel Fig.~\ref{fig:in&out} and in Fig.~\ref{fig:beaming} before including the relativistic beaming (blue curves). In particular, far from the cyclotron resonance, the intensity varies approximately as $\sin^2\theta$ yielding modest variability.  Similar results were found by \citet{1985ApJ...299..138M}, \citet{1982Ap&SS..86..249K} and \citet{1986PASJ...38..751K} for the cylinder model.  The only configurations in the context of these earlier models that yielded a strong pulsation were the so-called deep slabs, where a substantial intensity is beamed along the magnetic field direction, and only the deep slabs could hope to reproduce the large pulse fractions observed in X-ray pulsars such as Her X-1.

Turning to the polarisation parameters, in our models, in the direction perpendicular to the magnetic field (see Fig.\ref{fig:in&out}), the polarisation fraction is about eighty percent O-mode at low energy and changes rapidly at about 20~keV to be about eighty percent X-mode at higher energies except near to the cyclotron energy. At shallower angles with respect to the magnetic field, the transition as a function of energy is more gradual but both the polarised fractions at low and high energies approach eighty percent in O and X-mode respectively. The deep column model of \citet{1985ApJ...299..138M}, with a cyclotron energy of 38~keV and electron temperature of 8~keV exhibits a similar polarisation trend, except that the net polarisation vanishes at low energies (below 2~keV), and the total polarised flux peaks at fifty-five percent at about 6~keV. The model presented by \citet{1986PASJ...38..751K} following the slab configuration of \citet{1981ApJ...251..278N} exhibits a similar intensity profile as a function of viewing angle to our models, even exhibiting the flat-top peak at higher energies shown in the right panel of Fig.~\ref{fig:beaming}.  This model has strong polarisation even at low energies and switches modes at about 5~keV, a lower energy than our model, in spite of the fact that they assumed a cyclotron energy of 100~keV.  As the radiation leaves the column, we also find broad agreement with the results of \citet{1982Ap&SS..86..249K} for the beaming and the extent of polarisation as a function of energy and angle.  

The main difference between the model presented here for the radiation emerging from a layer in the column and the earlier models is the extent of pulsation of the emission near the cyclotron energy. Broadly speaking, the earlier models exhibit stronger dependence on emission angle near the cyclotron feature that we see in our model. Away from the cyclotron feature, the results are similar. Once the radiation leaves the column, however, the earlier models propagate the radiation to the observer only considering the geometry of the pulsar, so the translation between the properties of the radiation at the column and infinity is simply geometric.  Here, we consider several additional effects that have important consequences: the scattering electrons are moving relativistically at the top of the column and slow down at lower layers; the density is not constant, but increases down the column; the paths of photons leaving the column are bent by the gravitational field of the neutron star and their energy is gravitationally redshifted; the effect of vacuum birefringence as the photon propagates outside the column is also important. We will discuss each effect and the final results in the following section.

\section{Summary and Discussion}
\label{sec:discussion}
The production of polarised radiation in the column is dominated by the strong magnetic field, but other effects have to be taken into account to calculate the observed polarisation: relativistic beaming, gravitational lensing and QED. We here summarise how the different physics affect the polarisation and pulse fraction seen at the observer.

The very high average number of scatterings per photon ($\gtrsim250,000$) that is predicted by the B\&W model leads to an average polarisation state inside the column that is determined uniquely by the energy of the photon and by the strength of the magnetic field, and that is independent of the initial polarisation of the photon (Fig.~\ref{fig:in&out}, left panels). At low energies (far from the cyclotron energy) the average photon inside the column is nearly 100\% polarised in the ordinary mode, except for photons propagating in a direction almost parallel to the magnetic field ($\theta = 0$). In the direction parallel to the magnetic field (and to the column axis) the flux is drastically lowered, as the intensity peaks in the direction perpendicular to the column walls ($\theta = \pi/2$). Close to the cyclotron energy, there is an inversion in the polarisation direction (so that the linear polarisation fraction goes through a zero), and at the cyclotron energy photons are mostly polarised in the extraordinary mode. At low angles, close to the cyclotron energy, circular polarisation is predominant, as photons travelling parallel to the magnetic field can resonantly scatter off electrons, that receive enough energy to jump to the second Landau level. The circular polarisation fraction decreases with $\theta$.

As photons escape the column, the difference in scattering cross section between photons polarised in the different modes changes the picture (Fig.~\ref{fig:in&out}, right panels). At low energies, the scattering cross sections for light polarised in the O- or in the X-mode differ by several orders of magnitude (except for photons propagating parallel to the magnetic field, for which they are equal, see Fig.~\ref{fig:crosssections}). For this reason, photons in the extraordinary mode can escape freely, while photons in the ordinary mode are trapped. This difference causes the linear polarisation degree to drop from about 100\% at low energies to about 80\%. The increase of all the cross sections close to the cyclotron energy also reduces the emission at high energy (the relative intensity drops sharply above 20~keV). Please note that, as we are only calculating the effect of scattering on the Stokes vector and not solving the radiative transfer equation, the intensity shown in the figures is a relative intensity, which needs to be normalised as a function of energy against a phase-averaged spectrum.

The right plots of Fig.~\ref{fig:in&out} show the polarisation parameters of photons coming out of the column in the frame of the accreting gas. In order to calculate the parameters in the frame of the observer, the effects of relativistic aberration and beaming need to be included, especially for photons emitted at the top of the accretion column, because the electrons in the gas have a bulk downward speed that is as high as half the speed of light at the top of the column and decreases as the gas approaches the stellar surface. The radiation scattered by relativistic electrons is strongly beamed downward (Fig.~\ref{fig:beaming}) and the features described in the previous paragraph are shifted in energy by an amount that increases with the emission angle, (see Fig.~\ref{fig:beamStoke} for an electron velocity of $\sim0.4~c$). The amount of relativistic beaming depends on the velocity of the electrons and therefore changes along the column. Fig.s~\ref{fig:phase2col} and~\ref{fig:Stoke2col} show the effect of beaming when summing over the entire column and considering one or two accretion columns.

Another important factor that changes drastically what we expect to see at the observer is general relativity. Gravitational lensing bends the path of light and distorts the image of the star, while gravitational redshift lowers the energy of the photons as they leave the potential well of the neutron star. An important consequence is that part of the back of the star's surface becomes visible, and that the accretion column is visible also when it is in pointing away from us, even close to $\phi\sim0,~\pi$. If the height of the column is $\sim7$ km, as predicted by the B\&W model, at least part of the column is seen at all phases, and when it is exactly in the opposite direction with respect to the line of sight ($\phi=0,~\pi$), gravitational lensing generates a huge magnification, as it projects the column in an Einstein ring around the star (see Fig.~\ref{fig:lb2col}). Of course, if the back column is not perfectly aligned with the line of sight, the effect is reduced, and therefore, depending on the geometry of the system, very large pulse fractions can be achieved. 


The final effect to be considered is the effect of QED, and specifically the effect of vacuum birefringence. Inside the column, the main effect is a depolarisation around the narrow feature at the vacuum resonance, whose energy depends on the density of the gas. In our model, the column has a range of densities starting at the top of about $3 \times 10^{22}~\mathrm{cm}^{-3}$, a factor of ten lower than typically considered in earlier models, and increasing down the column. This changing density serves to dilute the vacuum resonance feature as discussed in \S~\ref{sec:QED}, so instead of seeing a sharp dip in polarisation at the vacuum resonance, there is a slight decrease in the extent of polarisation for viewing angles approximately along the magnetic field direction.  Because the bulk of the radiation emerges perpendicular to the field direction, this decrement does not have a significant effect on the observed polarisation. An additional, and more pronounced effect is caused by the polarisation of the magnetised vacuum in which the photons travel after they have left the column, and specifically by the quasi-tangential effect: when a photon crosses a region where its momentum is nearly aligned with the local magnetic field (the QT region), the polarisation direction of the photon can rotate. The overall effect is a reduction of the linear polarisation fraction and a complete destruction of the circular polarisation. The final linear polarisation fraction for the 1 and 2 columns possibilities is shown in Fig.~\ref{fig:finalQT}.

\begin{figure*}
    \centering
    \includegraphics[width=\textwidth]{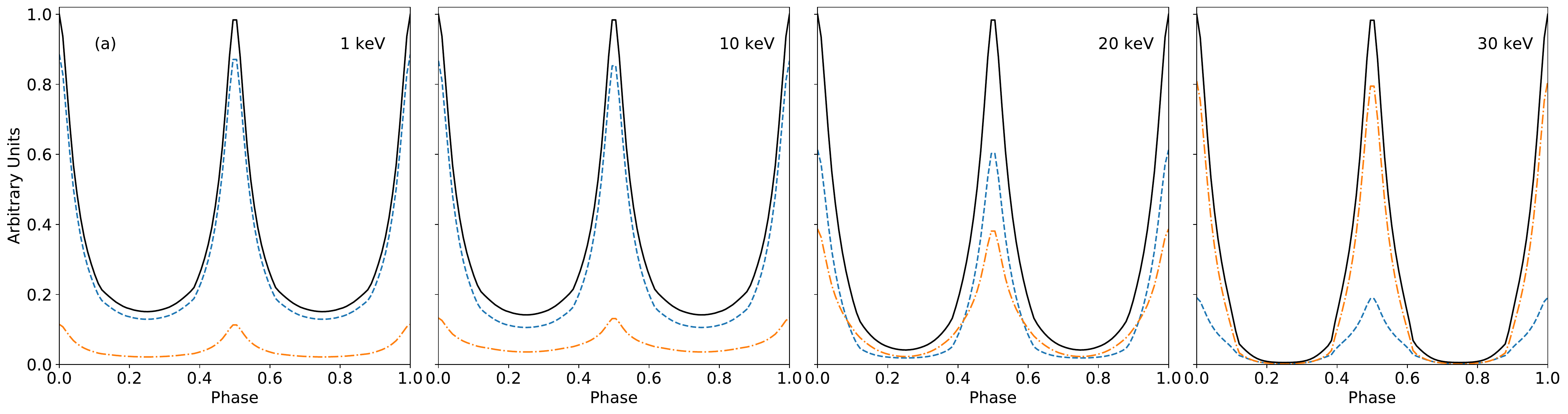}
    \includegraphics[width=\textwidth]{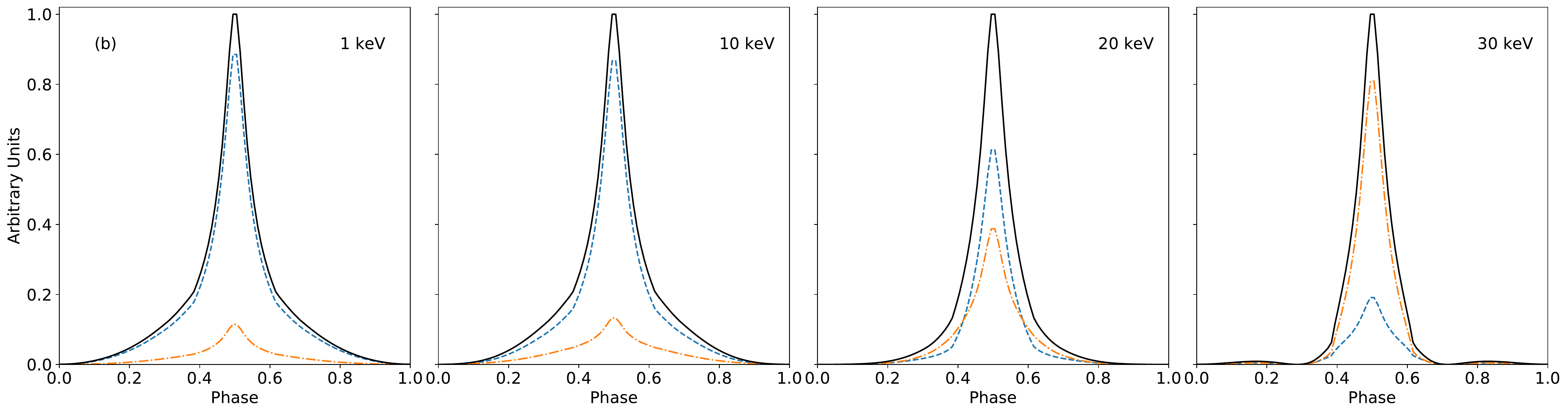}
    \includegraphics[width=\textwidth]{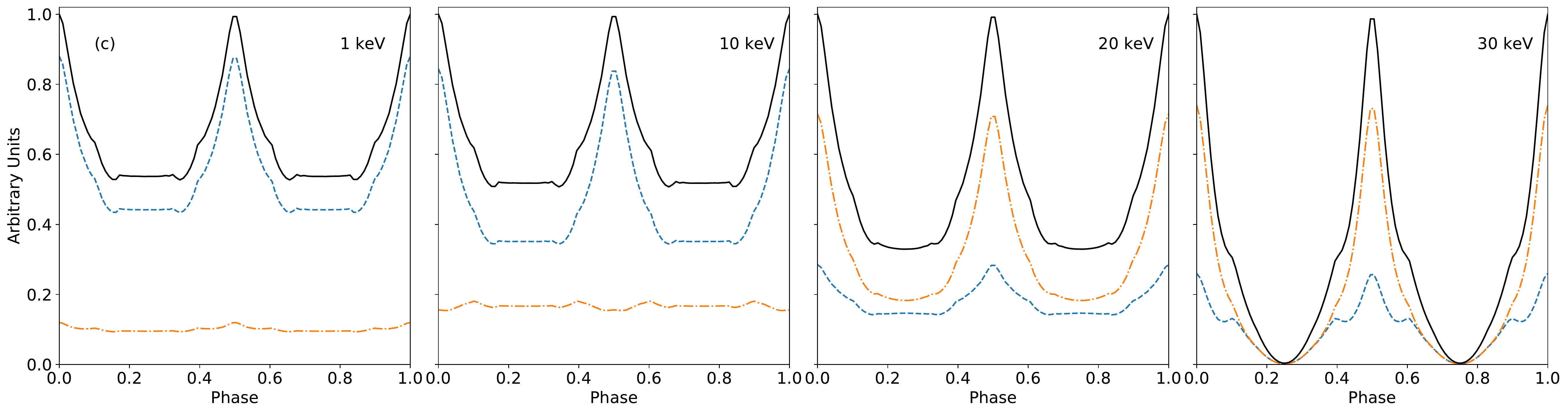}
    \includegraphics[width=\textwidth]{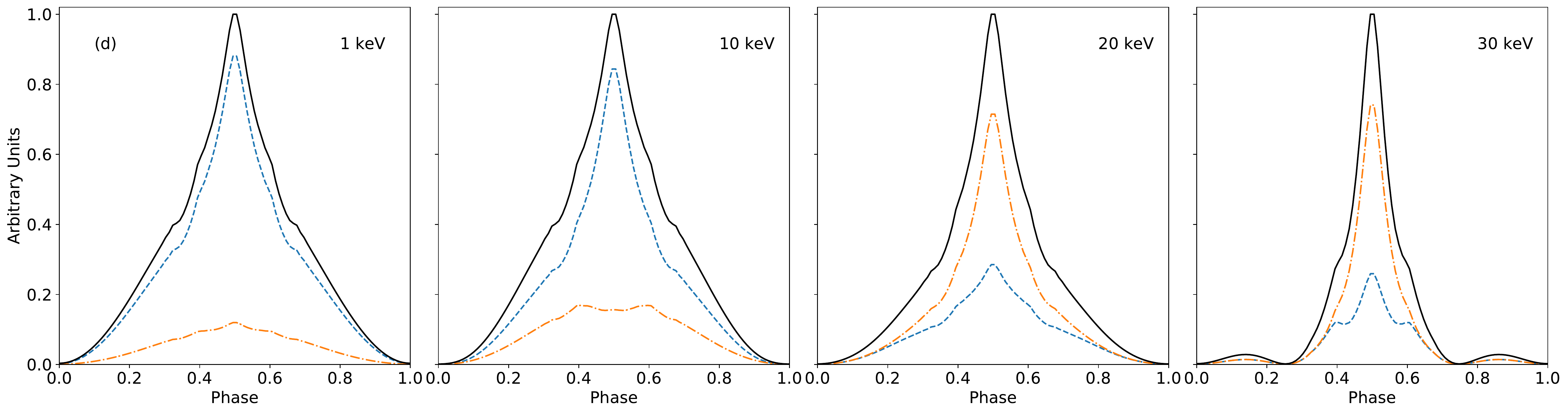}
    \caption{Polarisation parameters as a function of phase and energy for four different models: (a) two-column, $z_{\rm{max}}=6.6$~km, (b) one-column, $z_{\rm{max}}=6.6$~km, (c) two-column, $z_{\rm{max}}=1.4$~km, (d) one-column, $z_{\rm{max}}=1.4$~km. The angle between the line of sight and the rotation axis of the neutron star is $\alpha=90^\circ$ and the angle between the rotation axis and the magnetic field is $\beta=80^\circ$ (see Appendix~\ref{sec:geo}).  Black solid line: relative intensity $I$; blue dashed line: intensity in the $O$-mode, orange dash-and-dotted line: intensity in the $X$-mode. From left to right, photon energy at the observer: 1~keV, 10~keV, 20~keV, 30~keV.}
    \label{fig:finalpulse}
\end{figure*}
\begin{figure*}
    \centering
    \includegraphics[width=\textwidth]{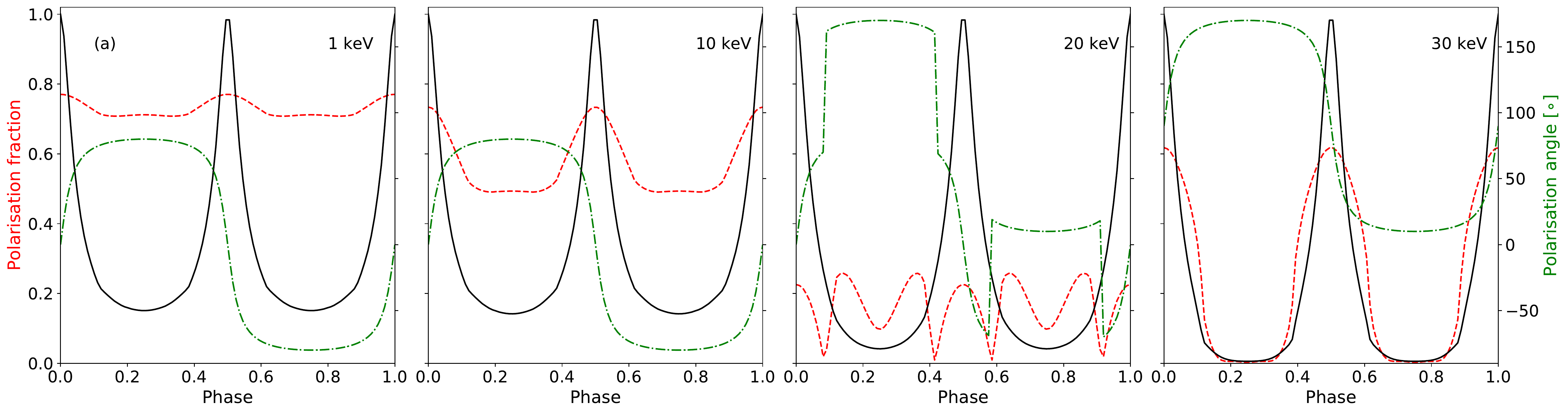}
    \includegraphics[width=\textwidth]{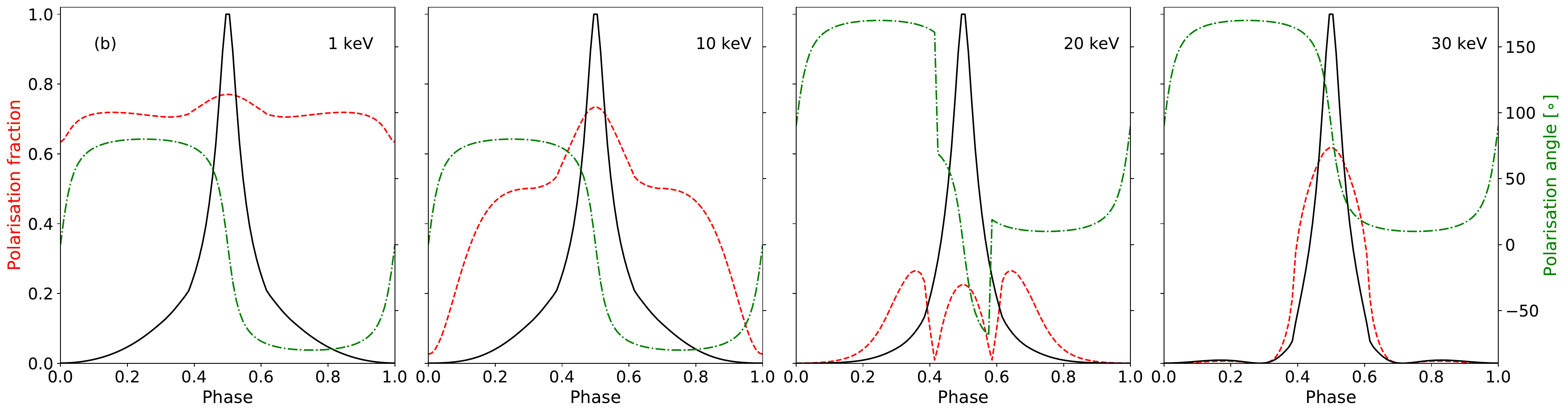}
    \includegraphics[width=\textwidth]{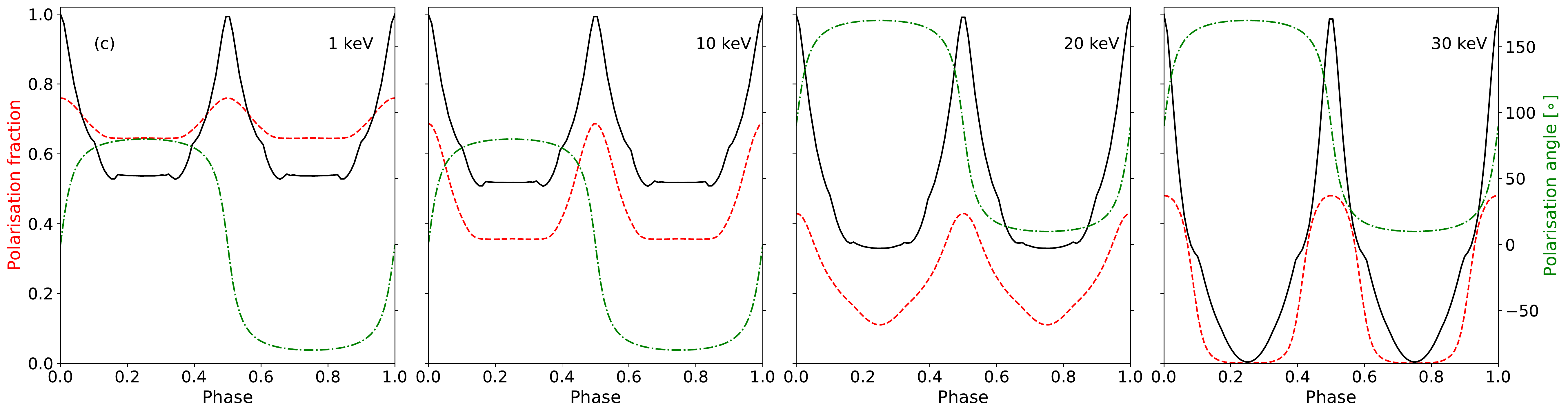}
    \includegraphics[width=\textwidth]{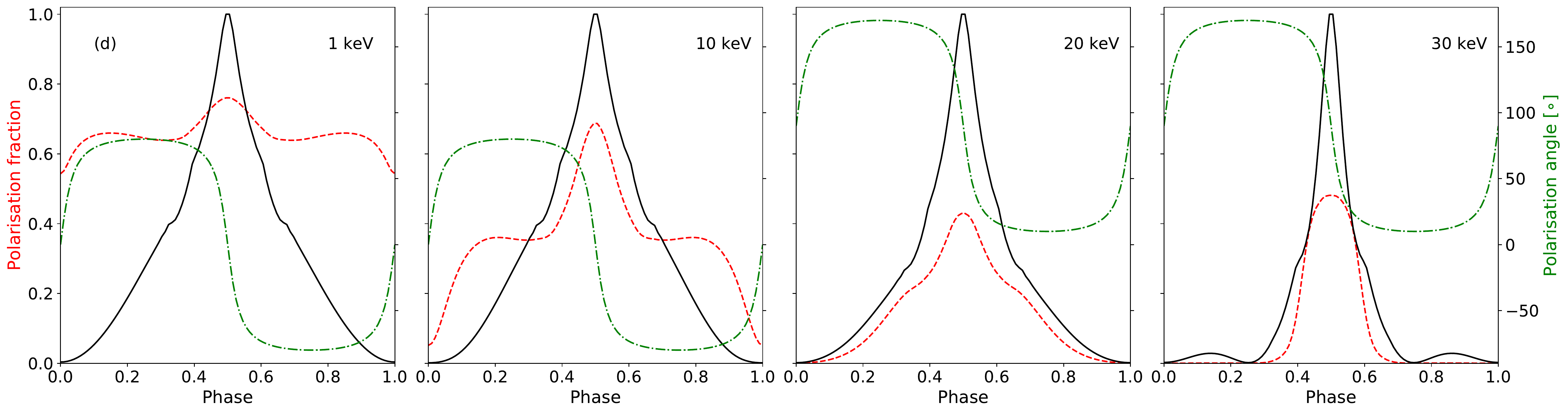}
    \caption{Polarisation parameters as a function of phase and energy for four different models: (a) two-column, $z_{\rm{max}}=6.6$~km, (b) one-column, $z_{\rm{max}}=6.6$~km, (c) two-column, $z_{\rm{max}}=1.4$~km, (d) one-column, $z_{\rm{max}}=1.4$~km. The angle between the line of sight and the rotation axis of the neutron star is $\alpha=90^\circ$ and the angle between the rotation axis and the magnetic field is $\beta=80^\circ$ (see Appendix~\ref{sec:geo}). Black solid line: relative intensity $I$; red dashed line and left y-axis: polarisation fraction, green dash-and-dotted line and right y-axis: polarisation degree. From left to right, photon energy at the observer: 1~keV, 10~keV, 20~keV, 30~keV.}
    \label{fig:finalpulse2}
\end{figure*}
\begin{figure*}
    \centering
    \includegraphics[width=\textwidth]{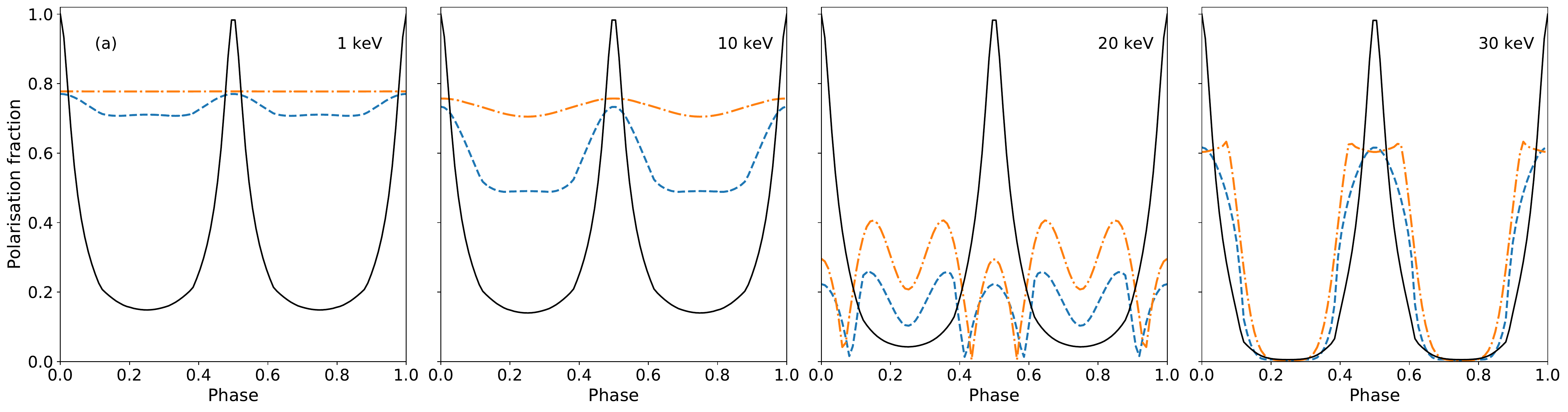}
    \includegraphics[width=\textwidth]{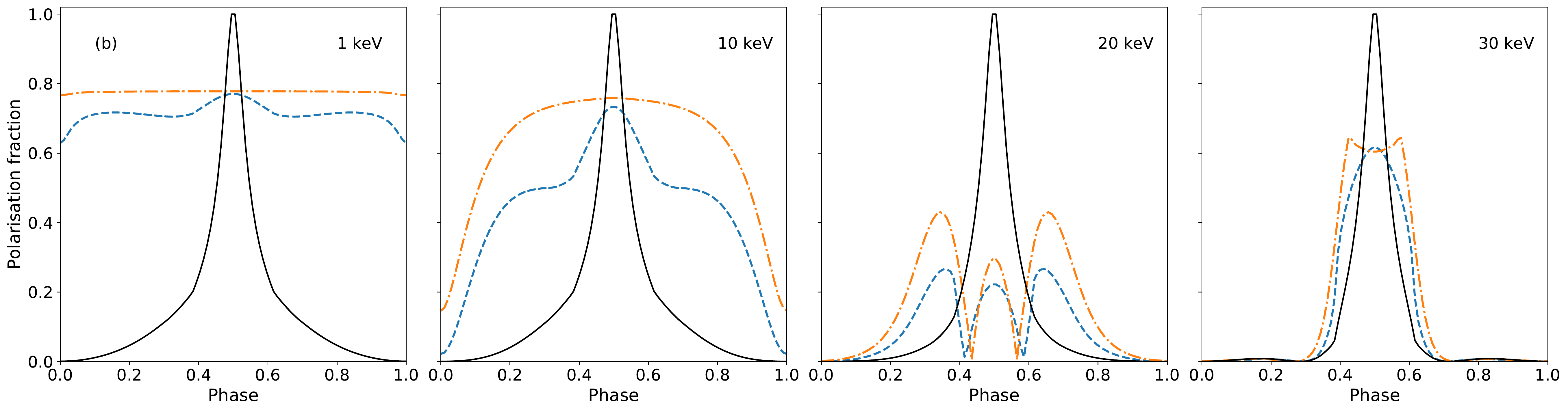}
    \includegraphics[width=\textwidth]{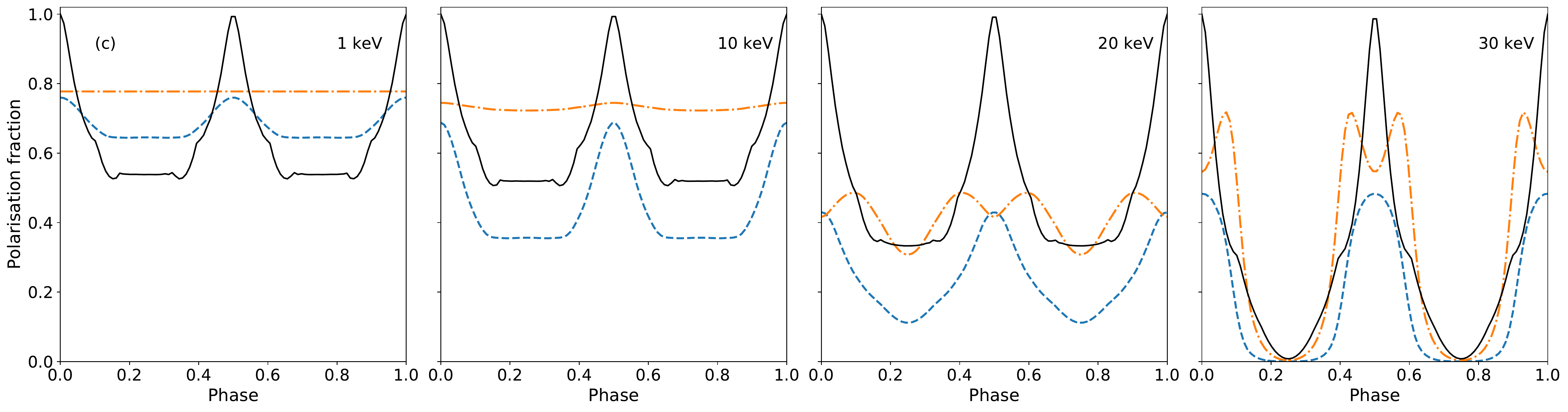}
    \includegraphics[width=\textwidth]{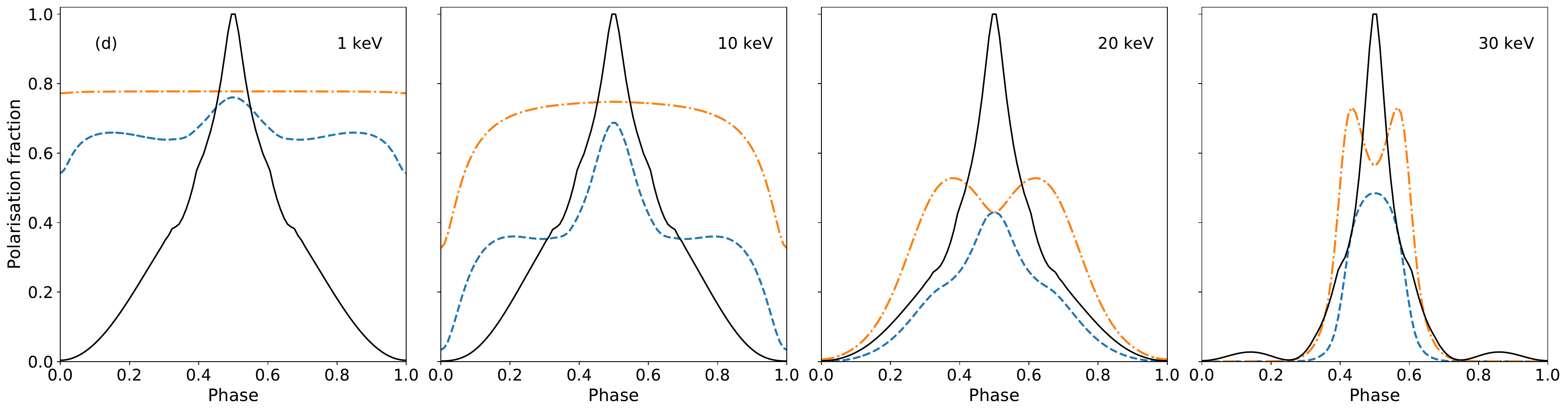}
    \caption{Comparison of the predicted polarisation fraction including and without QED for four different models: (a) two-column, $z_{\rm{max}}=6.6$~km, (b) one-column, $z_{\rm{max}}=6.6$~km, (c) two-column, $z_{\rm{max}}=1.4$~km, (d) one-column, $z_{\rm{max}}=1.4$~km. The angle between the line of sight and the rotation axis of the neutron star is $\alpha=90^\circ$ and the angle between the rotation axis and the magnetic field is $\beta=80^\circ$ (see Appendix~\ref{sec:geo}). Black solid line: relative intensity $I$; blue dashed: polarisation fraction including QED; orange dash-and-dotted line: polarisation fraction without QED. From left to right, photon energy at the observer: 1~keV, 10~keV, 20~keV, 30~keV.}
    \label{fig:finalpulse3}
\end{figure*}

In Figs.~\ref{fig:finalpulse},~\ref{fig:finalpulse2} and~\ref{fig:finalpulse3} we analyse how the predicted polarisation parameters at the observer vary with phase and energy. As in the perfectly orthogonal rotator geometry the accretion column points directly away from the observer, and the magnification of the column by gravitational lensing is huge, we choose to show a slightly different geometry because showing the orthogonal rotator would make interpreting the results difficult and also would represent an unlikely situation. In the figures, the the rotation axis lies in the plane of the sky (the angle $\alpha$ between the line of sight and the rotation axis is $90^\circ$, see Appendix~\ref{sec:geo}) and the angle between the rotation axis and the magnetic field ($\beta$ in Appendix~\ref{sec:geo}) is $80^\circ$.
We present four different configurations: emission from one or two symmetrically located columns for a short column (1.4~km, with an adiabatic shock at the top) or a tall column (6.6~km, as predicted by the B\&W model).  For the short column, the effects of relativity (special and general) are less pronounced.  First, the velocity of the flow at the top of column is reduced by a factor of seven from about half of the speed of light to less than one tenth.  Furthermore, because the column is short, it is more difficult to see it when it is pointing away from the observer, which is when the effects of gravitational lensing are the most dramatic.  A comparison of the second and fourth row of Fig.~\ref{fig:finalpulse} highlights the relativistic effects.  The pulse is much sharper and more compact for the taller column where both special and general relativity are more important.  For the tall column at 20~keV, the radiation in the pulse is mainly polarised in the O-mode similar to the situation at lower energies.  For the short column, at the same observed energy, the radiation is mainly in the X-mode at 20~keV.  For the short column, the gravitational redshift is larger and the Doppler boost is smaller, so an observed energy of 20~keV corresponds to a higher energy in the frame of the flow than in the case of the tall column. 

Turning to Fig.~\ref{fig:finalpulse2} we see that the extent of polarisation in the observed energy range of one to ten keV is between sixty and eighty percent in the pulse where the bulk of the emission from the column arrives. The direction of polarisation swings through the pulse.  In this particular geometry, the swing is nearly ninety degrees.In Paper II we will use the observations of Her X-1 to choose the most appropriate geometry for this particular source. For other pulsars, an examination of the pulse profile as a function of energy can be used to determine the likely geometries and make predictions of the observed polarisation. In general, the observed emission is more pulsed at higher energies and the two-column configurations are less pulsed than the single-column configurations; however, the geometry of the particular system will play a strong role in this as well.  Both Fig.~\ref{fig:finalpulse} and Fig.~\ref{fig:finalpulse2} include all the effects discussed in the paper, while the orange dash-and-dotted lines of Fig.~\ref{fig:finalpulse3} show the polarisation fraction predicted when we do not include QED. The effect of QED, which is almost only caused by the QT effect, reduces the observed polarisation outside of the main pulse.

So far we have discussed only the emission from the accretion column which clearly dominates the emission at certain phases of the rotation.  The column is not the only place on the star where we expect emission, and, furthermore, we expect emission from the accretion disc itself.  For the latter we encourage the reader to consult the literature  \citep[e.g.][and references therein]{1960ratr.book.....C,2009apj...703..569d,2009apj...701.1175s,2010apj...712..908s,2018phrvd..97h3001c} for details about the expected polarisation from the disc.  Generally speaking, for an edge-on configuration, the polarisation fraction peaks at about eleven percent with the polarisation direction parallel to the plane of the disc. At the rotation phases in which the emission from the disc is important compared to the emission from the star, it will therefore dilute the observed polarisation from the values shown in Fig.~\ref{fig:finalpulse2}. Much of the disk emission is expected to be constant in rotational phase; however, radiation may be scattered off the disc from the accretion column for certain geometries. In this case, we would expect the scattered radiation to be polarised in the same direction as the pulse for a scattered component close in phase to the pulse and polarised perpendicular to the disk for scattered components far from the pulse. The details will depend on the geometry of the pulsar and the accretion disc itself.

A second important contribution to the emission will come from the rest of the surface of the neutron star. As we have shown, the radiation from the accretion column is beamed downward by special relativity and pulled toward the stellar surface by gravity so we expect a substantial portion of the surface to be illuminated by the column.  For the taller column, this portion is larger.  \citet{1985ApJ...299..138M} examined a hot (7~keV), low-density slab illuminated by low energy photons (less than 1~keV).  The outgoing emission is strongly polarised in the O-mode up to 10~keV and essentially unpolarised above that energy until the cyclotron energy of 38~keV, where it becomes mildly polarised in the X-mode. In this case, the illuminating photons are boosted in energy through Comptonisation off of the slab. There is also the possibility of the opposite limit to be true, where the illuminating photons are absorbed and heat the slab. In this case, the emission is essentially the same as from an atmosphere heated from below, so we expect the opposite trend: strong polarisation in the X-mode at low energies, decreasing toward the cyclotron energy \cite[see][and references therein]{2019ASSL..460..301C}.  Thus, polarisation will yield a key diagnostic. Evidence that would hint for one situation over the other comes from the typical energy of the reprocessed radiation. If it is similar to or larger than for the radiation in the pulse, it means that the radiation has been Comptonised and we expect the polarisation to be in the O-mode below 10~keV.  If the reprocessed radiation is softer than the pulse, we expect a X-mode polarisation below 10~keV. Of course, the details of this effect will be different from source to source, and we will discuss the particular case of Her X-1 in further detail in Paper II.



\section*{Acknowledgements}

We would like to thank Michael Wolff, Peter Becker and the XMAG collaboration for valuable input, and Sterl Phinney for useful comments. The research was supported by NSERC Canada, Compute Canada, a Burke Fellowship at Caltech and a Four-Year Fellowship at UBC.




\bibliographystyle{mnras}
\bibliography{main} 

\begin{thebibliography}{}
\makeatletter
\relax
\def\mn@urlcharsother{\let\do\@makeother \do\$\do\&\do\#\do\^\do\_\do\%\do\~}
\def\mn@doi{\begingroup\mn@urlcharsother \@ifnextchar [ {\mn@doi@}
  {\mn@doi@[]}}
\def\mn@doi@[#1]#2{\def\@tempa{#1}\ifx\@tempa\@empty \href
  {http://dx.doi.org/#2} {doi:#2}\else \href {http://dx.doi.org/#2} {#1}\fi
  \endgroup}
\def\mn@eprint#1#2{\mn@eprint@#1:#2::\@nil}
\def\mn@eprint@arXiv#1{\href {http://arxiv.org/abs/#1} {{\tt arXiv:#1}}}
\def\mn@eprint@dblp#1{\href {http://dblp.uni-trier.de/rec/bibtex/#1.xml}
  {dblp:#1}}
\def\mn@eprint@#1:#2:#3:#4\@nil{\def\@tempa {#1}\def\@tempb {#2}\def\@tempc
  {#3}\ifx \@tempc \@empty \let \@tempc \@tempb \let \@tempb \@tempa \fi \ifx
  \@tempb \@empty \def\@tempb {arXiv}\fi \@ifundefined
  {mn@eprint@\@tempb}{\@tempb:\@tempc}{\expandafter \expandafter \csname
  mn@eprint@\@tempb\endcsname \expandafter{\@tempc}}}

\bibitem[\protect\citeauthoryear{{Amnu{\'e}l'} \& {Guseinov}}{{Amnu{\'e}l'} \&
  {Guseinov}}{1970}]{1970Ap......6..214A}
{Amnu{\'e}l'} P.~R.,  {Guseinov} O.~K.,  1970, \mn@doi [Astrophysics]
  {10.1007/BF01002657}, \href
  {https://ui.adsabs.harvard.edu/abs/1970Ap......6..214A} {6, 214}

\bibitem[\protect\citeauthoryear{{Basko} \& {Sunyaev}}{{Basko} \&
  {Sunyaev}}{1975}]{1975A&A....42..311B}
{Basko} M.~M.,  {Sunyaev} R.~A.,  1975, \aap, \href
  {https://ui.adsabs.harvard.edu/abs/1975A&A....42..311B} {42, 311}

\bibitem[\protect\citeauthoryear{{Basko} \& {Sunyaev}}{{Basko} \&
  {Sunyaev}}{1976}]{1976MNRAS.175..395B}
{Basko} M.~M.,  {Sunyaev} R.~A.,  1976, \mn@doi [\mnras]
  {10.1093/mnras/175.2.395}, \href
  {https://ui.adsabs.harvard.edu/abs/1976MNRAS.175..395B} {175, 395}

\bibitem[\protect\citeauthoryear{{Becker}}{{Becker}}{1998}]{1998ApJ...498..790B}
{Becker} P.~A.,  1998, \mn@doi [\apj] {10.1086/305568}, \href
  {https://ui.adsabs.harvard.edu/abs/1998ApJ...498..790B} {498, 790}

\bibitem[\protect\citeauthoryear{{Becker} \& {Wolff}}{{Becker} \&
  {Wolff}}{2005}]{2005ApJ...621L..45B}
{Becker} P.~A.,  {Wolff} M.~T.,  2005, \mn@doi [\apjl] {10.1086/428927}, \href
  {https://ui.adsabs.harvard.edu/abs/2005ApJ...621L..45B} {621, L45}

\bibitem[\protect\citeauthoryear{{Becker} \& {Wolff}}{{Becker} \&
  {Wolff}}{2007}]{2007ApJ...654..435B}
{Becker} P.~A.,  {Wolff} M.~T.,  2007, \mn@doi [\apj] {10.1086/509108}, \href
  {https://ui.adsabs.harvard.edu/abs/2007ApJ...654..435B} {654, 435}

\bibitem[\protect\citeauthoryear{{Beilicke} et~al.,}{{Beilicke}
  et~al.}{2014}]{2014JAI.....340008B}
{Beilicke} M.,  et~al., 2014, \mn@doi [Journal of Astronomical Instrumentation]
  {10.1142/S225117171440008X}, \href
  {http://adsabs.harvard.edu/abs/2014JAI.....340008B} {3, 1440008}

\bibitem[\protect\citeauthoryear{{Beloborodov}}{{Beloborodov}}{2002}]{2002ApJ...566L..85B}
{Beloborodov} A.~M.,  2002, \mn@doi [\apjl] {10.1086/339511}, \href
  {https://ui.adsabs.harvard.edu/abs/2002ApJ...566L..85B} {566, L85}

\bibitem[\protect\citeauthoryear{{Burnard}, {Arons}  \& {Klein}}{{Burnard}
  et~al.}{1991}]{1991ApJ...367..575B}
{Burnard} D.~J.,  {Arons} J.,   {Klein} R.~I.,  1991, \mn@doi [\apj]
  {10.1086/169653}, \href
  {https://ui.adsabs.harvard.edu/abs/1991ApJ...367..575B} {367, 575}

\bibitem[\protect\citeauthoryear{{Caiazzo} \& {Heyl}}{{Caiazzo} \&
  {Heyl}}{2018}]{2018phrvd..97h3001c}
{Caiazzo} I.,  {Heyl} J.,  2018, \mn@doi [\prd] {10.1103/PhysRevD.97.083001},
  \href {http://adsabs.harvard.edu/abs/2018PhRvD..97h3001C} {97, 083001}

\bibitem[\protect\citeauthoryear{{Caiazzo}, {Heyl}  \& {Turolla}}{{Caiazzo}
  et~al.}{2019}]{2019ASSL..460..301C}
{Caiazzo} I.,  {Heyl} J.,   {Turolla} R.,  2019, {Polarimetry of Magnetars and
  Isolated Neutron Stars}.
p.~301, \mn@doi{10.1007/978-3-030-19715-5_12}

\bibitem[\protect\citeauthoryear{{Chandrasekhar}}{{Chandrasekhar}}{1960}]{1960ratr.book.....C}
{Chandrasekhar} S.,  1960, {Radiative transfer}

\bibitem[\protect\citeauthoryear{{Chauvin} et~al.,}{{Chauvin}
  et~al.}{2018}]{2018MNRAS.tmpl..30C}
{Chauvin} M.,  et~al., 2018, \mn@doi [\mnras] {10.1093/mnrasl/sly027}, \href
  {http://adsabs.harvard.edu/abs/2018MNRAS.tmpL..30C} {77, L45–L49}

\bibitem[\protect\citeauthoryear{{Chiu} \& {Canuto}}{{Chiu} \&
  {Canuto}}{1969}]{1969PhRvL..22..415C}
{Chiu} H.-Y.,  {Canuto} V.,  1969, \mn@doi [\prl] {10.1103/PhysRevLett.22.415},
  \href {https://ui.adsabs.harvard.edu/abs/1969PhRvL..22..415C} {22, 415}

\bibitem[\protect\citeauthoryear{{Chou}}{{Chou}}{1986}]{1986Ap&SS.121..333C}
{Chou} C.~K.,  1986, \mn@doi [\apss] {10.1007/BF00653705}, \href
  {https://ui.adsabs.harvard.edu/abs/1986Ap&SS.121..333C} {121, 333}

\bibitem[\protect\citeauthoryear{{Daishido}}{{Daishido}}{1975}]{1975PASJ...27..181D}
{Daishido} T.,  1975, \pasj, \href
  {https://ui.adsabs.harvard.edu/abs/1975PASJ...27..181D} {27, 181}

\bibitem[\protect\citeauthoryear{{Davidson} \& {Ostriker}}{{Davidson} \&
  {Ostriker}}{1973}]{1973ApJ...179..585D}
{Davidson} K.,  {Ostriker} J.~P.,  1973, \mn@doi [\apj] {10.1086/151897}, \href
  {https://ui.adsabs.harvard.edu/abs/1973ApJ...179..585D} {179, 585}

\bibitem[\protect\citeauthoryear{{Davis}, {Blaes}, {Hirose}  \&
  {Krolik}}{{Davis} et~al.}{2009}]{2009apj...703..569d}
{Davis} S.~W.,  {Blaes} O.~M.,  {Hirose} S.,   {Krolik} J.~H.,  2009, \mn@doi
  [\apj] {10.1088/0004-637X/703/1/569}, \href
  {http://adsabs.harvard.edu/abs/2009ApJ...703..569D} {703, 569}

\bibitem[\protect\citeauthoryear{{Doroshenko}, {Tsygankov}, {Mushtukov},
  {Lutovinov}, {Santangelo}, {Suleimanov}  \& {Poutanen}}{{Doroshenko}
  et~al.}{2017}]{2017MNRAS.466.2143D}
{Doroshenko} V.,  {Tsygankov} S.~S.,  {Mushtukov} A. e.~A.,  {Lutovinov} A.~A.,
   {Santangelo} A.,  {Suleimanov} V.~F.,   {Poutanen} J.,  2017, \mn@doi
  [\mnras] {10.1093/mnras/stw3236}, \href
  {https://ui.adsabs.harvard.edu/abs/2017MNRAS.466.2143D} {466, 2143}

\bibitem[\protect\citeauthoryear{{Farinelli}, {Ferrigno}, {Bozzo}  \&
  {Becker}}{{Farinelli} et~al.}{2016}]{2016A&A...591A..29F}
{Farinelli} R.,  {Ferrigno} C.,  {Bozzo} E.,   {Becker} P.~A.,  2016, \mn@doi
  [\aap] {10.1051/0004-6361/201527257}, \href
  {https://ui.adsabs.harvard.edu/abs/2016A&A...591A..29F} {591, A29}

\bibitem[\protect\citeauthoryear{{Feng} et~al.,}{{Feng}
  et~al.}{2019}]{2019ExA....47..225F}
{Feng} H.,  et~al., 2019, \mn@doi [Experimental Astronomy]
  {10.1007/s10686-019-09625-z}, \href
  {https://ui.adsabs.harvard.edu/abs/2019ExA....47..225F} {47, 225}

\bibitem[\protect\citeauthoryear{{Feng} et~al.,}{{Feng}
  et~al.}{2020}]{2020NatAs.tmp..100F}
{Feng} H.,  et~al., 2020, \mn@doi [Nature Astronomy]
  {10.1038/s41550-020-1088-1}, \href
  {https://ui.adsabs.harvard.edu/abs/2020NatAs.tmp..100F} {4, 511}

\bibitem[\protect\citeauthoryear{{Gaenther}, {Egan}, {Heilmann}, {Heine},
  {Hellickson}, {Frost}, {Schulz}  \& {Theriault-Shay}}{{Gaenther}
  et~al.}{2017}]{SPIE_REDSoX}
{Gaenther} H.~M.,  {Egan} M.,  {Heilmann} R.~K.,  {Heine} S. N.~T.,
  {Hellickson} T.,  {Frost} J.and~{Marshall} H.~L.,  {Schulz} N.~S.,
  {Theriault-Shay} A.,  2017. pp 10399 -- 10399 -- 13,
  \mn@doi{10.1117/12.2273772}, \url {http://dx.doi.org/10.1117/12.2273772}

\bibitem[\protect\citeauthoryear{{Giacconi}, {Gursky}, {Paolini}  \&
  {Rossi}}{{Giacconi} et~al.}{1962}]{1962PhRvL...9..439G}
{Giacconi} R.,  {Gursky} H.,  {Paolini} F.~R.,   {Rossi} B.~B.,  1962, \mn@doi
  [Physical Review Letters] {10.1103/PhysRevLett.9.439}, \href
  {https://ui.adsabs.harvard.edu/abs/1962PhRvL...9..439G} {9, 439}

\bibitem[\protect\citeauthoryear{{Giacconi}, {Gursky}, {Kellogg}, {Schreier}
  \& {Tananbaum}}{{Giacconi} et~al.}{1971}]{1971ApJ...167L..67G}
{Giacconi} R.,  {Gursky} H.,  {Kellogg} E.,  {Schreier} E.,   {Tananbaum} H.,
  1971, \mn@doi [\apj] {10.1086/180762}, \href
  {https://ui.adsabs.harvard.edu/abs/1971ApJ...167L..67G} {167, L67}

\bibitem[\protect\citeauthoryear{{Gnedin} \& {Sunyaev}}{{Gnedin} \&
  {Sunyaev}}{1973}]{1973A&A....25..233G}
{Gnedin} Y.~N.,  {Sunyaev} R.~A.,  1973, \aap, \href
  {https://ui.adsabs.harvard.edu/abs/1973A&A....25..233G} {25, 233}

\bibitem[\protect\citeauthoryear{{Gnedin}, {Pavlov}  \& {Shibanov}}{{Gnedin}
  et~al.}{1978}]{1978SvAL....4..117G}
{Gnedin} Y.~N.,  {Pavlov} G.~G.,   {Shibanov} Y.~A.,  1978, Soviet Astronomy
  Letters, \href {https://ui.adsabs.harvard.edu/abs/1978SvAL....4..117G} {4,
  117}

\bibitem[\protect\citeauthoryear{{Heisenberg} \& {Euler}}{{Heisenberg} \&
  {Euler}}{1936}]{1936...heisen...euler}
{Heisenberg} W.,  {Euler} H.,  1936, {Zeitschrift f\"ur Physik}, 98, 714

\bibitem[\protect\citeauthoryear{{Herold}}{{Herold}}{1979}]{1979PhRvD..19.2868H}
{Herold} H.,  1979, \mn@doi [\prd] {10.1103/PhysRevD.19.2868}, \href
  {https://ui.adsabs.harvard.edu/\#abs/1979PhRvD..19.2868H} {19, 2868}

\bibitem[\protect\citeauthoryear{{Heyl} \& {Caiazzo}}{{Heyl} \&
  {Caiazzo}}{2018}]{2018galax...6...76h}
{Heyl} J.,  {Caiazzo} I.,  2018, \mn@doi [Galaxies] {10.3390/galaxies6030076},
  \href {http://adsabs.harvard.edu/abs/2018Galax...6...76H} {6, 76}

\bibitem[\protect\citeauthoryear{{Jahoda} et~al.,}{{Jahoda}
  et~al.}{2019}]{2019arXiv190710190J}
{Jahoda} K.,  et~al., 2019, arXiv e-prints, \href
  {https://ui.adsabs.harvard.edu/abs/2019arXiv190710190J} {p. arXiv:1907.10190}

\bibitem[\protect\citeauthoryear{{Kaminker}, {Pavlov}  \&
  {Shibanov}}{{Kaminker} et~al.}{1982}]{1982Ap&SS..86..249K}
{Kaminker} A.~D.,  {Pavlov} G.~G.,   {Shibanov} I.~A.,  1982, \mn@doi [\apss]
  {10.1007/BF00683336}, \href
  {https://ui.adsabs.harvard.edu/abs/1982Ap&SS..86..249K} {86, 249}

\bibitem[\protect\citeauthoryear{{Kii}}{{Kii}}{1987}]{1987pasj...39..781k}
{Kii} T.,  1987, \pasj, \href
  {http://adsabs.harvard.edu/abs/1987PASJ...39..781K} {39, 781}

\bibitem[\protect\citeauthoryear{{Kii}, {Hayakawa}, {Nagase}, {Ikegami}  \&
  {Kawai}}{{Kii} et~al.}{1986}]{1986PASJ...38..751K}
{Kii} T.,  {Hayakawa} S.,  {Nagase} F.,  {Ikegami} T.,   {Kawai} N.,  1986,
  \pasj, \href {https://ui.adsabs.harvard.edu/abs/1986PASJ...38..751K} {38,
  751}

\bibitem[\protect\citeauthoryear{{Klein}, {Arons}, {Jernigan}  \&
  {Hsu}}{{Klein} et~al.}{1996}]{1996ApJ...457L..85K}
{Klein} R.~I.,  {Arons} J.,  {Jernigan} G.,   {Hsu} J. J.~L.,  1996, \mn@doi
  [\apj] {10.1086/309897}, \href
  {https://ui.adsabs.harvard.edu/abs/1996ApJ...457L..85K} {457, L85}

\bibitem[\protect\citeauthoryear{{Krawczynski} et~al.,}{{Krawczynski}
  et~al.}{2019}]{2019arXiv190409313K}
{Krawczynski} H.,  et~al., 2019, arXiv e-prints, \href
  {https://ui.adsabs.harvard.edu/abs/2019arXiv190409313K} {p. arXiv:1904.09313}

\bibitem[\protect\citeauthoryear{{Lauer}, {Herold}, {Ruder}  \&
  {Wunner}}{{Lauer} et~al.}{1983}]{1983JPhB...16.3673L}
{Lauer} J.,  {Herold} H.,  {Ruder} H.,   {Wunner} G.,  1983, \mn@doi [Journal
  of Physics B Atomic Molecular Physics] {10.1088/0022-3700/16/19/024}, \href
  {https://ui.adsabs.harvard.edu/abs/1983JPhB...16.3673L} {16, 3673}

\bibitem[\protect\citeauthoryear{{M{\'e}sz{\'a}ros}}{{M{\'e}sz{\'a}ros}}{1992}]{1992hrfm.book.....M}
{M{\'e}sz{\'a}ros} P.,  1992, {High-energy radiation from magnetized neutron
  stars.}.
U. Chicago Press

\bibitem[\protect\citeauthoryear{{M{\'e}sz{\'a}ros} \&
  {Nagel}}{{M{\'e}sz{\'a}ros} \& {Nagel}}{1985a}]{1985ApJ...298..147M}
{M{\'e}sz{\'a}ros} P.,  {Nagel} W.,  1985a, \mn@doi [\apj] {10.1086/163594},
  \href {http://adsabs.harvard.edu/abs/1985ApJ...298..147M} {298, 147}

\bibitem[\protect\citeauthoryear{{M{\'e}sz{\'a}ros} \&
  {Nagel}}{{M{\'e}sz{\'a}ros} \& {Nagel}}{1985b}]{1985ApJ...299..138M}
{M{\'e}sz{\'a}ros} P.,  {Nagel} W.,  1985b, \mn@doi [\apj] {10.1086/163687},
  \href {https://ui.adsabs.harvard.edu/abs/1985ApJ...299..138M} {299, 138}

\bibitem[\protect\citeauthoryear{{Meszaros}, {Nagel}  \& {Ventura}}{{Meszaros}
  et~al.}{1980}]{1980ApJ...238.1066M}
{Meszaros} P.,  {Nagel} W.,   {Ventura} J.,  1980, \mn@doi [\apj]
  {10.1086/158073}, \href
  {https://ui.adsabs.harvard.edu/abs/1980ApJ...238.1066M} {238, 1066}

\bibitem[\protect\citeauthoryear{{Mihalas}}{{Mihalas}}{1978}]{1978stat.book.....M}
{Mihalas} D.,  1978, {Stellar atmospheres}

\bibitem[\protect\citeauthoryear{{Mushtukov}, {Suleimanov}, {Tsygankov}  \&
  {Poutanen}}{{Mushtukov} et~al.}{2015a}]{2015MNRAS.447.1847M}
{Mushtukov} A.~A.,  {Suleimanov} V.~F.,  {Tsygankov} S.~S.,   {Poutanen} J.,
  2015a, \mn@doi [\mnras] {10.1093/mnras/stu2484}, \href
  {https://ui.adsabs.harvard.edu/abs/2015MNRAS.447.1847M} {447, 1847}

\bibitem[\protect\citeauthoryear{{Mushtukov}, {Suleimanov}, {Tsygankov}  \&
  {Poutanen}}{{Mushtukov} et~al.}{2015b}]{2015MNRAS.454.2539M}
{Mushtukov} A.~A.,  {Suleimanov} V.~F.,  {Tsygankov} S.~S.,   {Poutanen} J.,
  2015b, \mn@doi [\mnras] {10.1093/mnras/stv2087}, \href
  {https://ui.adsabs.harvard.edu/abs/2015MNRAS.454.2539M} {454, 2539}

\bibitem[\protect\citeauthoryear{{Nagel}}{{Nagel}}{1981a}]{1981ApJ...251..278N}
{Nagel} W.,  1981a, \mn@doi [\apj] {10.1086/159463}, \href
  {https://ui.adsabs.harvard.edu/abs/1981ApJ...251..278N} {251, 278}

\bibitem[\protect\citeauthoryear{{Nagel}}{{Nagel}}{1981b}]{1981ApJ...251..288N}
{Nagel} W.,  1981b, \mn@doi [\apj] {10.1086/159464}, \href
  {https://ui.adsabs.harvard.edu/abs/1981ApJ...251..288N} {251, 288}

\bibitem[\protect\citeauthoryear{{Orlandini}}{{Orlandini}}{2006}]{2006AdSpR..38.2742O}
{Orlandini} M.,  2006, \mn@doi [Advances in Space Research]
  {10.1016/j.asr.2006.04.026}, \href
  {https://ui.adsabs.harvard.edu/abs/2006AdSpR..38.2742O} {38, 2742}

\bibitem[\protect\citeauthoryear{{Paul}, {Gopala Krishna}  \& {Puthiya
  Veetil}}{{Paul} et~al.}{2016}]{2016cosp...41E1533P}
{Paul} B.,  {Gopala Krishna} M.~R.,   {Puthiya Veetil} R.,  2016, in 41st
  COSPAR Scientific Assembly. pp E1.15--8--16

\bibitem[\protect\citeauthoryear{{Pringle} \& {Rees}}{{Pringle} \&
  {Rees}}{1972}]{1972A&A....21....1P}
{Pringle} J.~E.,  {Rees} M.~J.,  1972, \aap, \href
  {https://ui.adsabs.harvard.edu/abs/1972A%26A....21....1P} {21, 1}

\bibitem[\protect\citeauthoryear{{Reynolds}, {Quaintrell}, {Still}, {Roche},
  {Chakrabarty}  \& {Levine}}{{Reynolds} et~al.}{1997}]{1997MNRAS.288...43R}
{Reynolds} A.~P.,  {Quaintrell} H.,  {Still} M.~D.,  {Roche} P.,  {Chakrabarty}
  D.,   {Levine} S.~E.,  1997, \mn@doi [\mnras] {10.1093/mnras/288.1.43}, \href
  {https://ui.adsabs.harvard.edu/abs/1997MNRAS.288...43R} {288, 43}

\bibitem[\protect\citeauthoryear{{Rybicki} \& {Lightman}}{{Rybicki} \&
  {Lightman}}{1986}]{1986rpa..book.....R}
{Rybicki} G.~B.,  {Lightman} A.~P.,  1986, {Radiative Processes in
  Astrophysics}

\bibitem[\protect\citeauthoryear{{Schnittman} \& {Krolik}}{{Schnittman} \&
  {Krolik}}{2009}]{2009apj...701.1175s}
{Schnittman} J.~D.,  {Krolik} J.~H.,  2009, \mn@doi [\apj]
  {10.1088/0004-637X/701/2/1175}, \href
  {http://adsabs.harvard.edu/abs/2009ApJ...701.1175S} {701, 1175}

\bibitem[\protect\citeauthoryear{{Schnittman} \& {Krolik}}{{Schnittman} \&
  {Krolik}}{2010}]{2010apj...712..908s}
{Schnittman} J.~D.,  {Krolik} J.~H.,  2010, \mn@doi [\apj]
  {10.1088/0004-637X/712/2/908}, \href
  {http://adsabs.harvard.edu/abs/2010ApJ...712..908S} {712, 908}

\bibitem[\protect\citeauthoryear{Schwinger}{Schwinger}{1951}]{physrev.82.664}
Schwinger J.,  1951, \mn@doi [Phys. Rev.] {10.1103/PhysRev.82.664}, 82, 664

\bibitem[\protect\citeauthoryear{{She} et~al.,}{{She}
  et~al.}{2015}]{2015SPIE.9601E..0IS}
{She} R.,  et~al., 2015, in UV, X-Ray, and Gamma-Ray Space Instrumentation for
  Astronomy XIX. p. 96010I (\mn@eprint {arXiv} {1509.04392}),
  \mn@doi{10.1117/12.2186133}

\bibitem[\protect\citeauthoryear{{Simon}}{{Simon}}{1969}]{1969Natur.224...49S}
{Simon} M.,  1969, \mn@doi [\nat] {10.1038/224049b0}, \href
  {https://ui.adsabs.harvard.edu/abs/1969Natur.224...49S} {224, 49}

\bibitem[\protect\citeauthoryear{{Tananbaum}, {Gursky}, {Kellogg}, {Levinson},
  {Schreier}  \& {Giacconi}}{{Tananbaum} et~al.}{1972}]{1972ApJ...174L.143T}
{Tananbaum} H.,  {Gursky} H.,  {Kellogg} E.~M.,  {Levinson} R.,  {Schreier} E.,
    {Giacconi} R.,  1972, \mn@doi [\apj] {10.1086/180968}, \href
  {https://ui.adsabs.harvard.edu/abs/1972ApJ...174L.143T} {174, L143}

\bibitem[\protect\citeauthoryear{{Veetil} et~al.,}{{Veetil}
  et~al.}{2011}]{2011ASInC...3..165V}
{Veetil} R.~P.,  et~al., 2011, in Astronomical Society of India Conference
  Series. p.~165

\bibitem[\protect\citeauthoryear{{Wang} \& {Lai}}{{Wang} \&
  {Lai}}{2009}]{2009mnras.398..515w}
{Wang} C.,  {Lai} D.,  2009, \mn@doi [\mnras]
  {10.1111/j.1365-2966.2009.14895.x}, \href
  {http://adsabs.harvard.edu/abs/2009MNRAS.398..515W} {398, 515}

\bibitem[\protect\citeauthoryear{{Weisskopf}}{{Weisskopf}}{1936}]{1936...weisskopf}
{Weisskopf} V.,  1936, {Kongelige Danske Videnskabernes Selskab,
  Matematisk-fysiske Meddelelser}, 14, 714

\bibitem[\protect\citeauthoryear{{Weisskopf} et~al.}{{Weisskopf}
  et~al.}{2016}]{2016SPIE.9905E..17W}
{Weisskopf} M.~C.,  et~al., 2016, in Space Telescopes and Instrumentation 2016:
  Ultraviolet to Gamma Ray. p. 990517, \mn@doi{10.1117/12.2235240}

\bibitem[\protect\citeauthoryear{{West}, {Wolfram}  \& {Becker}}{{West}
  et~al.}{2017a}]{2017ApJ...835..129W}
{West} B.~F.,  {Wolfram} K.~D.,   {Becker} P.~A.,  2017a, \mn@doi [\apj]
  {10.3847/1538-4357/835/2/129}, \href
  {https://ui.adsabs.harvard.edu/abs/2017ApJ...835..129W} {835, 129}

\bibitem[\protect\citeauthoryear{{West}, {Wolfram}  \& {Becker}}{{West}
  et~al.}{2017b}]{2017ApJ...835..130W}
{West} B.~F.,  {Wolfram} K.~D.,   {Becker} P.~A.,  2017b, \mn@doi [\apj]
  {10.3847/1538-4357/835/2/130}, \href
  {https://ui.adsabs.harvard.edu/abs/2017ApJ...835..130W} {835, 130}

\bibitem[\protect\citeauthoryear{{Wolff} et~al.,}{{Wolff}
  et~al.}{2016}]{2016ApJ...831..194W}
{Wolff} M.~T.,  et~al., 2016, \mn@doi [\apj] {10.3847/0004-637X/831/2/194},
  \href {https://ui.adsabs.harvard.edu/abs/2016ApJ...831..194W} {831, 194}

\bibitem[\protect\citeauthoryear{{Yahel}}{{Yahel}}{1980}]{1980ApJ...236..911Y}
{Yahel} R.~Z.,  1980, \mn@doi [\apj] {10.1086/157818}, \href
  {https://ui.adsabs.harvard.edu/abs/1980ApJ...236..911Y} {236, 911}

\bibitem[\protect\citeauthoryear{{Zhang} et~al.}{{Zhang}
  et~al.}{2016}]{2016SPIE.9905E..1QZ}
{Zhang} S.~N.,  et~al., 2016, in Space Telescopes and Instrumentation 2016:
  Ultraviolet to Gamma Ray. p. 99051Q (\mn@eprint {arXiv} {1607.08823}),
  \mn@doi{10.1117/12.2232034}

\makeatother
\end{thebibliography}




\appendix

\section{The Quasi-Tangential Effect}
\label{sec:QT}
We here expand the treatment of \citet{2009mnras.398..515w} of the quasi-tangential (QT) effect to apply it to our case. Not all light rays go through a real tangential point, where the photon wavevector $\mathbf{k}$ is perfectly aligned with the local magnetic field direction; however, there is always a point in the photon path, called QT point, where the angle between $\mathbf{k}$ and $\mathbf{B}$, $\theta_B$, reaches a minimum. The magnetic field around the QT point can be expressed, without loss of generality, as
\begin{equation}
    B_X = \frac{B}{\mathcal{R}}s, \quad B_Y = \varepsilon B
\end{equation}
in the fixed $XYZ$ frame where $\mathbf{\hat{Z}}\parallel\mathbf{\hat{k}}$. Here, $\mathcal{R}$ is the curvature radius of the projected magnetic field line in the $X-Z$ plane and $s$ measures the distance from the QT point along the $Z-$ axis. At the QT point, $s=0$ and $\varepsilon = \sin\theta_B$.
Depending on the strength of the vacuum birefringence at the QT point, the outcome for the polarisation of the photon crossing the point can be different. 

As the photon propagates in the neutron star's magnetosphere, whenever the vacuum birefringence dominates, the photon polarisation modes are decoupled and evolve independently following the local magnetic field lines \citep[see][and references therein]{2019ASSL..460..301C}. \citet{2009mnras.398..515w} introduced a condition which states that the photon modes are decoupled if  $\Gamma_{\rm{ad}} \gg 1$ \citep[eq.~2.12 in][]{2009mnras.398..515w}, where they call $\Gamma_{\rm{ad}}$ the adiabaticity parameter. The value of $\Gamma_{\rm{ad}}$ at the QT point is given by \citep[eq.~3.22 in][]{2009mnras.398..515w}
\begin{equation}
    \Gamma_t \simeq 1.0 \times 10^8 E_1 B^2_{13} \varepsilon^3\mathcal{R}_1
    \label{eq:gammat}
\end{equation}
where $E_1 = E_p/$(1 keV) is the photon energy at the QT point (and therefore it is lower than the energy at emission because of gravitational redshift), $B_{13} = B/(10^{13}$ G) is the magnetic field at the QT point and $\mathcal{R}_1 = \mathcal{R}/$(10 km). \citet{2009mnras.398..515w} show that in both limiting cases of adiabatic ($\Gamma_t\gg1$) and non-adiabatic ($\Gamma_t\ll1$) propagation, the polarisation direction is unchanged when the photon traverses the QT point. The only interesting effect is for the intermediate case, $\Gamma_t\sim1$. In this latter case, even if a photon is in a pure mode prior to QT crossing, it will come out of the QT region in a mixture of the two modes.

\begin{figure}
    \centering
    \includegraphics[width=0.98\columnwidth]{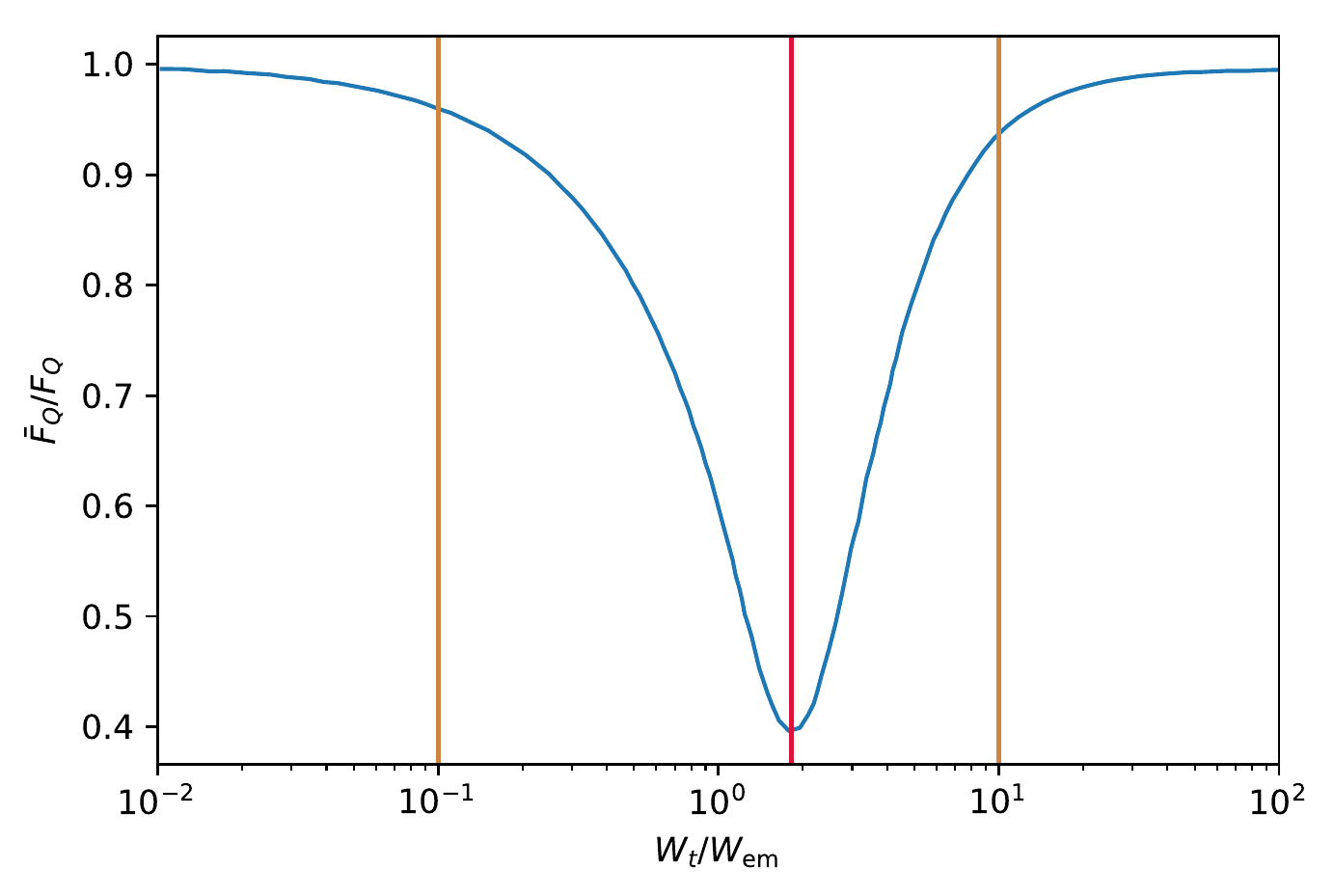}
    \includegraphics[width=0.98\columnwidth]{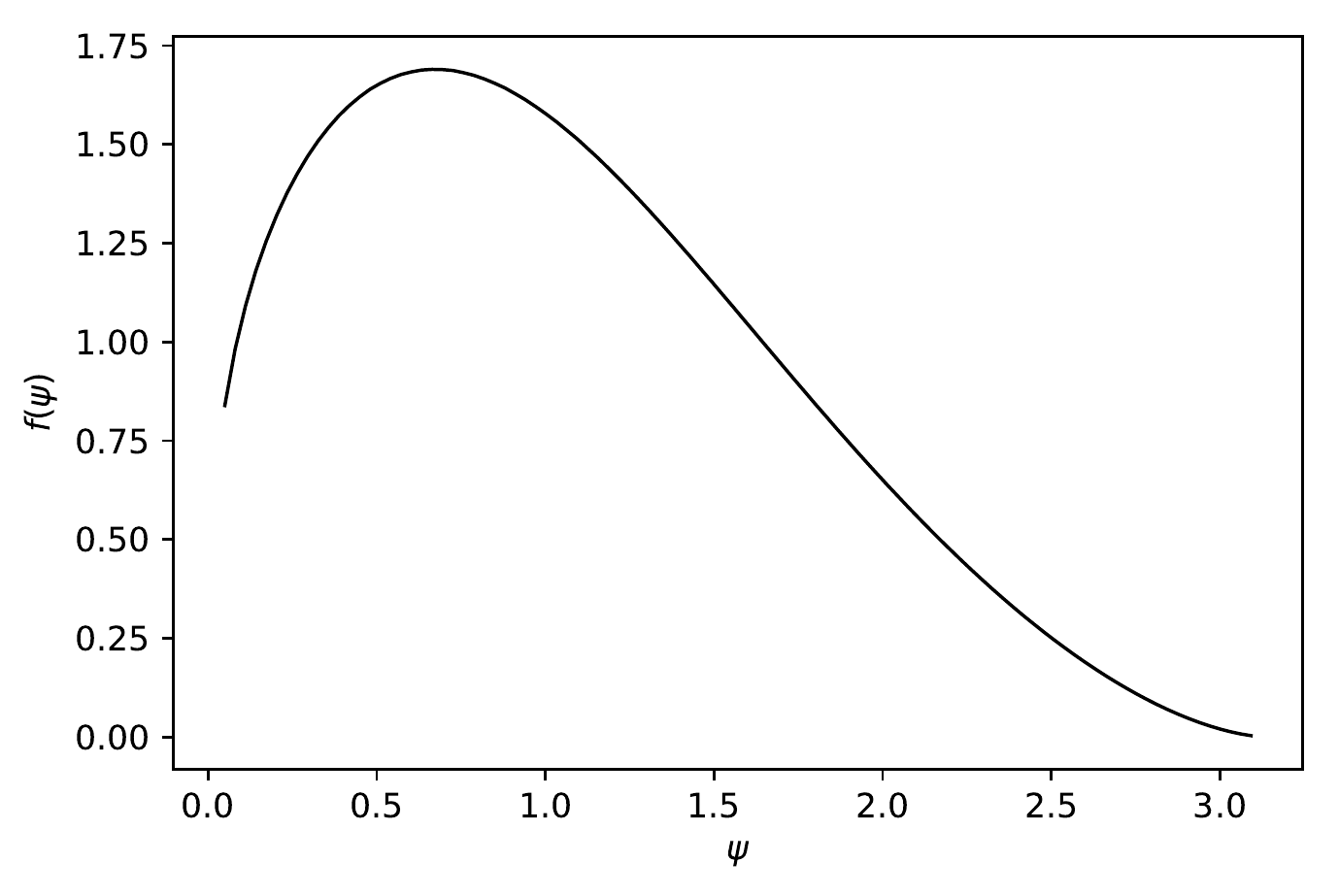}
    \caption{Upper panel: the depolarisation effect of QT propagation on linear polarisation; $W_t$ is the width of the QT effective region, $W_{\rm{em}}$ is the width of the emission region and $F_Q$ ($\bar{F}_Q$) is the polarised radiation flux before (after) traversing the QT region. Same as Figure~11 in \citet{2009mnras.398..515w}. The vertical beige lines highlight the region where the effect is strongest, and the red vertical line pinpoints the peak of the effect, at $W_t/W_{\rm{em}} \sim 1.82$. Lower panel: the function $f(\psi)$, which in \citet{2009mnras.398..515w} is called $f(\theta_{\mu_i})$. }
    \label{fig:QT}
\end{figure}

In the same paper, \citet{2009mnras.398..515w} give a prescription to account for the QT effect in case of a dipolar field. In particular, they calculate the effect on the emission coming from the polar cap. They find that the region in which the QT effect is important is the region where $\Gamma_t\lesssim3$ and the width of this region can be expressed as \citep[eq.~4.32 in][]{2009mnras.398..515w}
\begin{equation}
    W_t \simeq 2.7\times10^{-2} (B_{*13}^2E_1)^{-1/3}f(\psi) R_*
    \label{eq:Wt}
\end{equation}
where $B_*$ is the magnetic field at the pole and $f(\psi)$ is a dimensionless function of the angle between the magnetic axis and the line of sight, which they call $\theta_{\mu_i}$ and we call $\psi$ in this paper (see Fig.~\ref{fig:lensingT}). Once the width of the QT effective region has been calculated, the linearly polarised radiation flux ($\bar{F}_Q$) can be obtained from the ratio between the width of the QT effective region ($W_t$) and the emission region ($W_{\rm{em}}$); the numerical result, taken from Figure~11 of \cite{2009mnras.398..515w}, is shown in the upper panel of Figure~\ref{fig:QT}, where $F_Q$ is the flux of linearly polarised radiation prior to passing the QT region.

In \citet{2009mnras.398..515w}, the authors are only interested in $\psi<90^\circ$, while in the case of an accretion column, $\psi$ can be higher than that. For this reason, we have calculated $f(\psi)$ for all angles. First, we define $r_{\rm{qt}}$ as the distance from the centre of the star to the QT point. We indicate with $\delta$ the angle between the magnetic axis and $r_{\rm{qt}}$, and since the magnetic field is dipolar, the relation between $\delta$ and $\psi$ is given by
\begin{equation}
    \cos^2(\psi - \delta) = \frac{4\cos^2\delta}{3\cos^2\delta+1}
    \label{eq:deltapsi}
\end{equation}
The relation between the impact parameter $b$ and  $r_{\rm{qt}}$ is given by
\begin{equation}
    r_{\rm{qt}}\sin(\psi-\delta) = b = (R_*+z)\sin\psi
\end{equation}
where $z$ is the height above the star of the part of the column that we are considering. This yields
\begin{equation}
    r_{\rm{qt}} = \frac{(R_*+z)\sin\psi}{\sin(\psi-\delta)} \,.
\end{equation}
We can write the dipolar field as \citep[eq.~4.28 of][]{2009mnras.398..515w}
\begin{equation}
    \mathbf{B} = -\frac{\bm{\mu}}{r_{\rm{qt}}^3} + \frac{3\mathbf{r}_{\rm{qt}}}{r_{\rm{qt}}^5}(\bm{\mu}\cdot\mathbf{r}_{\rm{qt}})
\end{equation}
where $r$ is the distance from the centre of the star, and since at the QT point $\mu_y = 0$
\begin{equation}
    B_y = \frac{3y_{\rm{qt}}}{r_{\rm{qt}}^5}\mu r_{\rm{qt}}\,. \cos\delta
\end{equation}
The field strength at the QT point is related to the field strength at the pole by the relation
\begin{equation}
    B = B_*\left( \frac{R_*}{r_{\rm{qt}}} \right)^3 \left( \frac{3\cos^2\delta+1}{4} \right)^{1/2}, \quad \textrm{and}\quad B_* = \frac{2\mu}{r_{\rm{qt}}^3}
\end{equation}
which yields
\begin{equation}
    \varepsilon = \frac{B_y}{B} = \frac{3y_{\rm{qt}}}{r_{\rm{qt}}}\frac{\cos\delta}{(3\cos^2\delta+1)^{1/2}}
\end{equation}
From $\varepsilon$, we can find the width of the QT region to be
\begin{equation}
    \frac{W_t}{R_*} = \frac{2y_{\rm{qt}}}{R_*}=\varepsilon\frac{2r_{\rm{qt}}}{3R_*}\frac{(3\cos^2\delta+1)^{1/2}}{\cos\delta}\,.
\end{equation}
From eq.~\ref{eq:gammat}, we can determine the value of $\varepsilon$ for which $\Gamma_t\lesssim 3$:
\begin{equation}
    \epsilon(\Gamma_t=3) = (3\times10^{-8})^{1/3}*(E_1B^2_{13} \mathcal{R}_1)^{-1/3}\,.
\end{equation}
We still need the radius of curvature $\mathcal{R}$, which, for a dipolar magnetic field, reads
\begin{equation}
    \mathcal{R} = \frac{r_{\rm{qt}}}{3}\frac{(3\cos^2\delta+1)^{3/2}}{|\sin\delta|(\cos^2\delta+1)}\,.
\end{equation}
We finally have all the ingredients to find $f(\psi)$ of eq.~\ref{eq:Wt}
\begin{equation}
    f(\psi) = 7.7\times10^{-2} \left(\frac{r_{\rm{qt}}}{R_*}\right)^{8/3} \left(\frac{10\,\rm{km}}{R_*}\right)^{1/3}\left(\frac{12|\sin\delta|(\cos^2\delta+1)}{\cos^3\delta(3\cos^2\delta+1)}\right)^{1/3}
\end{equation}
where the relation between $\psi$ and $\delta$ is given in eq.~\ref{eq:deltapsi}. This result is shown in the lower panel of Figure~\ref{fig:QT} and it reproduces the function $f(\theta_{\mu_i})$ shown in Figure~10 of \citet{2009mnras.398..515w} for $ \psi = \theta_{\mu_i}  < 90^\circ$, but it extends it to higher angles.


Following this approach, we can calculate the QT effect from the ratio $W_t/W_{\rm{em}}$ for the radiation coming from the column(s), where $W_t$ is the width of the region in which the QT effect is important, given by eq.~\ref{eq:Wt} and $W_{\rm{em}}$ is the width of the emission region. The strength of the effect depends on how far from the star the light ray crosses the QT region, as the magnetic field scales as the distance from the star to the power of $-3$. For this reason, $W_t/W_{\rm{em}}$ depends on both the $z$ coordinate along the column of the emitting region and on the position of the column with respect to the line of sight (which for one column is simply indicated by $\phi$). Additionally,  $W_t/W_{\rm{em}}$ decreases with energy to the power of $-1/3$ (eq.~\ref{eq:Wt}).

\begin{figure*}
    \centering
    \includegraphics[width=0.48\textwidth]{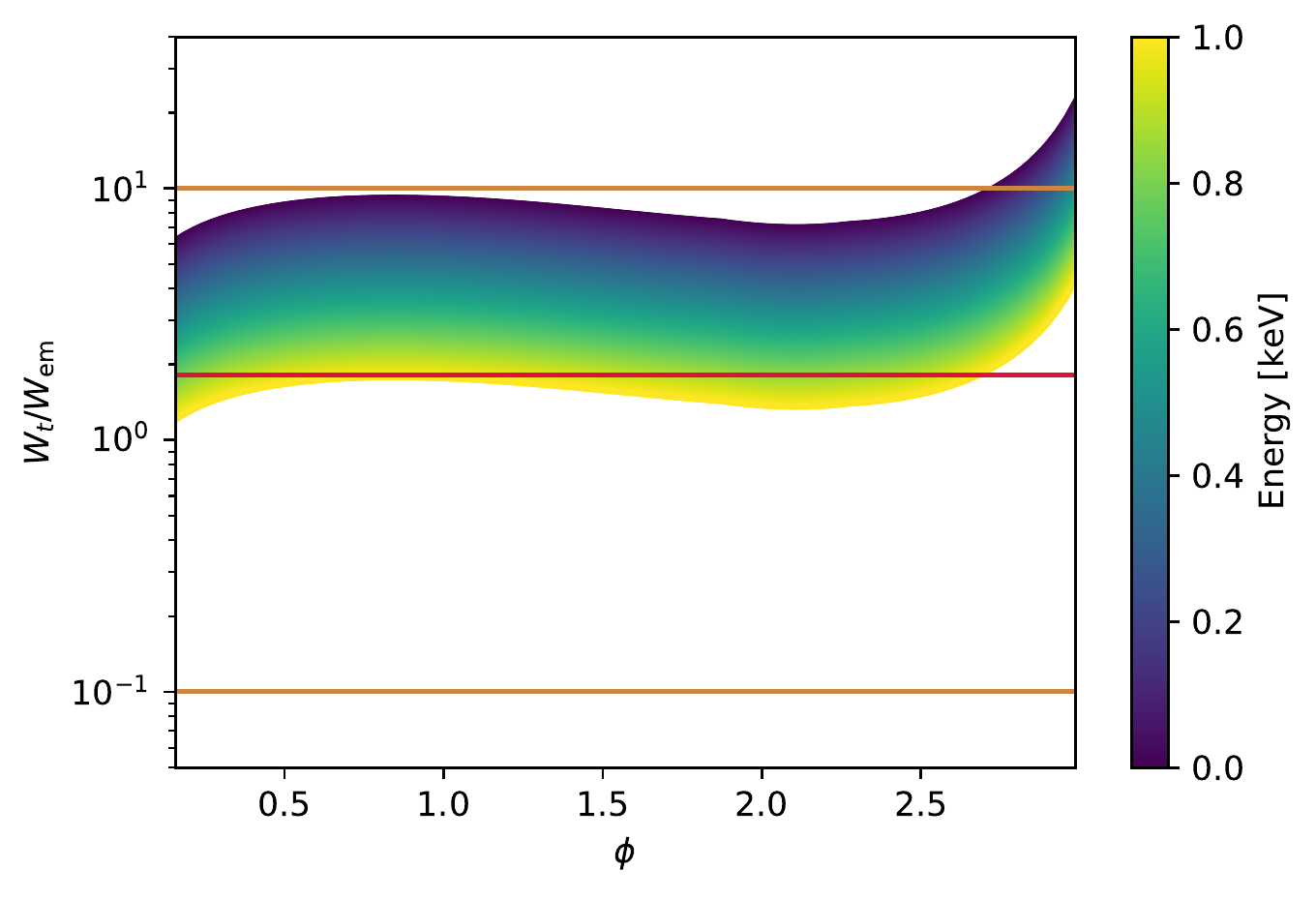}
    \includegraphics[width=0.48\textwidth]{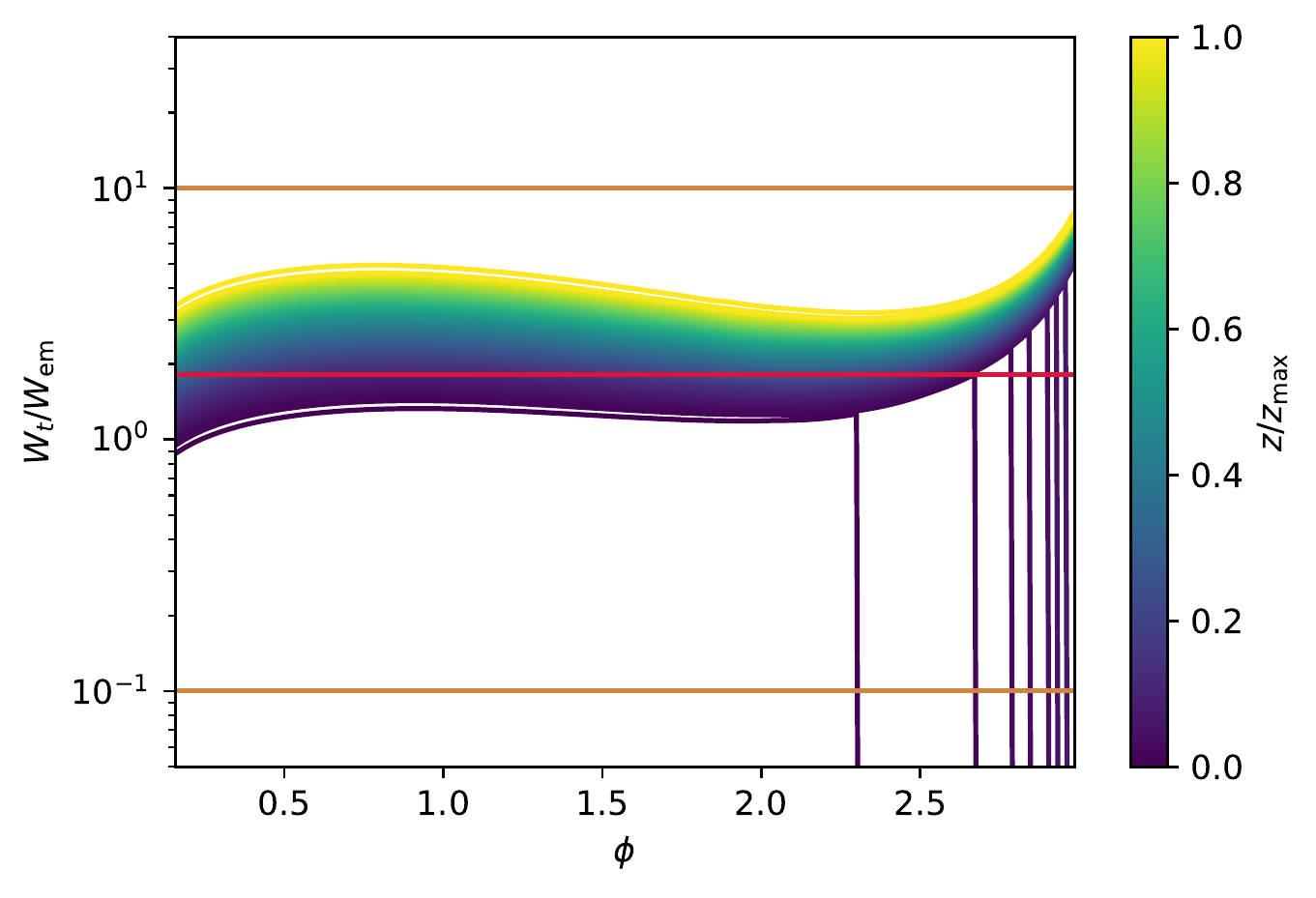}
    \caption{Both panels show $W_t/W_{\rm{em}}$ versus $\phi$. The horizontal lines are the same as the vertical lines in Figure~\ref{fig:QT}: the beige lines delimit the region where the QT effect is strong and the red line indicates the value at which it is the strongest. Left panel: photons are coming from $z = 0.4 z_{\rm{max}} = 2.6$~km above the stellar surface; the different colours represent different photon energies. Right panel: the photon energy is 30~keV; the different colours represent different position in the column, $z$, of the emitting region.}
    \label{fig:WtWem}
\end{figure*}

These dependencies are shown in Figure~\ref{fig:WtWem}. On both panels, the $x$ axis represents the phase $\phi$ and the $y$ axis is $W_t/W_{\rm{em}}$. The horizontal lines are the same as the vertical lines in Figure~\ref{fig:QT}: the beige lines delimit the region where the QT effect is important and the red line indicates the value at which it is the strongest. In the left panel, radiation is coming from a region of the column at $z=0.4~z_{\rm{max}}$ and the different colours depict photons of different energies; we can see the dependence on energy, with higher energies being more affected by the QT propagation effect. On the right panel, photons have the same energy, 30~keV, but come from different altitudes along the column, with yellow lines representing photons coming from the top of the column, and blue lines coming from the bottom. The lower parts of the column are more affected by the QT effect, but are blocked by the star at high $\phi$ (in the plot, it is shown by $W_t/W_{\rm{em}}$ dropping abruptly to zero). 

The effect of QT crossing on the linear polarisation of radiation coming from the accretion column is therefore a reduction of the total linear polarisation fraction that depends on the energy of the photon and on the location along the column of the photon's emission region. The circular polarisation of each photon receives instead a random rotation, and therefore the average circular polarisation of the emission is completely destroyed by the vacuum birefringence.

\section{Changing the geometry}
\label{sec:geo}
\begin{figure}
    \centering
    \includegraphics[width=0.8\columnwidth]{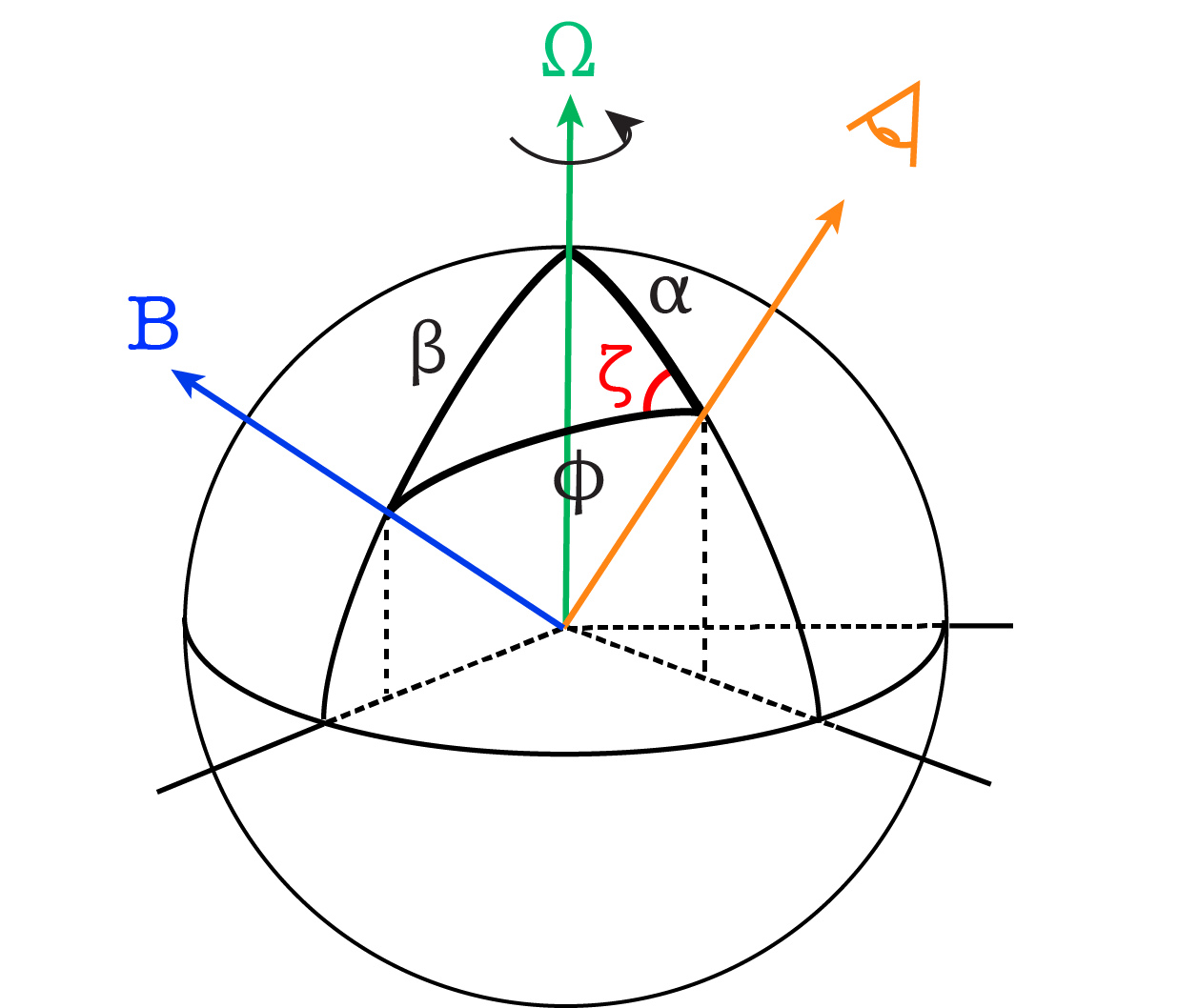}
    \caption{Generic geometry identified by the angle $\alpha$ between the line of sight (in orange) and the rotation axis (in green) and the angle $\beta$ between the rotation axis and the magnetic axis (in blue). The angle $\phi$ between the line of sight and the magnetic axis is the same as in Fig.~\ref{fig:lensingT}. The angle $\zeta$ (in red) between the magnetic axis and the rotation axis projected in the plane of the sky determines the polarisation angle. }
    \label{fig:Geometry}
\end{figure}

From the orthogonal rotator geometry, it is straightforward to derive the emission properties in other geometries. The important angles are illustrated in Fig.~\ref{fig:Geometry}. We have indicated the rotation axis $\hat{\mathbf{\Omega}}$ in green, the magnetic axis in in blue and the line of sight in orange. The angle between the line of sight and the rotation axis $\alpha$ and the angle between the rotation axis and the magnetic axis $\beta$ remain constant as the star rotates. The angle between the line of sight and the magnetic field is what we call $\phi$ in the paper. Additionally, the angle between the magnetic axis and the rotation axis projected in the plane of the sky, $\zeta$, determines the polarisation angle.

The geometry of interest is specified by the angles $\alpha$ and $\beta$. Using the rule of cosines on the sphere, we can derive $\phi$ as a function of the phase $\Omega t$:
\begin{equation}
    \cos\phi=\cos\alpha\cos\beta + \sin\alpha\sin\beta\cos\Omega t\,.
\end{equation}
For the polarisation angle, a positive $Q$ corresponds to linear polarisation in the direction of the magnetic field in the sky, and therefore at an angle $\zeta$ with respect to the projection of the rotation axis on the sky. The polarisation angle in the specific geometry identified by the angles $\alpha$ and $\beta$ can be found as:
\begin{equation}
    \sin\zeta = \frac{\sin\beta \sin\Omega t}{\sin\phi} = \frac{\sin\beta \sin\Omega t}{[1-(\cos\alpha\cos\beta + \sin\alpha\sin\beta\cos\Omega t)^2]^{1/2}}\,.
\end{equation}

\bsp	
\label{lastpage}
\end{document}